\documentclass[fleqn,usenatbib,useAMS]{mnras}

\usepackage{tabularx}

\usepackage{pifont}
\newcommand{\cmark}{\ding{51}}%
\newcommand{\xmark}{\ding{55}}%

\usepackage{tikz}
\usetikzlibrary{arrows,calc,positioning}
\tikzstyle{intt}=[draw,text centered,minimum size=6em,text width=5.25cm,text height=0.34cm]
\tikzstyle{intl}=[draw,text centered,minimum size=2em,text width=2.75cm,text height=0.34cm]
\tikzstyle{int}=[draw,minimum size=2.5em,text centered,text width=3.5cm]
\tikzstyle{intr}=[draw,minimum size=2.5em,text centered,text width=2.5cm,rounded corners]
\tikzstyle{intg}=[draw,minimum size=3em,text centered,text width=6.cm]
\tikzstyle{sum}=[draw,shape=circle,inner sep=2pt,text centered]
\tikzstyle{summ}=[drawshape=circle,inner sep=4pt,text centered]

\usepackage{graphicx}	
\usepackage{caption}
\usepackage{subcaption} 
\usepackage{xcolor}
\usepackage{soul}
\usepackage{amsmath}	
\usepackage{amssymb}	
\usepackage{multicol}        
\usepackage{bm}		
\usepackage{pdflscape}	
\usepackage[detect-all,per-mode=symbol]{siunitx}
\usepackage{ulem}

\usepackage{multirow} 

\DeclareSIUnit\year{yr}

\usepackage[T1]{fontenc}
\usepackage{ae,aecompl}

\usepackage{newtxtext,newtxmath}

\title[Noise and precision pulsar timing]{Identifying and mitigating noise sources in precision pulsar timing data sets}

\author[B. Goncharov et al.]{Boris Goncharov,$^{1,2}$\thanks{E-mail: \href{mailto:boris.goncharov@me.com}{boris.goncharov@me.com}},
D.~J.~Reardon$^{3,2}$,
R.~M.~Shannon$^{3,2}$,
Xing-Jiang Zhu$^{1,2}$,
\newauthor Eric Thrane$^{1,2}$,
M.~Bailes$^{3,2}$,
N.~D.~R.~Bhat$^{4}$,
S.~Dai$^{5}$,
G.~Hobbs$^{5,2}$,
M.~Kerr$^{6}$,
\newauthor R.~N.~Manchester$^{5}$,
S.~Os{\l}owski$^{3,2}$,
A.~Parthasarathy$^{3,2,7}$,
C.~J.~Russell$^{8}$,
\newauthor R.~Spiewak$^{3,2}$,
N.~Thyagarajan$^{9}$,
J.~B.~Wang$^{10}$
\\\\
$^{1}$ School of Physics and Astronomy, Monash University, Clayton, VIC 3800, Australia\\
$^{2}$ Australian Research Council Centre of Excellence for Gravitational Wave Discovery (OzGrav)\\
$^{3}$ Centre for Astrophysics and Supercomputing, Swinburne University of Technology, P.O. Box 218, Hawthorn, Victoria 3122, Australia\\
$^{4}$ International Centre for Radio Astronomy Research, University of Western Australia, Crawley, WA 6009, Australia\\
$^{5}$ CSIRO Astronomy and Space Science, Australia Telescope National Facility, PO~Box~76, Epping NSW~1710, Australia\\
$^{6}$ Space Science Division, Naval Research Laboratory, Washington, DC 20375-5352, USA\\
$^{7}$ Max-Planck-Institut fur Radioastronomie, Auf dem H\"{u}ugel 69, D-53121 Bonn, Germany\\
$^{8}$ CSIRO Scientific Computing, Australian Technology Park, Locked Bag 9013, Alexandria, NSW 1435, Australia\\
$^{9}$ National Radio Astronomy Observatory, 1003 Lopezville Rd, Socorro, NM 87801, USA\\
$^{10}$ Xinjiang Astronomical Observatory, Chinese Academy of Sciences, 150 Science 1-Street, Urumqi, Xinjiang 830011, China\\
}

\pubyear{2020}

\begin{document}
\label{firstpage}
\pagerange{\pageref{firstpage}--\pageref{lastpage}}
\maketitle

\begin{abstract}

Pulsar timing array projects measure the pulse arrival times of millisecond pulsars for the primary purpose of detecting nanohertz-frequency gravitational waves.
The measurements include contributions from a number of astrophysical and instrumental processes, which can either be deterministic or stochastic.
It is necessary to develop robust statistical and physical models for these noise processes because incorrect models diminish sensitivity and may cause a spurious gravitational wave detection. 
Here we characterise noise processes for the $26$ pulsars in the second data release of the Parkes Pulsar Timing Array using Bayesian inference.
In addition to well-studied noise sources found previously in pulsar timing array data sets such as achromatic timing noise and dispersion measure variations, we identify new noise sources including time-correlated chromatic noise that we attribute to variations in pulse scattering.
We also identify ``exponential dip'' events in four pulsars, which we attribute to magnetospheric effects as evidenced by pulse profile shape changes observed for three of the pulsars.
This includes an event in PSR~J1713$+$0747, which had previously been attributed to interstellar propagation.
We present noise models to be used in searches for gravitational waves.
We outline a robust methodology to evaluate the performance of noise models and identify unknown signals in the data.
The detection of variations in pulse profiles highlights the need to develop efficient profile domain timing methods. 
\end{abstract}




\begingroup
\let\clearpage\relax
\endgroup
\newpage

\section{Introduction}
\label{sec:intro}

Pulsar Timing Arrays (PTA) perform measurements of pulse arrival times from millisecond pulsars over the time scales on the order of years~\citep{foster1990pta}.
Benefiting from the long-term timing stability of millisecond pulsars, the arrays are the most sensitive detectors of nanohertz gravitational waves.
\cite{taylor2016timetodetection} predicts that a stochastic gravitational wave background from supermassive black hole binaries will be detected and studied  with pulsar timing arrays in the following decade.
The background would manifest as a red noise process that is  correlated between pulsars \cite[][]{rajagopal1995}. 
In addition to gravitational waves, pulsar timing arrays are sensitive to other correlated signals, including errors in terrestrial time standards~\citep{hobbs2012clock,hobbs2020clock} and solar system ephemerides~\citep{championplanets,iptaplanet,vallisneri2020bayesephem}, and, potentially, ultralight dark matter~\citep{porayko2018darkmatter}.
Data from pulsar timing arrays is used to study a wide range of astrophysical topics including: neutron-star interiors \citep{shannon2010spin,9yrnanograv2016noise} and magnetospheres \citep{shannon2016disturbance}, the interstellar medium \citep{coles2015exscat,nanogravism}, and the solar wind~\citep{you2007solarwind,dustysun}.

There are a number of pulsar timing array projects underway, utilising the most sensitive metre- and centimetre-wavelength radio telescopes. 
The Parkes Pulsar Timing Array~\citep{manchester2013pptadr1} utlises the 64-m Parkes telescope in Australia to monitor $24$ millisecond pulsars.
The first data release of the Parkes Pulsar Timing Array (DR1) has been described in~\cite{manchester2013pptadr1}, while timing properties of DR1 pulsars have been described in~\cite{reardon2016pptatiming}.
The first data release comprises observations between 1994 and 2011.
The project has recently completed a second data release \cite[DR2, ][]{kerr2020pptadr2}, which extends beyond DR1 by 7 years.
Other timing array projects include the European Pulsar Timing Array \cite[EPTA,][]{kramer2013epta} and North-American Nanohertz Gravitational-wave observatory \cite[NANOGrav,][]{mclaughlin2013nanograv}. 
Together, PPTA, EPTA and NANOGrav form the International Pulsar Timing Array \cite[IPTA,][]{hobbs2010ipta}.

In order to effectively search for spatially correlated signals, it is necessary to provide complete models for the arrival time variations of the pulsars.
This includes both deterministic processes encapsulated in the pulsar ephemerides, and stochastic processes.
Otherwise, the estimate of the gravitational-wave signal or other correlated signal could absorb unaccounted features in pulsar noise.
In \cite{shannon2016disturbance}, it was  found that a dip in timing residuals in PSR~J1643$-$1224, when not modelled, affects upper limits on the stochastic gravitational-wave background with 4 years of data by an order of magnitude.
The dip itself is associated with a sudden change of pulse profile.
Moreover, \cite{hazboun2020timeslicing} and \cite{tiburzi2016spacecorr} studied cases where incorrect noise models led to false positives in gravitational wave searches.

Deterministic processes include the non-linear change in the relative distance between the pulsar and the Earth, relativistic propagation effects in the solar system or binary (if the pulsar has a companion) \cite[][]{edwardstempo2}. 
Stochastic terms can be divided phenomenologically into two classes: temporally uncorrelated or correlated  processes.  
The two classes are often referred to respectively as white noise and red noise,  reflecting the shape of their Fourier spectra.
On short time scales (at high fluctuation frequency), pulsar timing observations are dominated by white noise.
The main sources of white noise are radiometer noise and pulse-to-pulse variations of profile shape, referred to as pulse jitter.
\cite{shannon2014jitter} found that the brightest observations of the brightest PPTA pulsars are dominated by jitter noise.
In~\cite{9yrnanograv2016noise} jitter was studied in the context of pulse phase and amplitude, and it has also been found that jitter noise evolves with radio frequency.

On longer time scales (lower flucutation frequencies), pulsar timing arrays are dominated by time-correlated  red noise.
Achromatic red noise, which is not dependent on radio frequency, is referred to as timing noise or spin noise because it is thought to largely be associated with irregularities in the rotation of the neutron star.
\cite{shannon2010spin} suggested that more millisecond pulsars are expected to be dominated by spin noise when observed over longer observing spans, and that scaling relations for spin noise in millisecond pulsars are consistent with those for regular pulsars.
Later, \cite{caballero2016eptanoise} estimated power law parameters of the timing noise in the first data release of the EPTA and found that timing noise reduces the sensitivity of the EPTA to stochastic gravitational waves by a factor of $> 9$.
Low-frequency turnover in the power-law timing noise could potentially stop the deterioration of timing precision on long time scales.
Although marginal evidence for the low-frequency turnover has been found in a power-law spectrum of canonical pulsars~\citep{parthasarathy2019timing}, no presence of a spectral turnover has been found in 49 millisecond pulsars from the first data release of the IPTA~\citep{goncharov2019turnover}.

There is also evidence for chromatic red noise processes in pulsars. The strongest red noise source is thought to be dispersion-measure variations~\citep{keith2013dmvar}, a manifestation of changing column density of ionised plasma along the pulsar-Earth line of sight.
However, other forms of chromatic noise have been identified. In the first data release of the IPTA, \cite{lentati2016iptanoise} identified new band-dependent and system-dependent red noise processes.
Ignoring these components resulted in 60\% less stringent upper limits on the gravitational-wave background.
The origins of these components are unclear, and were speculated to be either related to propagation effects in the interstellar medium or instrument-based systematic errors \citep{cs2010,shannon2017}. 

The PPTA DR2 data set comprises observations for as long as 15 years from $26$ pulsars \citep{kerr2020pptadr2}.
At each epoch (with epochs typically having a three-week cadence) the pulsars were usually observed in three bands:  the 10-cm, 20-cm,  and either the 40-cm or 50-cm. The central radio frequencies of the observations in these bands were close to 3100 MHz, 1370 MHz, 730 MHz and 680 MHz, respectively.   
Before mid-2009, the low frequency observations were conducted at 680~MHz (50-cm band). 
However the presence of digital television necessitated adjusting the observations to shorter-wavelength 40-cm band.    
Additionally, each observation has been performed with one of the following observing processing systems (referred to as backends or signal processors): \texttt{CASPSR}, \texttt{CPSR2},  \texttt{PDFB1}, \texttt{PDFB2}, \texttt{PDFB3}, \texttt{PDFB4}, and \texttt{WBCORR}.
The most notable difference between DR1 and DR2 is the presence of sub-banded arrival times.
Pulse arrival times are provided not only averaged across each band, but also for between 2 and 32 dynamically chosen sub-bands, with the level of sub-banding determined by the signal to noise ratio of the observation.     
The sub-banding allows us to account for profile evolution, which is known to bias arrival time measurements \cite[][]{demorest2013}.
It also allows us to examine chromatic processes in greater detail. 

In this work, we characterise sources of noise in the second data release of the PPTA.
We outline the Bayesian approach to analysis of the data in Section~\ref{sec:method}.
In Section~\ref{sec:models}, we describe noise models in our analysis.
We present our results in Section~\ref{sec:results} and our conclusions in Section~\ref{sec:conclusion}.

\section{Bayesian inference}
\label{sec:method}

Our methods follow those described in sections 2.1 and 2.2 of~\cite{goncharov2019turnover}.
We assume that the data is represented by contributions from deterministic signals, included in  the timing model, and Gaussian stochastic processes.
We Taylor-expand the timing model for each time of arrival (ToA), keeping the linear term $\bm{M \xi}$ and assuming non-linear terms to be negligible.
Here, $\bm{\xi}$ is the vector of timing model parameters and $\bm{M}$ is the design matrix, which represents contribution of the timing model to each measured ToA.
Following~\cite{van2009bayesian1,van2013bayesian2, taylor2017allcorrelations}, we employ the Gaussian likelihood. 
We use the Bayes factor to select which of two given models (A and B), with parameters $\bm{\theta}_{\text{A}}$ and $\bm{\theta}_{\text{B}}$, better explains the data:
\begin{equation}\label{eq:logbf}
\mathcal{B}_{\text{A},i}^{\text{B}} = \frac{\mathcal{Z}^{\text{B}}_i(\bm{\theta}_\text{B}, \bm{\delta t})}{\mathcal{Z}^{\text{A}}_i(\bm{\theta}_{\text{A}}, \bm{\delta t})},~i \in [1,N_\text{psr}]~,
\end{equation}
where $N_{\text{psr}}$ is the number of pulsars, and the function $\mathcal{Z}(\bm{\theta}, \bm{\delta t})$ is the Bayesian evidence for the model,
\begin{equation}\label{eq:evidence}
\mathcal{Z}(\bm{\theta}, \bm{\delta t}) = \int \mathcal{L}(\bm{\delta t}| \bm{\theta}) \pi(\bm{\theta}) d\bm{\theta}.
\end{equation}
It is an integral over the prior volume of the product of the likelihood and the prior probability.

We perform parameter estimation using Markov-chain Monte-Carlo methods.
To calculate the Bayesian evidence for a given model, we perform nested sampling~\citep{skilling2006nested} using \textsc{polychordlite}~\citep{handley2015polychord1,handley2015polychord2}.
For parameter estimation, we sample the likelihood function using \textsc{ptmcmcsampler}~\citep{justin_ellis_ptmcmcsampler}.
We employ \textsc{tempo2}~\citep{edwardstempo2} to fit the deterministic timing model parameters and use \textsc{enterprise}~\citep{ellis2019enterprise} and \textsc{libstempo}~\citep{libstempo} to perform likelihood evaluations.
The \textsc{bilby} package \citep{ashton2019bilby} is used to access \textsc{polychordlite}.
The \textsc{chainconsumer} package, developed by~\cite{Hinton2016Chainconsumer}, is used to plot posterior distributions.

\section{Signal models}
\label{sec:models}
In the following subsections, we describe families of signal models we considered.
The empirical prior distributions are listed in Table~\ref{tab:priors}.
We reference pulse arrival times at the position of Solar System barycentre using ephemeris DE436 and clock TT(BIPM18), which were used in the PPTA DR2 publication.

\subsection{White noise}
\label{sec:white}

We model white noise to be diagonal components $\sigma_j$ of the covariance matrix $\bm{C}$, which contains known contributions from ToA uncertainties $\sigma_j^{\text{ToA}}$ and unknown contributions that we take into account by introducing parameters EFAC, EQUAD and ECORR.
The parameter EFAC modifies the TOA uncertainty while EQUAD adds in quadrature an extra term that is independent of the formal TOA uncertainty. The modified white noise component to the timing noise is then
\begin{equation}\label{eq:efacequad}
\sigma_j^2 = (\text{EFAC} ~\sigma^{\text{ToA}}_j)^2 + \text{EQUAD}^2~.
\end{equation}
Because sub-banded times of arrival are, essentially, observations within a given observation, the parameter ECORR is introduced as an analogue of EQUAD, only to describe the excess variance for groups of sub-banded observations.
ECORR models the white noise in sub-banded data points in one observation independently of the white noise in sub-banded data points in other observations.
The formalism behind ECORR is described in the Appendix C of~\cite{arzoumanian2015nanograv}.
The ECORR parameter can model some of the noise attributed to pulse jitter~\citep{9yrnanograv2016noise}
As these values are expected to be signal-processor- and band-dependent, we assume different white noise terms for each band and all backends.
The exception is for \texttt{PDFB2}, \texttt{PDFB3}, and \texttt{PDFB4}, which have similar hardware architectures (digital polyphase filter banks); for these we assume to have the same white noise properties within a band.


\subsection{Red noise}
\label{sec:rednoise}
We implement frequency-domain models of time-correlated stochastic proceses in the time-domain likelihood function, using a Fourier-sum approach, described in~\cite{lentati2013hyperefficient}.
The red noise component of the covariance matrix $\bm{C}$ is represented in a reduced order as
\begin{equation}\label{eq:getresiduals}
\bm{K} = {\bm{F \Phi  F}}^{\text{T}} ~\Delta f~,
\end{equation}
where $\Phi_i = P(f_i)$ is the power spectral density model of the red noise for each frequency $f_i$ that we include in our model, $\bm{F}$ is the Fourier basis, the matrix that Fourier-transforms frequency-domain power spectral density model into the time domain covariance.
The size of a frequency bin, $\Delta f$, is equal to the inverse of the total observation time for a given pulsar.
The exact form of $\bm{F}$ that we use can be found in Equation 9 in \cite{goncharov2019turnover}.
A Woodbury lemma is then used to simplify the inversion of a covariance matrix, decomposed into $\bm{N}$ and $\bm{K}$ \citep{hager1989woodburylemma,vanhaasteren2014newadvances}.
We use the frequency-domain model for pulsar red noise with power spectral density following a power law in units of [$s^3$]:
\begin{equation}\label{eq:redpl}
    P_{\text{PL}}(f | A, \gamma) = \frac{A^2}{12\pi^2}\left(\frac{f}{ \text{yr}^{-1}}\right)^{-\gamma} \text{yr}^3,
\end{equation}
where $A$ quantifies the amplitude\footnote{The scaling for $A$ is chosen such that it represents the amplitude of the strain spectrum of a stochastic gravitational wave background, measured a frequency of 1\,yr$^{-1}$. } of the power-law, $\gamma$ is the slope of the power-law and yr is the number of seconds in a year.
We discuss three subsets of red noise in this work:
\begin{itemize}
    \item Achromatic spin noise (SN)
    \item Frequency-dependent dispersion measure (DM) noise
    \item Achromatic band noise (BN) and system (``group'') noise (GN)
    \item Frequency-dependent chromatic noise (CN)
\end{itemize}

For each pulsar, the spin noise is a common red-noise process  in all observing systems and bands, across all radio-frequencies.
There are several potential origins for spin noise.
Some studies suggest spin noise to be the consequence of the interaction between the crust and the superfluid core of a neutron star \citep{alpar1986spinnoisevortexcreep, jones1990spinnoisesuperfluid}.
A model that links power-law parameters of a spin noise to physical parameters of such systems has been derived in ~\cite{melatos2013superfluid}.
Other studies attempt to link spin noise and pulsar glitches, sudden jumps in rotational frequency of pulsars \citep{johnston1999glitchrecovery, cordes1985microglitches,dAlessandro1995timingnoise,melatos2008microglitches}.
In \cite{lyne2010switchpdot}, the authors suggested switching between two different spin-down rates as the origin of spin noise.
Some models suggest that the influence of planets \citep{cordes1993spinnoiseplanet}, asteroids \citep{shannon2013spinnoiseasteroid} and possibly unmodeled binary motion~\citep{bassa1024,kaplan1024}.
We denote spin noise parameters $A_\text{SN}$ and $\gamma_\text{SN}$.
We do not attempt to model uncertainties in the solar system ephemeris, which are known to contribute red noise to pulsar timing data sets.
It is likely to affect red noise in the most stable pulsars.
In particular, some of the red noise  in PSR~J1909$-$3744 can be attributed to this, as discussed below.

Stochastic variations in DM~\cite[][]{phillips1991timevardm} are another source of red noise.
We model DM noise as a power-law with parameters $A_\text{DM}$ and $\gamma_\text{DM}$, with Fourier basis components $\bm{F} \propto \kappa_j$, where $\kappa_j = K^2 \nu_j^{-2}$ is introduced to model the dependency of DM noise amplitude on the radio frequency $\nu_j$ of ToA $j$.
We choose $K = 1400$ MHz to be the reference frequency.
A Kolmogorov spectrum for turbulence in a neutral gas is used as a standard model to describe DM variations, $\text{DM}(t)$, in the interstellar medium.
In the case of Kolmogorov turbulence, we would expect  $\gamma_\text{DM} = 8/3$~\citep{rickett1990dmreview}.

We also search for a more general form of red noise which we refer to as chromatic red noise.  
In this case, we have $\kappa_j = K^\chi \nu_j^{-\chi}$, where $\chi$ is a value other than $2$. 
We may refer to $\chi$ as the chromaticity of a red process.
Numerous astrophysical mechanisms can potentially  introduce chromatic red noise.
Scattering variations in the interstellar medium change widths of radio pulses by $\Delta \nu \propto \nu^{4}$~\citep{lyne2012pulsar}.
While a template for pulse profiles does not account for this, the arrival times would temporal variations that would scale with radio frequency proportional to $\nu^{-4}$.
\cite{shannon2017} show through simulation how refractive propagation effects can potentially introduce correlations in arrival times which can have frequency dependencies as steep as $\nu^{-6.4}$.
Simulations of scattering of pulsar radio emission by the interstellar plasma have also been performed by~\cite{coles2010scattering}.

Band noise and system noise are separate red noise processes in a given band or system. 
Both were introduced and discussed in modelling of the first IPTA data release ~\citep{lentati2016iptanoise}.
System noise is attributed to instrumental artifacts, including polarisation calibration errors.
Band noise could potentially be produced by processes incoherent between bands in the interstellar medium, such as frequency dependent dispersion measure variations~\citep{cordes2016freqdependentdm}, frequency-dependent calibration errors~\citep{vanstraten2013polarimetry}, or radio frequency interference.


\begin{figure}
\centering
\includegraphics[width=1.0\linewidth]{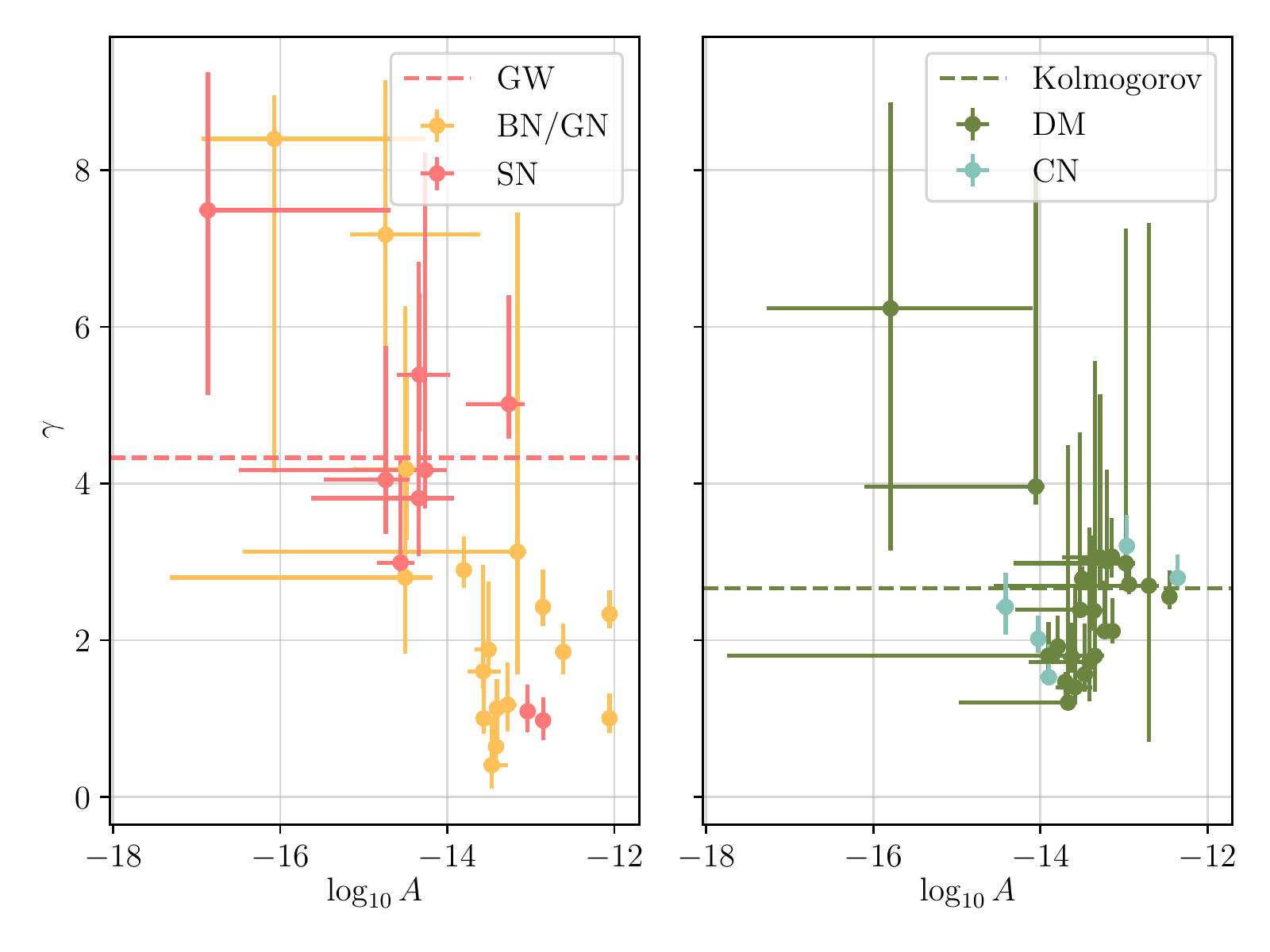}
\caption{
Strength and spectral index for red noise processes for the PPTA-DR2 pulsars. Left panel: spin noise (SN), band noise (BN) and system noise (GN). Right panel: DM noise and chromatic noise (CN) with strength referenced to $K = 1400$ MHz. The main feature of the left panel is the clustering of red noise parameters around two areas of the parameter space: where $\gamma$ is between $3$ and $10$ (mostly spin noise), and where $\gamma$ is between $0$ and $3$ (mostly band noise and system noise). For some pulsars, we found only marginal preference to choose between competing noise models with band and system noise, see Section~\ref{sec:results_spin} for more details.
The green dashed line in the right panel highlights $\gamma = 8/3$, predicted for the standard model of DM variations from Kolmogorov turbulence.
The red dashed line (GW) highlights the spectral index $\gamma = 13/3$, predicted for a red noise process induced by the stochastic gravitational-wave background.
The three pulsars with spin-noise power-law index closest to $13/3$ correspond to the top strongest contributors to the common red noise in~\protect\cite{arzoumanian2020limits}, which are visible from Parkes.
}
\label{fig:red_population}
\end{figure}

\subsection{Deterministic signals}
\label{sec:determ}

To fully model the data we identified  new deterministic signals in the timing model.  We describe these in the following subsections. 

\subsubsection{Chromatic exponential dips}
\label{sec:dmdipn}
Some pulsars show evidence of frequency-dependent events in timing residuals on time scales of a few months.
Some events have been identified  as a sudden advancement in apparent pulse arrival time, followed by an exponential relaxation.
PSR~J1713+0747 shows an exponential discontinuity in timing residuals at around MJD 54757, which has been attributed to the sudden drop in dispersion measure~\citep{coles2015exscat,desvignes2016timing,lentati2016iptanoise}.
In~\cite{lam2018j1713event2}, a second exponential event in PSR~J1713+0747 was reported and also attributed to the interstellar medium.
In~\cite{shannon2016disturbance}, an exponential timing event in PSR~J1643$-$1224 was reported, which had the most pronounced effect at high radio frequency.
This event was connected with a sudden change in the pulse profile.  

We model exponential events in the time domain to be
\begin{equation}\label{eq:model_expdip}
    s_{\text{E}}( t | A_{\text{E}}, t_{\text{E}}, \tau_{\text{E}}, \chi_\text{E}) = K^\chi \nu^{-\chi}
        \begin{cases}
            0 , ~t < t_{\text{E}}~; \\
            A_{\text{E}}~e^{- \frac{t-t_{\text{E}}}{\tau_{\text{E}}} }, ~t \geq t_{\text{E}}~;
        \end{cases}
\end{equation}
where $A_{\text{E}}$ is the amplitude of the event in seconds, $t_\text{E}$ is the time of the event, $\tau_\text{E}$ is the characteristic relaxation time.
The radio-frequency dependence, as for the case of chromatic noise, is treated by the parameter $\chi$, and the amplitude of the event is at a frequency of $1400$ MHz.

\subsubsection{Extreme scattering events}
\label{sec:dmgaussn}

Extreme scattering events (ESEs) have been observed in the direction of a number of pulsars.   The events are manifested as increase in the electron density along the line of sight and diffractive scattering strength. This suggests that the line of sight to the pulsar passed through an over-dense region of the interstellar medium~\citep{coles2015exscat,keith2013dmvar}.
In our sample PSR~J1603$-$7202 has been observed to have experienced an ESE \cite[][]{coles2015exscat}.
We model the dispersion measure variations associated with the event using a Gaussian function: 
\begin{equation}\label{eq:model_gaussian}
    s_{\text{G}}( t | A_{\text{G}}, t_{\text{G}}, \sigma_{\text{G}}) = K^2 \nu^{-2} A_{\text{G}}~e^{- \frac{(t-t_{\text{G}})^2}{2 \sigma_{\text{G}}^2} },
\end{equation}
where $A_\text{G}$ is the amplitude of the Gaussian in seconds at $K = 1400$ MHz, $t_\text{G}$ is the time of the event, $\sigma_\text{G}$ is the width.
There are no measurable arrival time variations from diffractive scintillation during this event. 


\subsubsection{Annual dispersion measure variations}
\label{sec:dmannual}

In case there is a strong gradient in electron column density between the pulsar and the Earth, the motion of the Earth around the sun will cause the gradient to manifest as annual DM variations.
In~\cite{keith2013dmvar} a clear annual modulation in DM  was identified for J0613$-$0200.
We model this effect by including the deterministic yearly sinusoids:
\begin{equation}\label{eq:dmannual}
    s_{\text{Y}}(t | A_\text{Y}, \phi_\text{Y}) = A_\text{Y} K^2 \nu^{-2} \sin( 2 \pi t \times \text{yr} + \phi_\text{Y}),
\end{equation}
with amplitude $A_\text{Y}$ in seconds at $K = 1400$ MHz and dimensionless phase $\phi_\text{Y}$.
The strength of annual DM variations will depend on the two main factors.
The first is the persistence of the annual gradient over time scales much longer than a year.
The second is the mutual orientation of the gradient and the velocity of a pulsar.
The orthogonal orientation of the velocity of the pulsar with respect to the gradient provide the strongest annual DM signal, while the parallel orientation will diminish the signal.
Additionally, contributions to DM from the heliosphere could potentially show up in annual DM.


\subsubsection{System dependent profile evolution}
\label{sec:fdlin}

In PPTA-DR2, arrival times were measured using standard techniques. While different standards were used for each band and most subsystems \cite[see][ for more information]{kerr2020pptadr2}, the templates themselves were one-dimensional.  This necessitated the use of FD parameters, where each parameter represents a log-polynomial term of radio-frequency-dependence of timing residuals for the whole data span~\citep{arzoumanian2015nanograv}.
These terms arise from profile templates not being tailored enough for each radio frequency and pulse profile evolution.
This effect is most prominent in PSR~J0437$-$4715, which exceptional brightness significantly biases estimates of the pulse arrival times.
To account for this, six \texttt{FD} parameters are introduced to the timing model of PSR~J0437$-$4715.
Up to two FD parameters are included in timing models of the remaining DR2 pulsars.
For PSR~J0437$-$4715,~\cite{kerr2020pptadr2} also noticed that pulse profile evolution depends on a given observing system.
In DR2, the dependence of timing residuals on radio-frequency for PSR~J0437-4715 has been subtracted using the model with three \texttt{FD} parameters for specific systems and sub-systems: \texttt{CPSR2\_50CM}, \texttt{CPSR2\_20CM} above $1370$ MHz, \texttt{CPSR2\_20CM} below $1370$ MHz, \texttt{PDFB1\_1433}, \texttt{PDFB1\_10CM}, \texttt{PDFB1\_early\_10CM}, \texttt{CPSR2\_10CM} between $2970$ MHz and $3030$ MHz, \texttt{CPSR2\_10CM} between $3100$ MHz and $3160$ MHz, \texttt{CPSR2\_10CM} between $3230$ MHz and $3290$ MHz, \texttt{20CM\_H-OH\_PDFB1}, \texttt{20CM\_MULTI\_PDFB1}, \texttt{WBCORR\_10CM} with $512$ MHz bandwidth, \texttt{WBCORR\_10CM} with $1024$ MHz bandwidth.
During the course of our work, we found that the above procedures did not eliminate all unphysical frequency-dependence of timing residuals in a few observing systems for PSR J0437$-$4715.
So, we perform additional model comparisons using a linear function:
\begin{equation}\label{eq:fdlin}
    s_{\text{F}}(\nu | \alpha) = \alpha (\nu - \Tilde{\nu}),
\end{equation}
where $\alpha$  determines the tilt in timing residuals in the radio frequency domain, while $\Tilde{\nu}$ is the median radio frequency in the given system.
More details and examples are provided in Section~\ref{sec:j0437}.



\section{Results}
\label{sec:results}
We perform our analysis in four steps.
For each pulsar, we  first establish a base model, which contains white noise, spin noise (common red noise process in all observing bands and systems) and DM noise, and perform parameter estimation for the noise processes while marginalising over the timing model.
Until the fourth step, our white noise model does not include $\text{ECORR}$ parameters.
In the second step, we start with the base model but perform model selection for the possible additional band/system noise components described in Section~\ref{sec:models}.
Model selection for red noise in all possible bands and systems is computationally expensive, so we fixed white noise parameters at maximum-posterior values that we obtained in the first step.
In the third step, after finding the most probable band/system noise combination, we perform model selection for spin noise and DM noise in all pulsars, with free white noise parameters.
Including white noise in parameter estimation increases the computation time by the order of magnitude.
Similarly, if frequency-dependent index $\chi$ for chromatic red noise is a free parameter, the calculation takes by one order of magnitude more time than when $\chi$ is fixed.
In the third step we fix $\chi$ at the maximum-posterior values.
On the fourth step, we perform model selection for excess white noise in sub-banded observations.
We describe this noise by the $\text{ECORR}$ parameter, introduced in Section~\ref{sec:white}.

We chose to incorporate additional terms if the more complicated model resulted in an increase in the $\ln \mathcal{B}$ of 3.
Then, we perform model selection for spin noise and DM noise.
All posteriors for red noise power-law parameters are presented in Figure~\ref{fig:red_population}.
The DM and spin noise processes are described  Table~\ref{tab:results_wrd}.
Band-dependent and system-dependent red noise processes are described in Table~\ref{tab:results_band_system_noise}.
Chromatic noise processes are described in Table~\ref{tab:results_chromatic}.
We clarify that, although the noise models presented in the above tables are valid for use in gravitational-wave searches, some of the red processes are only attributed to a certain class (i.e., spin noise or system noise) due to a marginal preference over competing hypotheses. 
We provide more details and explain the caveats below.

\begin{figure*}
    \centering
    \begin{subfigure}[b]{0.46\textwidth}
        \includegraphics[width=\textwidth]{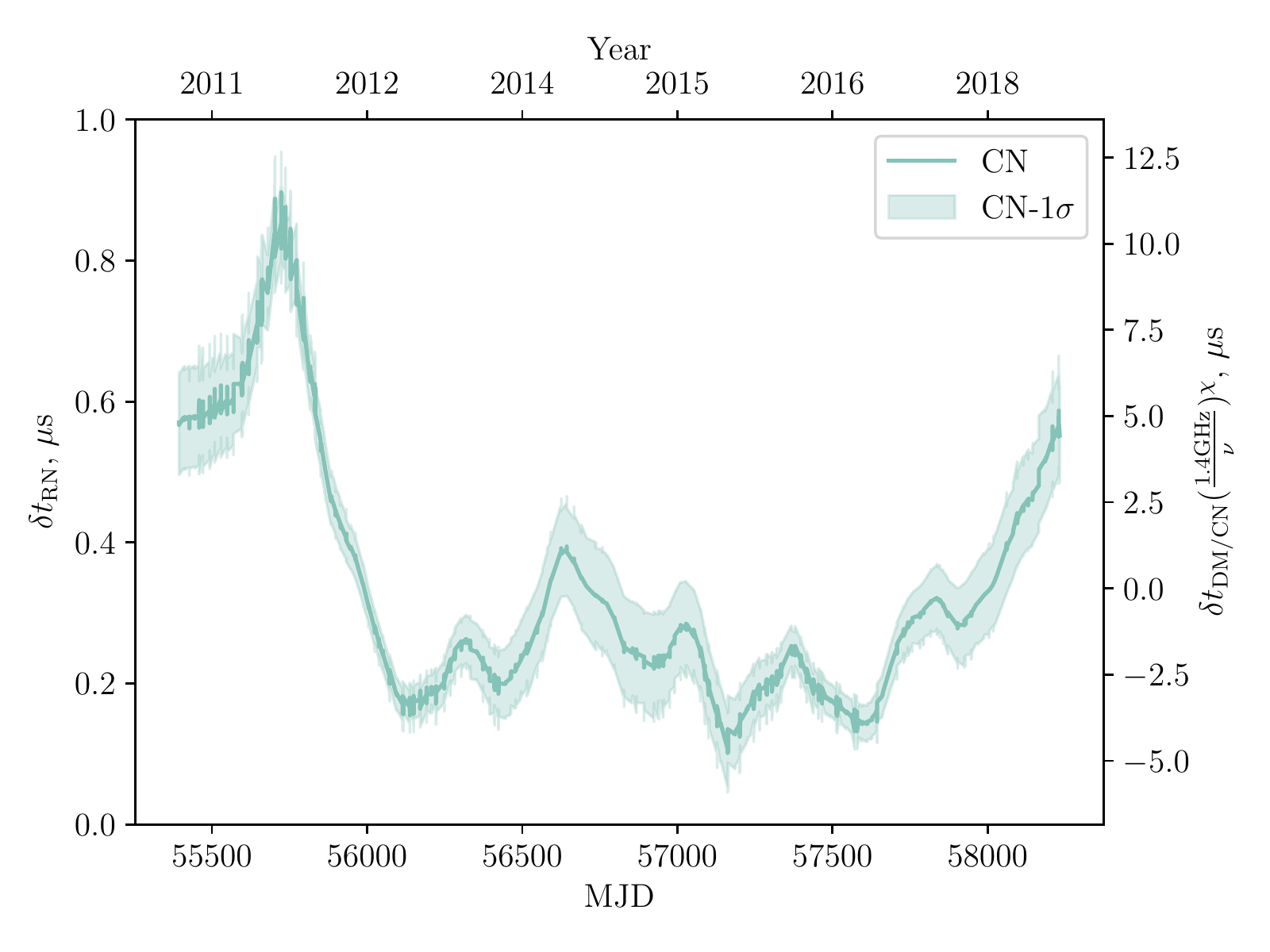}
       \caption{PSR~J1017$-$7156: red noise reconstruction}
        \label{fig:j1017_1}
    \end{subfigure}
    \begin{subfigure}[b]{0.34\textwidth}
        \includegraphics[width=\textwidth]{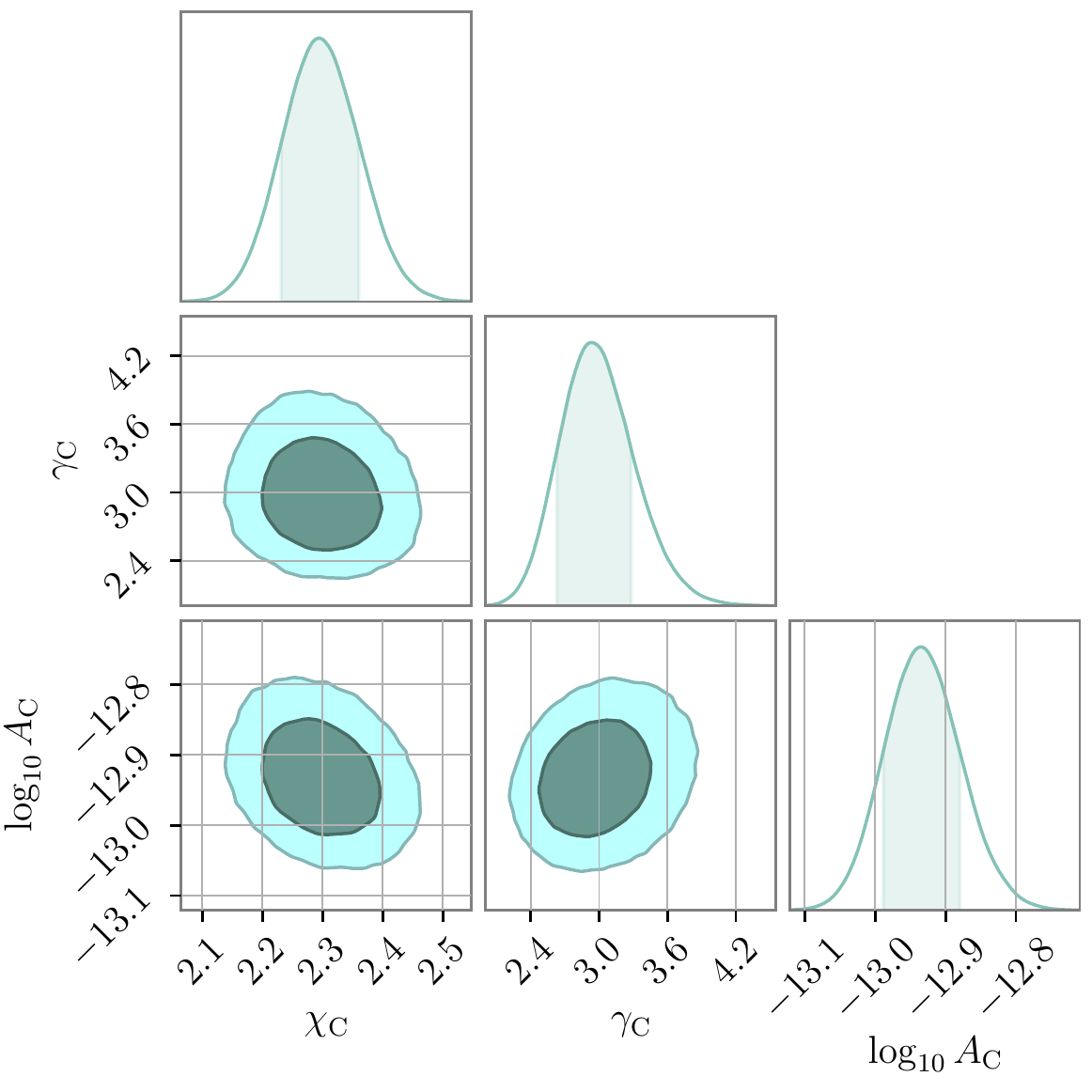}
       \caption{PSR~J1017$-$7156: chromatic noise parameters}
        \label{fig:j1017_2}
    \end{subfigure}
    \begin{subfigure}[b]{0.46\textwidth}
        \includegraphics[width=\textwidth]{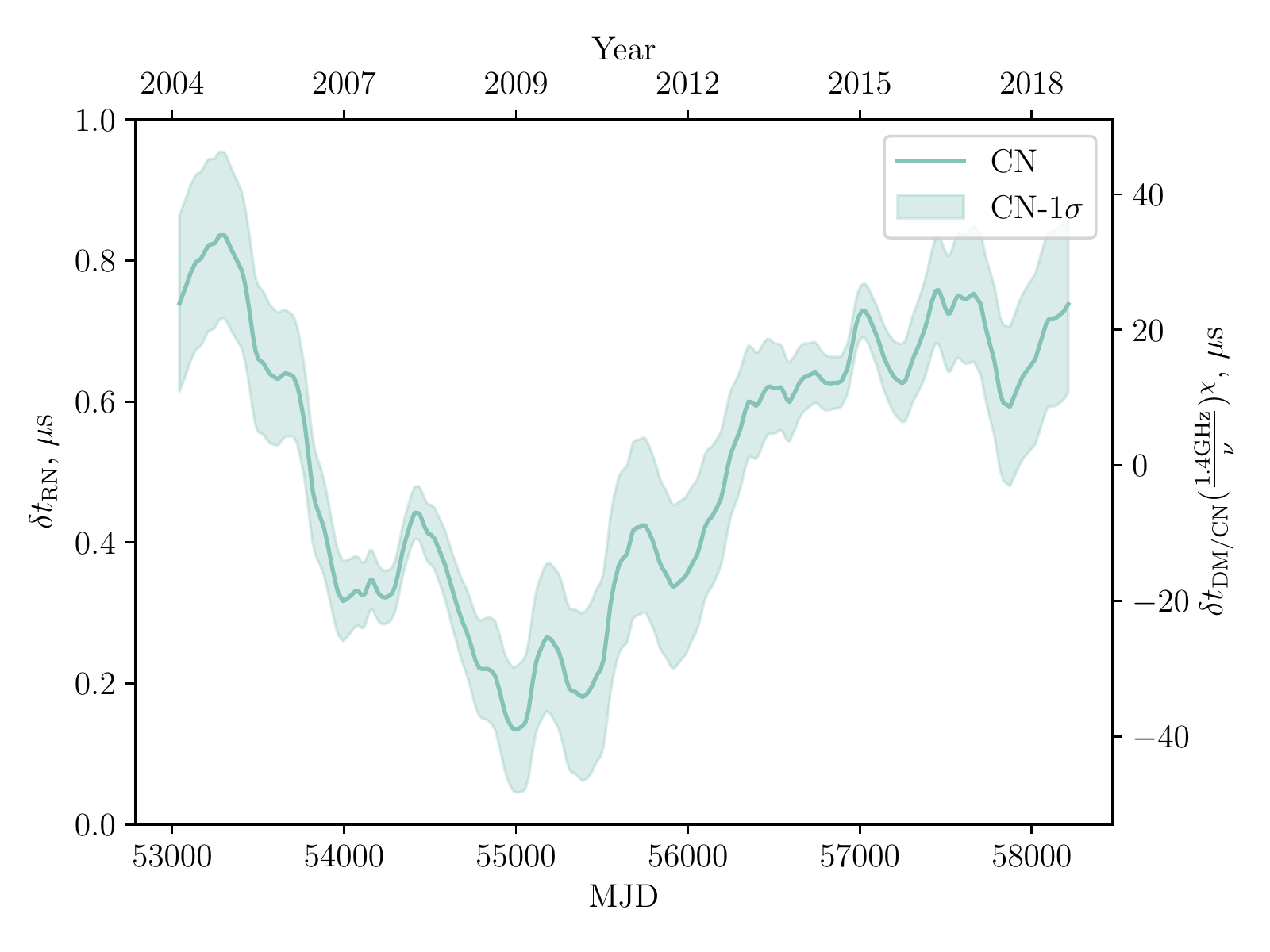}
        \caption{PSR~J1045$-$4509: red noise reconstruction}
        \label{fig:j1045_1}
    \end{subfigure}
    \begin{subfigure}[b]{0.34\textwidth}
        \includegraphics[width=\textwidth]{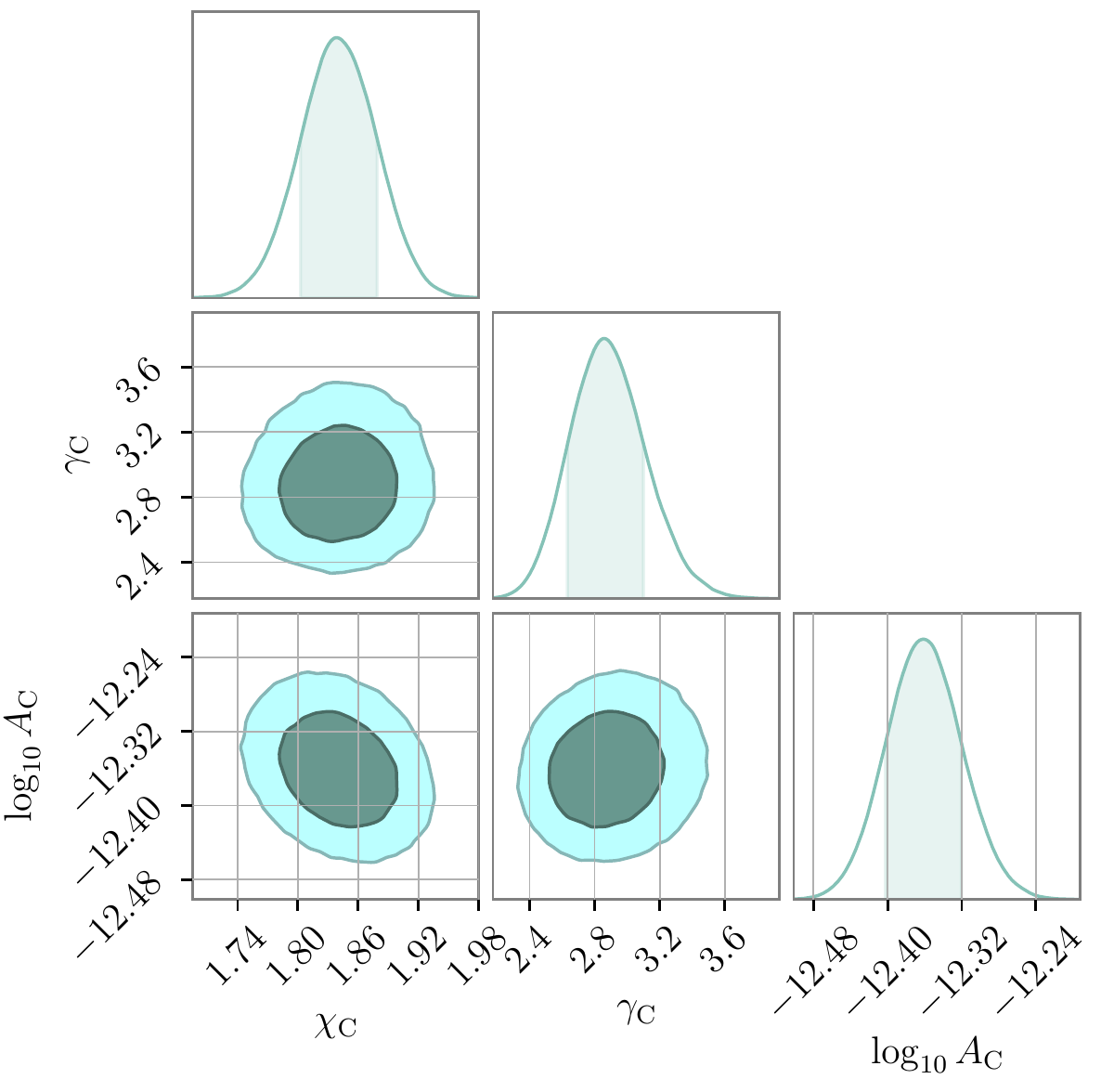}
        \caption{PSR~J1045$-$4509: chromatic noise parameters}
        \label{fig:j1045_2}
    \end{subfigure}
    \caption{Noise properties for PSR~J1017$-$7156 and PSR~J1045$-$4509. These pulsars show evidence for  chromatic noise (turquoise), with chromaticity close to what is expected for dispersion measure variations. Noise realizations are displayed on the left two panels (\ref{fig:j1017_1}, \ref{fig:j1045_1}), posterior distributions for chromatic noise parameters are on the right two panels (\ref{fig:j1017_2}, \ref{fig:j1045_2}). The shaded regions correspond to 1-dimensional 1-$\sigma$ credible levels and 2-dimensional 1-$\sigma$ and 2-$\sigma$ credible levels.}
    \label{fig:j1017j1045}
\end{figure*}

Additionally, we provide maximum-likelihood realisations of red noise processes in pulsars.
We obtain them in two steps.
Firstly, we perform red noise power-law parameter estimation, marginalizing over Fourier coefficients that determine the time evolution of red noise~\citep{vanhaasteren2014newadvances}.
Secondly, with \texttt{TEMPO2}, we estimate maximum-likelihood values of Fourier coefficients and hence the time evolution of red noise.
We provide maximum-likelihood noise realisations for pulsars with chromatic noise in figures~\ref{fig:j1017j1045},~\ref{fig:j0437},~\ref{fig:j0613j1939}.
The maximum-likelihood noise realisations for the remaining pulsars are shown in Figure \ref{fig:noiserealizations} in the Appendix.

\begin{table*}
\caption{\label{tab:results_wrd}
The presense of spin noise and DM noise in the PPTA DR2 pulsars.
There are four vertical line-separated sections in this table.
The first vertical section contains pulsar names and Bayes factors in favour of the presence of spin noise and/or DM noise against the null hypothesis ($\varnothing$).
The second vertical section contains spin noise parameters.
The third vertical section contains DM noise parameters.
The last vertical section contains the observation time $T_\text{obs}$ in years and the number of Fourier components, explained in Section~\ref{sec:rednoise}.
For PSR~J1024$-$0719, we present two models.
The ``$\times$'' symbol identifies a model in which a second spin frequency derivative is not included into the timing model to account for the influence of a possible companion star~\citep{bassa1024,kaplan1024}.
Parameters with asterisks are estimated from the median marginalised posterior values, while other parameter estimates are calculated from the maximum-posterior values.
}
\renewcommand{\arraystretch}{1.4}
\begin{tabular}{| l c c c | c c c | c c c | c c | }
\hline

Pulsar & $\text{ln} \mathcal{B}^{\text{DM,SN}}_\varnothing$ & $\text{ln} \mathcal{B}^{\text{DM}}_\varnothing$ & $\text{ln} \mathcal{B}^{\text{SN}}_\varnothing$ & SN & $\log_{10}A_\text{SN}$ & $\gamma_\text{SN}$ & DM & $\log_{10}A_\text{DM}$ & $\gamma_\text{DM}$ & $T_\text{obs}$ [yr] & $n_\text{c}$ \\ [1ex] \hline \hline 
J0437$-$4715 & - & - & - & \cmark & $-14.56^{+0.16}_{-0.29}$ & $2.99^{+1.36}_{-0.18}$ & \cmark & $-13.50^{+0.04}_{-0.04}$ & $2.78^{+0.16}_{-0.14}$ & 15.0 & 91\\ [1pt] \hline 
J0613$-$0200 & 57.3 & 52.5 & 57.6 & \cmark & $-14.26^{+0.26}_{-2.23}$ & $4.17^{+4.06}_{-0.49}$ & \xmark & & & 14.2 & 85\\ [1pt] \hline 
J0711$-$6830 & 75.6 & 64.6 & 68.3 & \cmark & $-13.04^{*,+0.07}_{-0.07}$ & $1.09^{+0.34}_{-0.27}$ & \cmark & $-15.79^{*,+1.70}_{-1.49}$ & $6.24^{*,+2.63}_{-3.10}$ & 14.2 & 86\\ [1pt] \hline 
J1017$-$7156 & -0.7 & -0.4 & -0.2 & \xmark & & & \xmark & & & 7.8 & 46\\ [1pt] \hline 
J1022+1001 & 52.1 & 49.1 & 46.0 & \xmark & & & \cmark & $-13.41^{+0.06}_{-0.73}$ & $1.72^{+1.72}_{-0.50}$ & 14.2 & 85\\ [1pt] \hline 
J1024$-$0719 & 45.9 & 46.1 & 1.9 & \xmark & & & \cmark & $-14.05^{+0.11}_{-2.05}$ & $3.96^{+4.01}_{-0.23}$ & 14.1 & 85\\ [1pt]
J1024$-$0719$^\times$ & 474.0 & 259.1 & 435.1 & \cmark & $-14.62^{+0.43}_{-0.95}$ & $6.39^{+2.16}_{-0.71}$ & \cmark & $-13.98^{+0.17}_{-1.50}$ & $4.04^{+2.99}_{-0.51}$ & 14.1 & 85\\ [1pt] \hline 
J1045$-$4509 & -1.0 & -0.1 & -1.2 & \xmark & & & \xmark & & & 14.2 & 85\\ [1pt] \hline 
J1125$-$6014 & 170.9 & 171.3 & 124.2 & \xmark & & & \cmark & $-13.14^{+0.09}_{-0.09}$ & $3.07^{+0.49}_{-0.27}$ & 12.3 & 74\\ [1pt] \hline 
J1446$-$4701 & 3.1 & 2.8 & 2.5 & \xmark & & & \cmark & $-13.35^{+0.11}_{-4.40}$ & $1.80^{+3.77}_{-0.46}$ & 7.4 & 44\\ [1pt] \hline 
J1545$-$4550 & 31.3 & 31.6 & 20.6 & \xmark & & & \cmark & $-13.29^{+0.14}_{-0.46}$ & $3.06^{+2.09}_{-0.38}$ & 7.0 & 41\\ [1pt] \hline 
J1600$-$3053 & 58.5 & 52.5 & 32.3 & \cmark & $-14.34^{+0.42}_{-1.29}$ & $3.81^{+3.02}_{-0.74}$ & \cmark & $-13.23^{+0.08}_{-0.11}$ & $2.12^{+0.64}_{-0.08}$ & 14.2 & 86\\ [1pt] \hline 
J1603$-$7202 & 457.7 & 458.0 & 341.7 & \xmark & & & \cmark & $-13.14^{+0.08}_{-0.08}$ & $2.12^{+0.43}_{-0.15}$ & 14.2 & 86\\ [1pt] \hline 
J1643$-$1224 & 275.4 & 260.6 & 235.2 & \cmark & $-12.85^{+0.07}_{-0.06}$ & $0.98^{+0.29}_{-0.25}$ & \cmark & $-13.20^{+0.14}_{-0.31}$ & $3.01^{+1.17}_{-0.29}$ & 14.2 & 85\\ [1pt] \hline 
J1713+0747 & 34.8 & 34.8 & 6.3 & \xmark & & & \cmark & $-13.79^{+0.05}_{-0.09}$ & $1.92^{+0.39}_{-0.21}$ & 14.2 & 86\\ [1pt] \hline 
J1730$-$2304 & 73.7 & 74.2 & 34.2 & \xmark & & & \cmark & $-13.52^{+0.13}_{-0.78}$ & $2.39^{+2.26}_{-0.10}$ & 14.2 & 86\\ [1pt] \hline 
J1732$-$5049 & 2.8 & 3.2 & 1.4 & \xmark & & & \cmark & $-12.70^{+0.12}_{-1.86}$ & $2.70^{*,+4.63}_{-1.99}$ & 7.2 & 43\\ [1pt] \hline 
J1744$-$1134 & 161.4 & 161.8 & 118.0 & \xmark & & & \cmark & $-13.62^{+0.05}_{-0.11}$ & $1.78^{+0.44}_{-0.19}$ & 14.2 & 86\\ [1pt] \hline 
J1824$-$2452A & 4328.9 & 3859.5 & 4204.2 & \cmark & $-13.26^{+0.19}_{-0.52}$ & $5.02^{+1.38}_{-0.44}$ & \cmark & $-12.46^{+0.05}_{-0.05}$ & $2.56^{+0.33}_{-0.16}$ & 13.8 & 83\\ [1pt] \hline 
J1832$-$0836 & 36.7 & 36.7 & 29.8 & \xmark & & & \cmark & $-12.98^{+0.11}_{-1.34}$ & $2.98^{+4.28}_{-0.24}$ & 5.4 & 32\\ [1pt] \hline 
J1857+0943 & 65.7 & 48.0 & 42.6 & \cmark & $-16.86^{+2.19}_{-0.10}$ & $7.49^{*,+1.76}_{-2.35}$ & \cmark & $-13.36^{+0.10}_{-0.25}$ & $2.38^{+0.96}_{-0.25}$ & 14.2 & 86\\ [1pt] \hline 
J1909$-$3744 & 121.4 & 113.4 & 9.2 & \cmark & $-14.74^{+0.28}_{-0.74}$ & $4.05^{+1.71}_{-0.69}$ & \cmark & $-13.70^{+0.04}_{-0.04}$ & $1.47^{+0.13}_{-0.21}$ & 14.2 & 85\\ [1pt] \hline 
J1939+2134 & 3834.5 & 3263.1 & 2507.9 & \cmark & $-14.33^{+0.36}_{-0.27}$ & $5.39^{+1.03}_{-0.73}$ & \cmark & $-12.94^{+0.04}_{-0.03}$ & $2.71^{+0.19}_{-0.13}$ & 14.1 & 85\\ [1pt] \hline 
J2124$-$3358 & 4.1 & 4.3 & 2.5 & \xmark & & & \cmark & $-13.59^{+0.21}_{-0.23}$ & $1.40^{+1.28}_{-0.18}$ & 14.2 & 85\\ [1pt] \hline 
J2129$-$5721 & 43.2 & 43.7 & 35.5 & \xmark & & & \cmark & $-13.47^{+0.08}_{-0.10}$ & $1.57^{+0.65}_{-0.22}$ & 13.9 & 83\\ [1pt] \hline 
J2145$-$0750 & 98.3 & 82.0 & 91.0 & \xmark & & & \cmark & $-13.67^{+0.11}_{-1.31}$ & $1.20^{+3.29}_{-0.04}$ & 14.1 & 85\\ [1pt] \hline 
J2241$-$5236 & 104.7 & 103.4 & 88.4 & \xmark & & & \cmark & $-13.90^{+0.09}_{-0.06}$ & $1.80^{+0.43}_{-0.33}$ & 8.2 & 49\\ [1pt] \hline 

\end{tabular}
\end{table*}

\subsection{Spin noise}\label{sec:results_spin}

We identify an increased sample of millisecond pulsars showing evidence for red noise.  Longer datasets are more sensitive to low-frequency red noise.
For example, for PSR~J1909$-$3744 we find evidence for steep spin noise with $\gamma_{\rm SN} \approx 4$ in PPTA DR2, while in \cite{pptalimit2}, with an 11 year dataset, there was no evidence for red noise. 
In DR2, we find evidence for spin noise in 10 pulsars, while \cite{lentati2016iptanoise} found spin noise only in 6 PPTA DR1 pulsars.
In two of these six pulsars, J0613$-$0200 and PSR~J1939+2134, we measure spin noise parameters consistent with~\cite{lentati2016iptanoise}.
For PSR~J1024$-$0719, we find that $\ddot{\nu}$ is preferred over spin noise.
When we do not include $\ddot{\nu}$, our measurement of $A_\text{SN}$ is consistent with \cite{lentati2016iptanoise}, while for $\gamma_\text{SN}$ only 1-$\sigma$ uncertainties overlap, so that the values are approximately within 2-$\sigma$ agreement.
For PSR~J2145$-$0750, we find that the spin noise with a shallow spectrum disappears after we include ECORR in the noise model. Prior to including ECORR, our measurement of $\gamma_\text{SN}$ is consistent with \cite{lentati2016iptanoise}, while values of $A_\text{SN}$ are within 2-$\sigma$ agreement.
For PSR~J1824$-$2452A, we find evidence for steep spin noise, as in \cite{lentati2016iptanoise}, although our measurements of spin noise parameters are not consistent.
\cite{lentati2016iptanoise} finds stronger and more shallow spin noise.
However, additional shallow band noise we identify in PSR~J1824$-$2452A could be the reason for discrepancy.
Similarly, for PSR~J0613$-$0200 and PSR~J1024$-$0719, where we find no evidence of band noise, spin noise parameters that we measure without $\ddot{\nu}$ are consistent with red timing noise parameters in \cite{caballero2016eptanoise} within 1-2 $\sigma$ credible levels.
Since \cite{caballero2016eptanoise} do not model band noise and system noise, in other pulsars they identify examples of shallow red noise.
Additionally, the sixth pulsar with spin noise in \cite{lentati2016iptanoise} is PSR~J1713+0747, which is discussed below.


Some pulsars with band and system noise have only marginal evidence in favour of a hypothesis that excludes spin noise, and vice versa.
For example, in PSR~J1713+0747, with steep band noise in 10 cm and 20 cm data ($\gamma \approx 4$), the current model is preferred over a model with spin noise and 40-cm band noise  with only a $\ln \mathcal{B}$ of 2.
Generally speaking,if one is interested in studying spin noise to infer inner the workings of neutron stars, we recommend strengthening the requirement for the evidence in favour of the absence of chromaticity of the red process, while also assuming that band and system noise could contain the missing spin noise just because some other marginally-preferred term is not in the noise model.
Otherwise, the noise might not be intrinsic to the pulsar.
For example, one could require a minimum threshold on Bayes factor in favour of spin noise over each band noise component.
Another suggestion would be to choose different priors for spin noise and band noise, as well as to fine-tune model selection by establishing prior odds.
At the same time, for the purpose of gravitational-wave searches, it is justified to use Table~\ref{tab:results_wrd} as a guide on whether to include spin noise in pulsar noise models.
We also find that spin noise processes in PSR~J0711$-$6830 and PSR~J1643$-$1224 have $\gamma_\text{SN} \approx 1$, although there is a significant evidence that this noise process is not band-specific.

For the remaining pulsars with spin noise, we find evidence for steep red spectra, with power-law indices between 4 and 10.
For example, we measure $\gamma_\text{SN}$ consistent with 7 for PSR~J1857$+$0943.
Similarly, for young pulsars,~\cite{parthasarathy2019timing} measured steep red noise with power-law indices between 3 and 10.
In the timing analysis of the NANOGrav 11 year data set~\cite{11yrnanogravtiming}, the spectral index of observed red noise slope ranged between 1 and 3, which is why the authors suggested that the physical process is distinct from spin noise.
According to \cite{lentati2016iptanoise}, these results could be influenced by the absence of band-specific and system-specific red noise processes in pulsar noise models.








\subsection{Dispersion measure variations}\label{sec:results_dm}

We find evidence for stochastic DM noise in 23 of the pulsars.
PSR~J1017$-$7156 and PSR~J1045$-$4059 favour chromatic noise with a frequency-dependent index $\chi \approx 2$, the value expected from DM noise.
More details will be provided in Section~\ref{sec:j1017j1045} below.
In PSR~J0613$-$0200, we do not find evidence with $\ln \mathcal{B} > 3$ in favour of stochastic DM noise.
We do find evidence of annual DM variations in PSR~J0613$-$0200, which is consistent with previous studies by~\cite{keith2013dmvar}.
Moreover, in PSR~J0613$-$0200 we find evidence for chromatic noise with $\chi \approx 4$, which we attribute to scattering variations in the interstellar medium. 
Given that the chromatic index $\chi$ has not been widely explored as a free parameter in the past, we suspect that DM noise previously observed in PSR~J0613$-$0200 (i.e., in~\citealt{lentati2016iptanoise}), had been, in fact, a chromatic noise.
Moreover, the spectral index, $\gamma_\text{DM}$, for PSR~J0613$-$0200, found in~\cite{lentati2016iptanoise}, is consistent with $\gamma_\text{CN}$ for PSR~J0613$-$0200 in our work.

In PSR~J1939+2134 and PSR~J0437$-$4715, we find evidence for both stochastic DM noise and chromatic noise. Chromatic noise for these pulsars will be discussed in more detail the following subsections.

In the NANOGrav 11-year timing analysis by~\cite{11yrnanogravtiming}, DM variations were modeled by independently measuring dispersion for each epoch.   
We defer a comparison of these DM  time series to our  maximum-likelihood DM noise reconstructions to future work.
Annual DM variations of PSR~J0613$-$0200, low-frequency band noise in PSR~J1939+2134 (PSR~B1937+21), and a DM event for PSR~J1713+0747 are all clearly seen in the dispersion measure time series presented in \cite{11yrnanogravtiming}. 

In addition to DM variations, DR2 pulsar timing models include first and second time-derivatives of dispersion measure (\texttt{DM1} and \texttt{DM2}), which absorb long-term variations in dispersion measure.
We measure significant \texttt{DM1} values $\sim 10^{-3}$ [cm$^{-3}$ pc yr$^{-1}$] in PSRs J0613$-$0200, J1017$-$7156, J1045$-$4509, J1643$-$1224 and J1824$-$2452A,  and significant \texttt{DM2} values on the order of $\sim 10^{-4}$ [cm$^{-3}$ pc yr$^{-2}$] in PSRs J1732$-$5049 and J1832$-$0836.
For PSR J1824$-$2452A we measure the highest value of DM and the highest amplitude of stochastic DM variations.


\begin{table}
\caption{\label{tab:results_chromatic}Results for chromatic noise parameter estimation and model selection. The second column is the chromaticity of the red process, the third and the fourth columns are parameters of the power-law spectrum. The last column is the $\ln \mathcal{B}$ in favor of chromatic noise over the hypothesis of just white noise, spin noise and DM noise. The asterisk means that the parameter is estimated from the median marginalised posterior values. The other parameter estimates are calculated from the maximum-posterior values.}
\renewcommand{\arraystretch}{1.4}
\begin{tabular}{ | l c c c c | }
\hline

PSR & $\chi$ & $\log_{10}A$ & $\gamma$ & $\ln \mathcal{B}$  \\ [1ex] \hline \hline 
J0437$-$4715 & 4 & $-14.41^{+0.06}_{-0.12}$ & $2.43^{+0.43}_{-0.36}$ & 286.5 \\
J0613$-$0200 & 4 & $-14.03^{+0.07}_{-0.05}$ & $2.02^{+0.30}_{-0.18}$ & 11.1 \\ 
J1017$-$7156 & 2.29 & $-12.96^{+0.07}_{-0.03}$ & $3.20^{+0.39}_{-0.24}$ & 3.9 \\ 
J1045$-$4509 & 1.82 & $-12.36^{+0.04}_{-0.04}$ & $2.80^{+0.29}_{-0.16}$ & 3.2 \\ 
J1939+2134 & 4 & $-13.90^{+0.05}_{-0.11}$ & $1.53^{+0.42}_{-0.10}$ & 4.7 \\ 
\hline

\end{tabular}

\end{table}

\subsection{Band and system noise}
\label{sec:bandsystem}
In an analysis of IPTA DR1, \cite{lentati2016iptanoise} found evidence for band and system noise terms.
We also find evidence for these processes in PPTA DR2.
In particular, \cite{lentati2016iptanoise} found that PSRs J0437$-$4715, J1600$-$3053, J1643$-$1224 and J1939$+$2134 have band noise processes at low frequencies, which we also see in PPTA DR2.
Unlike \cite{lentati2016iptanoise}, we do not find evidence for band noise in 10-cm band and system noise in \texttt{CPSR2\_20CM} for PSR~J1939+2134; however, we observe new chromatic noise in this pulsar, as discussed below.
In PPTA DR2, we also find more pulsars to show evidence of band and system noise.
Our measurements of band noise are provided in Table~\ref{tab:results_band_system_noise}.
Initially, we found evidence for 32 band- and system-dependent red noise terms.
However, the number of these terms dropped to 17 after we introduced ECORR parameters.
As one would expect, most of the terms that disappeared were described by $\gamma < 1$, which means that the noise was nearly white.
On average, band and system noise processes are found to have more shallow power-law spectra than spin noise processes, as seen in Figure~\ref{fig:red_population}.
For PSR~J1022+1001, if we do not include ECORR, we detect weak low-frequency spin noise term with a power-law index of 7.
For this pulsar the ECORR is larger than other pulsars by an order of magnitude.
We find it plausible that variations in the amplitude of the ECORR over the observation span are modelled as spin noise.


There are a few potential sources of systematic error that may contribute to this noise. It is possible that it arises from polarisation calibration~\citep{kramer1999profile}.
Highly polarised pulsars are more susceptible to this form of noise.
Radio frequency interference is another possible origin.
One possible origin  for the high-frequency system noise in \texttt{CPSR2\_50CM}, is residual radio-frequency interference.
Additionally, the solar wind is known to contribute to pulsar timing noise~\citep{you2007solarwind,dustysun}.
Given the time scale on which angle between the line of sight to the pulsar and the Sun changes, some contributions to high-frequency band and system noise could be influenced by the solar wind if DM is not properly modeled.
Stochasticity in the solar wind unaccounted for in noise noise models  \cite[][]{tiburzisun} introduces another potential source of noise.

\subsection{Updated timing model parameters}
\label{sec:timing}
For PSRs J1017--7156, J1024--0719, J1125--6014, J1939+2134 and J2241--5236 we include updated timing models, which will be discussed in more detail in the upcoming publication Reardon et al. (in prep.).
In particular, strong Shapiro delay due to a binary companion was found in PSR~J1125$-$6014.
Orbital-frequency derivatives \texttt{FB} were added to describe tidal interactions between PSR~J2241$-$5236 and its binary companion.
If not included in timing models, these deterministic processes are detected as excess red noise.

For PSR J1024--0719, \cite{kaplan1024} identified a presence of the second spin-frequency derivative $\ddot{\nu}$, suggesting it to be due to the presence of a stellar companion in a wide orbit.
We performed a model comparison between the $\ddot{\nu}$ hypothesis and the spin noise hypothesis and found $\text{ln} \mathcal{B} = 12$ in favour of $\ddot{\nu}$.
We find no additional spin noise in the presence of $\ddot{\nu}$ in PPTA DR2.

%

\subsection{PSR~J1643--1224: profile event, band noise}
\label{sec:1643}

The  DM ($\approx 62.4$\,pc\,cm$^{-3}$) makes PSR~J1643$-$1224 one of the more susceptible to noise introduced by propagation effects.
However the most unusual feature in its arrival time is unlikely to have originated in the interstellar medium.
Between 2015 and 2016, PSR~J1643$-$1224 exhibited a sudden change in pulse shape \cite[][]{shannon2016disturbance}, which caused the same evolution of timing residuals.
Interestingly, the event was most pronounced at high radio frequency.  
\cite{shannon2016disturbance} suggested that this event originated in the pulsar magnetosphere, because the spectral properties were inconsistent with being a propagation effect, and the presence in multiple bands and instruments made it inconsistent with being a telescope dependent effect.
In our work, we estimate the chromaticity of the event to be $\chi = -0.99^{+0.10}_{-0.11}$, consistent with the inverted spectrum noted by \cite{shannon2016disturbance}.
In this work, we find that the profile event in PSR~J1643$-$1224 must be taken into account, in order not to be confused with a red process in 10-cm and 20-cm data.
The event was also identified in the NANOGrav 11-year data set.
However because of their lower frequency data and dual-frequency observations, they were unable to establish its chromaticity~\cite[][]{brooknanogravprofilevariability}.
Additionally, we find that J1643$-$1224 shows evidence of red noise process in 40-cm and 50-cm observations, which might be attributed to propagation effects.

 \begin{figure*}
     \centering

    \begin{subfigure}[b]{0.46\textwidth}
         \includegraphics[width=\textwidth]{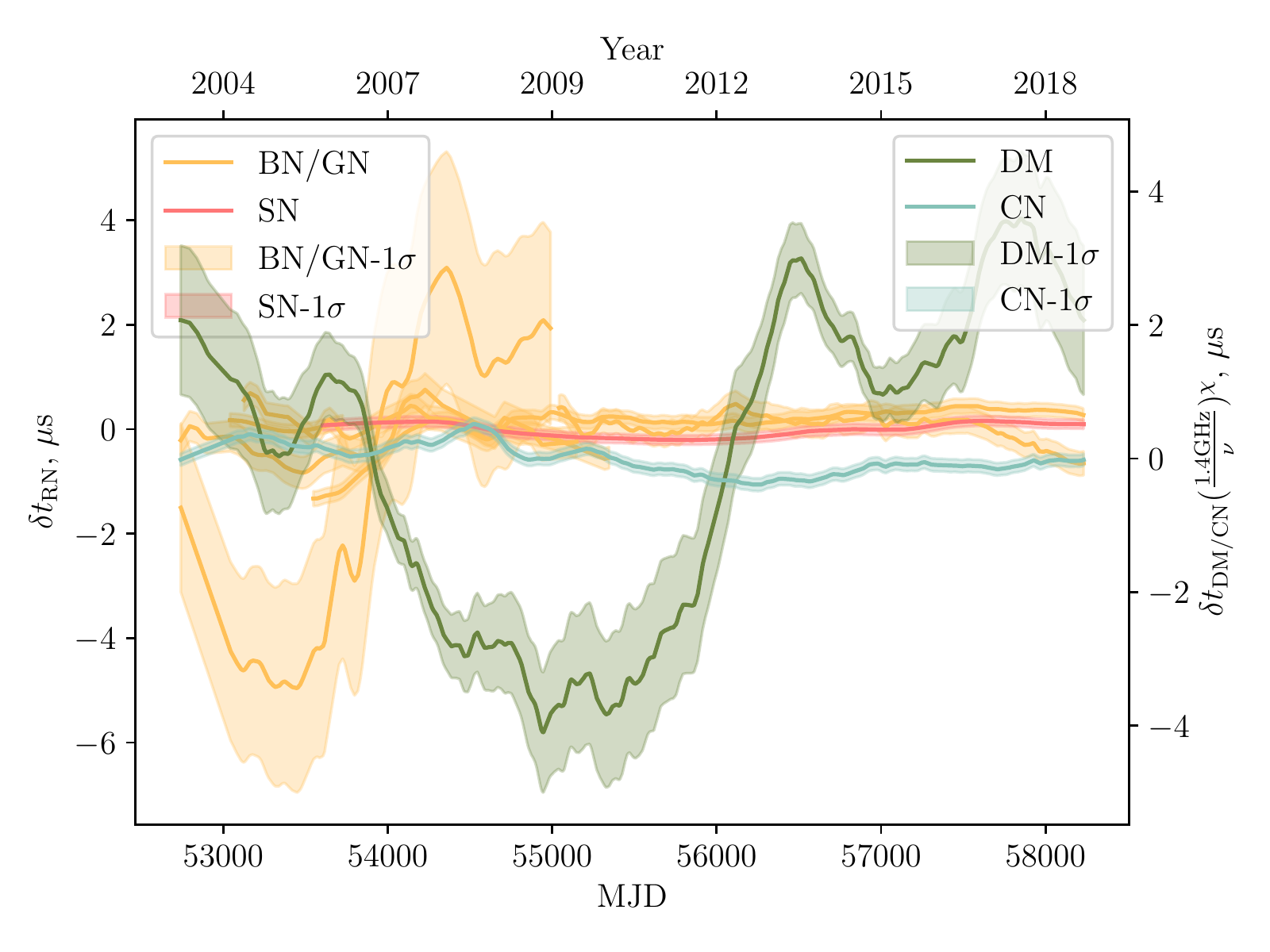}
       \caption{PSR~J0437$-$4715: red noise reconstruction}
         \label{fig:j0437_1}
     \end{subfigure}
     ~ 
     \begin{subfigure}[b]{0.34\textwidth}
         \includegraphics[width=\textwidth]{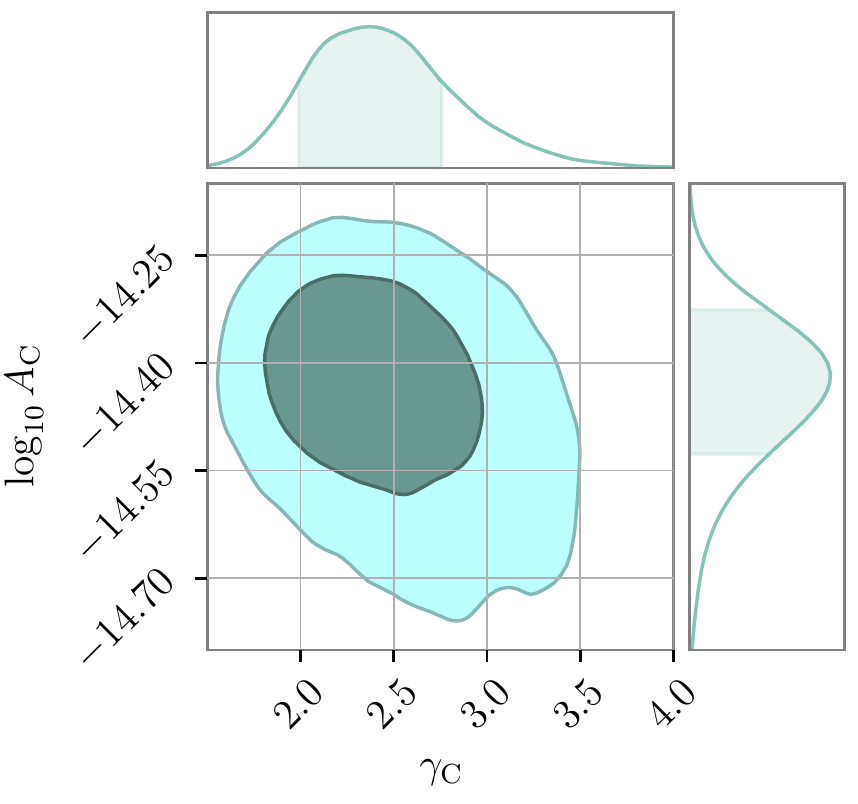}
         \caption{PSR~J0437$-$4715: chromatic noise parameters}
         \label{fig:j0437_2}
     \end{subfigure}
     
    \caption{Noise properties of PSR~J0437$-$4715. Left panel (\ref{fig:j0437_1}): maximum-likelihood realisation of spin noise (red), band and system noise (yellow), chromatic noise (turquoise) and DM noise (green) for PSR J0437$-$4715. Right panel (\ref{fig:j0437_2}): posterior distribution for chromatic noise power-law parameters $A_\text{C}$ and $\gamma_\text{C}$. The shaded regions correspond to 1-dimensional 1-$\sigma$ credible levels and 2-dimensional 1-$\sigma$ and 2-$\sigma$ credible levels. Due to the largest amount of data and the most complicated noise model for PSR~J0437$-$4715 in DR2, it is computationally challenging to produce a posterior distribution that would include the chromatic index $\chi$. We found $\chi$ consistent with 4 for this pulsar empirically.}
     \label{fig:j0437}
 \end{figure*}

\subsection{PSR~J1713+0747: DM event, profile event, system noise, band noise}
\label{sec:j1713}

Previous studies have found that PSR~J1713+0747 has shown two exponential events.
The first event, which started at approximately MJD 54750,  was observed as a sudden decrease in dispersion measure by ~\cite{keith2013dmvar}, and discussed in detail in  ~\cite{desvignes2016timing, jones2017dmvar,coles2015exscat}.
A second event, at approximately MJD 57510, has been reported in~\cite{lam2018j1713event2}.
It shows deviations from $\chi = 2$, which is the index for DM-related radio pulse time delays.
We include both of these events into the noise model of PSR~J1713+0747.
We found the chromatic index of the second event to be $\chi=1.15^{+0.18}_{-0.19}$.
This, and the observation of a pulse shape change at the time of the event (Figure~\ref{fig:event_j1713}), points to the magnetospheric origin.
We do not find evidence for pulse shape changes for PSR~J1713+0747 at the time of the first exponential event.
In addition to exponential events, we find evidence of system noise in \texttt{CPSR2\_20CM}.
We also find evidence of band noise in 10-cm and 20-cm data.

 \begin{figure*}
     \centering
     
    \begin{subfigure}[b]{0.46\textwidth}
         \includegraphics[width=\textwidth]{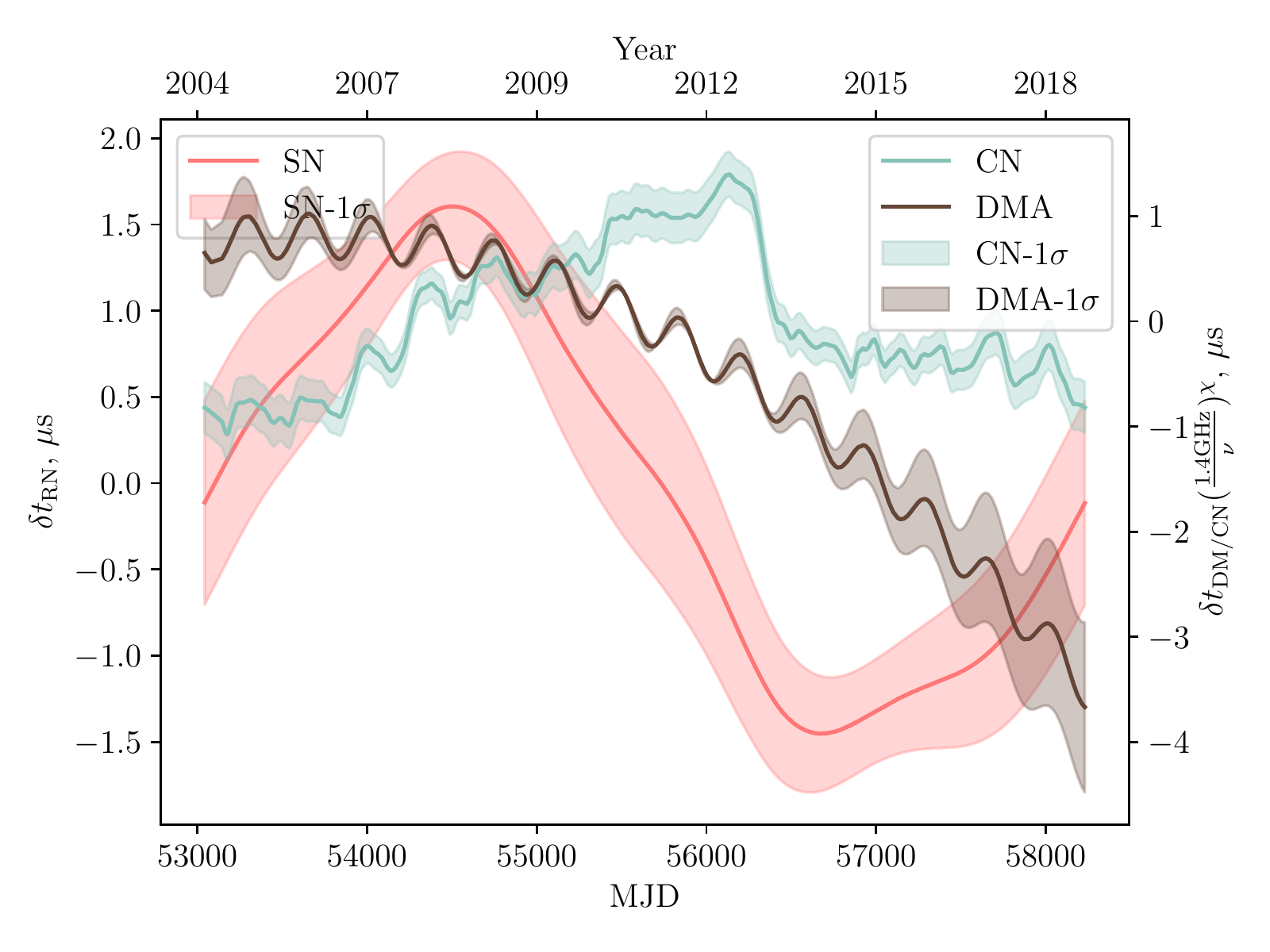}
       \caption{PSR~J0613-0200: red noise reconstruction}
         \label{fig:j0613_1}
     \end{subfigure}
     ~ 
     \begin{subfigure}[b]{0.34\textwidth}
         \includegraphics[width=\textwidth]{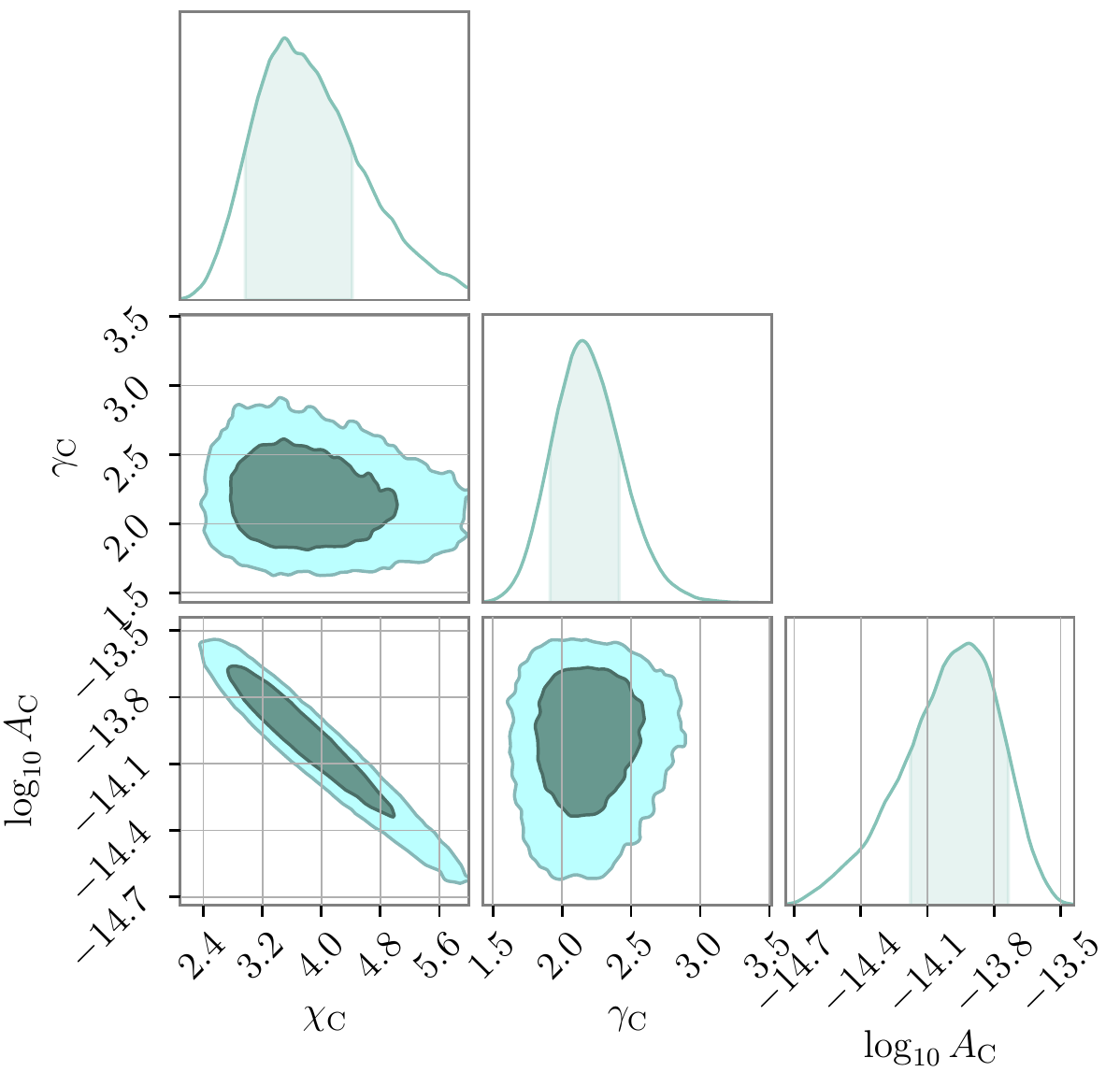}
         \caption{PSR~J0613-0200: chromatic noise parameters}
         \label{fig:j0613_2}
     \end{subfigure}
     
    \begin{subfigure}[b]{0.46\textwidth}
        \includegraphics[width=\textwidth]{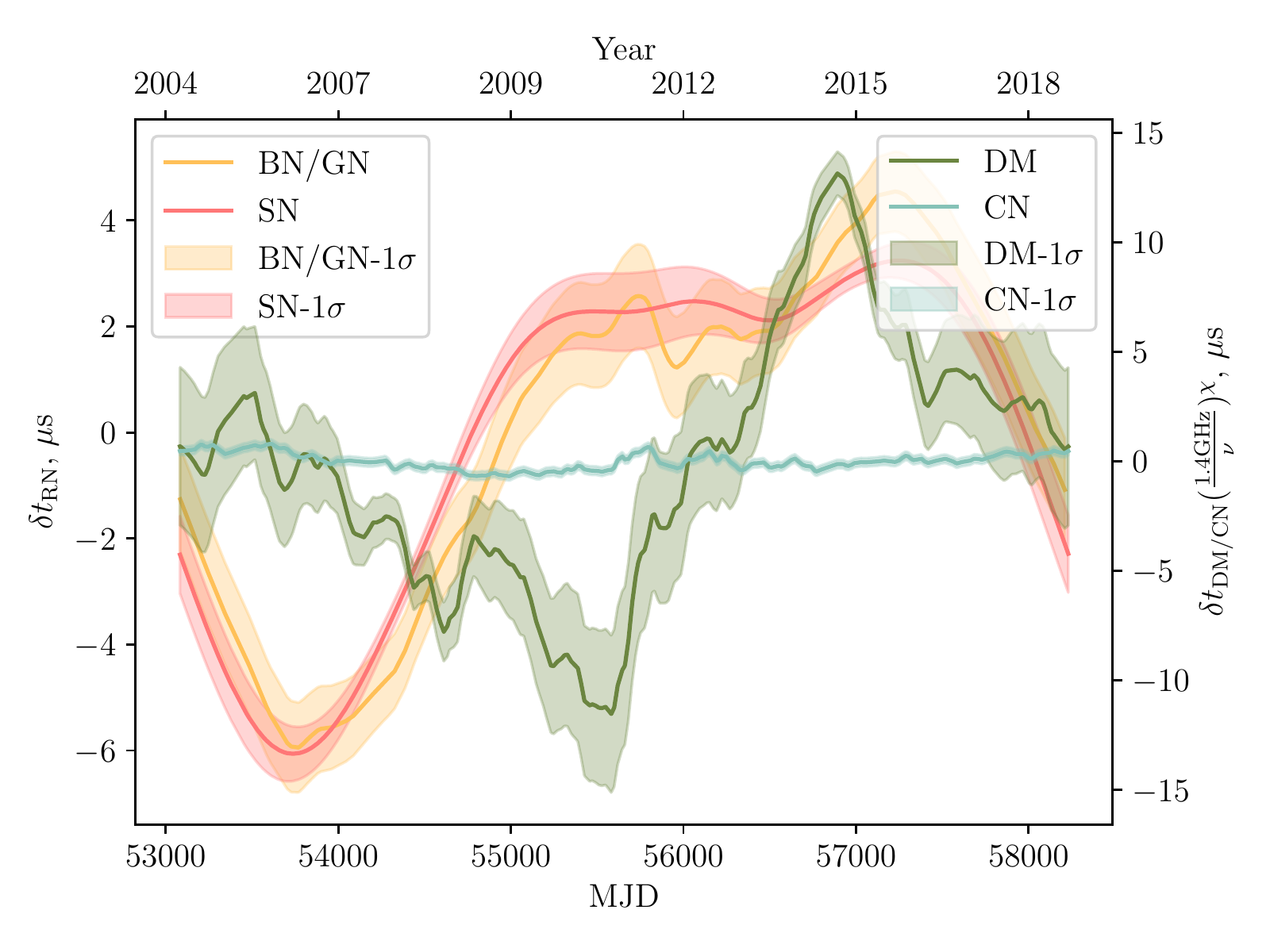}
       \caption{PSR~J1939+2134: red noise reconstruction}
        \label{fig:j1939_1}
    \end{subfigure}
    ~ 
    \begin{subfigure}[b]{0.34\textwidth}
        \includegraphics[width=\textwidth]{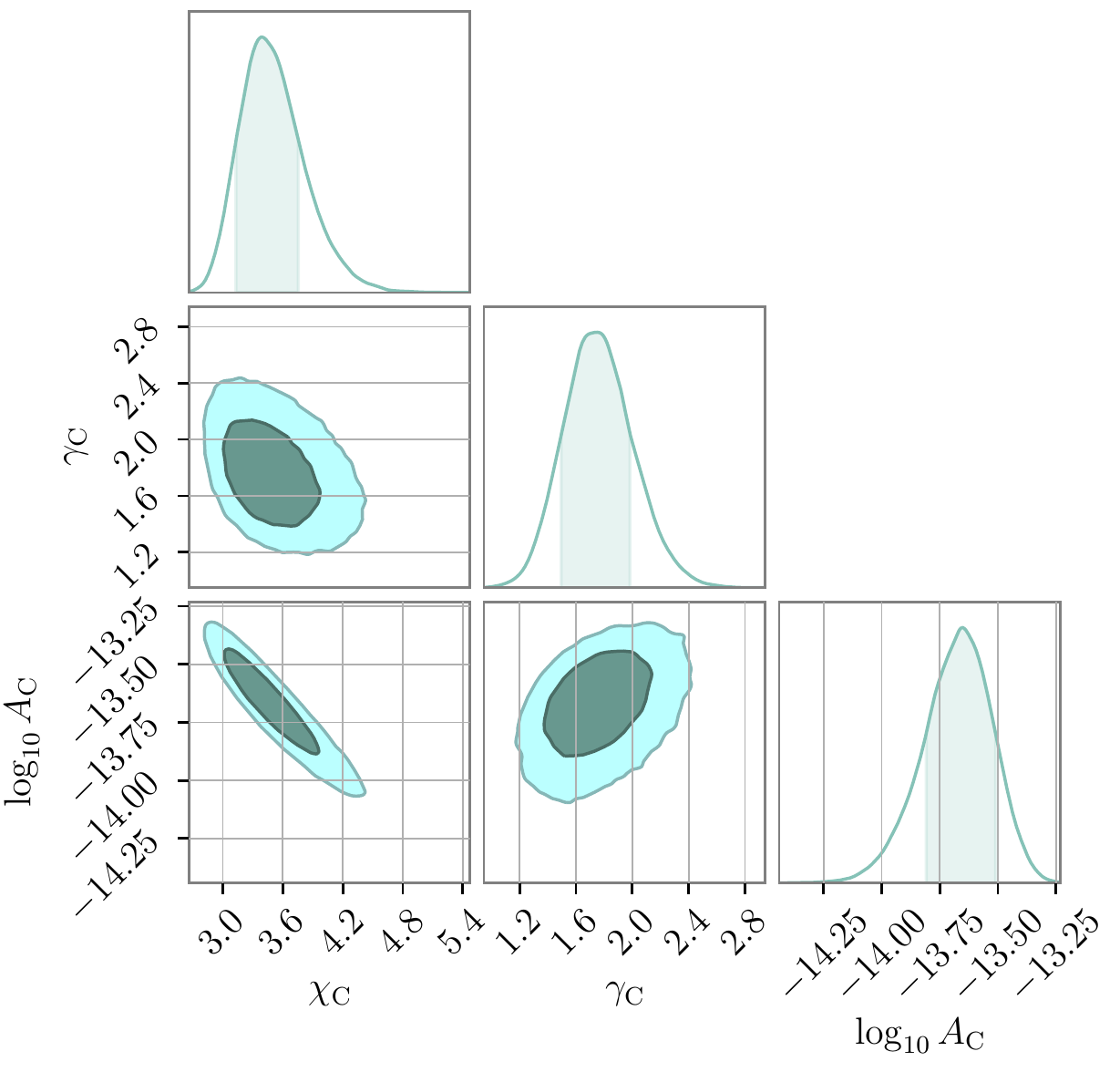}
        \caption{PSR~J1939+2134: chromatic noise parameters}
        \label{fig:j1939_2}
    \end{subfigure}
     ~ 
     \caption{Noise models for pulsars with chromatic red noise with $\chi \approx 4$. Noise with this chromatic index would be expected from scattering variations in the interstellar medium. On the left two panels (\ref{fig:j0613_1} and \ref{fig:j1939_1}) we present maximum-likelihood realization of spin noise (red), band and system noise (yellow), DM noise (green), chromatic noise (turquoise) and annual DM noise on top of the first and the second time derivatives of the dispersion measure (brown). We do not find evidence with $\ln \mathcal{B}$ of above three for stochastic DM noise in PSR~J0613$-$0200, but we do find evidence of annual DM variations in this pulsar. On the right two panels (\ref{fig:j0613_2} and \ref{fig:j1939_2}) we present posterior distributions for chromatic noise power-law parameters $A_\text{C} \propto \nu^{-\chi_\text{C}}$ and $\gamma_\text{C}$, where $\nu$ is a radio frequency. The shaded regions correspond to 1-dimensional 1-$\sigma$ credible levels and 2-dimensional 1-$\sigma$ and 2-$\sigma$ credible levels.}
     \label{fig:j0613j1939}
 \end{figure*}

\subsection{PSR~J0613--0200: scattering variations, annual dispersion measure variations}
\label{sec:J0163}

In PSR~J0613$-$0200, we find evidence of scattering variations in the interstellar medium, as a red process, with an amplitude roughly proportional to radio-frequency to the power of $-4$: $A_\text{C} \propto \nu^{-4}$.
We also find evidence of the annual DM signal, described by Equation~\ref{eq:dmannual}.
The detection of annual DM variations is consistent with~\cite{keith2013dmvar}.
We do not find any additional evidence for stochastic DM variations in PSR~J0613$-$0200, suggesting the DM(t) is well modelled by the annual and quadratic terms.
Reconstruction of the red noise in PSR~J0613$-$0200 using the maximum-likelihood method is provided in Figure~\ref{fig:j0613_1}.
Posterior distribution for chromatic noise parameters is in Figure~\ref{fig:j0613_2}.


\subsection{PSR~J1939+2134: scattering variations, band noise}
\label{sec:1939}
Pulsar PSR~J1939+2134 is known as a pulsar with strong DM variations and spin noise.
In this study, we find that PSR~J1939+2134, like PSR~J0613$-$0200, exhibits chromatic noise with an amplitude: $A_\text{C} \propto \nu^{-4}$.
This, again, suggests the cause may be  scattering variations towards the line of sight to the pulsar.
With a narrow pulse profile and high dispersion, the pulsar is expected to show multipath propagation effects \cite[][]{ramachandran1937}. 
Posterior distribution for chromatic noise parameters is presented in Figure~\ref{fig:j1939_2}.
Additionally, we find evidence of band noise in 40-cm and 50-cm observations.
We reconstruct red noise processes in PSR~J1939+2134 in Figure~\ref{fig:j1939_1}.

\subsection{PSR~J1017--7156 and PSR~J1045$-$4509: chromatic noise, dominated by DM variations}
\label{sec:j1017j1045}
In PSR~J1017$-$7156, we find a chromatic noise with $\chi = 2.29$, which is close to what we expect from DM variations.
In PSR~J1045$-$4509 we also find a similar chromatic noise with $\chi = 1.82$.
When chromatic noise is included in the noise model, we find no evidence of DM noise in these two pulsars.
We find $\ln \mathcal{B}$ in favour of excess chromatic noise in addition to DM are 3.9 and 3.2 for PSR~J1017$-$7156 and PSR~J1045$-$4509, respectively.
We interpret these noise processes, as dispersion measure variations on a similar time scale as scattering of radio pulses in the interstellar medium.
In Figure~\ref{fig:j1017j1045}, we provide maximum-likelihood reconstructions of the chromatic noise in PSR~J1017$-$7156 and PSR~J1045$-$4509, as well as posterior distributions for chromatic noise parameters.

\subsection{PSR~J2145--0750: an achromatic exponential dip}
\label{sec:j2145}
Initially, we found a presense of spin noise in PSR~J2145$-$0750, which later disappeared after ECORR parameters were introduced.
The red noise spectrum of this pulsar was shallow, as found by~\cite{lentati2016iptanoise}.
The maximum-likelihood red noise realisation of PSR~J2145$-$0750 contained a dip in timing residuals, like the ones found in PSRs J0437$-$4715, J1713+0747 and J1643$-$1224.
We found $\ln \mathcal{B} \approx 30$ in favour of the exponential dip on top of known red noise in PSR~J2145$-$0750.
The dip persisted in the presence of ECORR.
Results of parameter estimation for exponential dip parameters are presented in Figure~\ref{fig:event_j2145}.
Unlike the other exponential dip events, the chromatic index $\chi$ for the one in PSR~J2145$-$0750 is mostly consistent with zero.
Additionally, we do not see any significant pulse profile residuals for PSR~J2145$-$0750 at around the time of the event, which we show in Figure~\ref{fig:pulse_profiles}.
The absence of pulse profile residuals could be either the consequence of its short spin period (16 ms) or a different origin of the exponential dip.
Although evidence for spin noise is diminished after the inclusion of ECORR, it seems possible to us that the exponential dip in PSR~J2145$-$0750 is a feature of a more general non-power-law spin noise process in the pulsar, distinct from other exponential dips that we find.
Additionally, whereas we measure the chromaticity $\chi_\text{P}$ of other exponential dips to be non-zero, the exponential dip in PSR~J2145$-$0750 is completely achromatic.


\subsection{PSR~J0437--4715: profile evolution, profile events, scattering variations, band and system noise}\label{sec:j0437}

PSR~J0437$-$4715 is by far the brightest millisecond pulsar at metre and centimetre wavelengths.
Its short term timing is severely limited by pulse jitter and self noise effects \cite[][]{oslowskijitter}.
The pulsar is also susceptible to  additional instrumental noise, introduced through pulse profile distortions.
Because of its brightness, the profile can experience distortions, particularly when observed with the early backends which had low bit depth. The level of the distortions would depend on both frequency and date as the pulsar is subject to diffractive and refractive scintillation. 
The polarisation shows rapid changes in time in the region of pulse phase close to the peak of the pulse profile \cite[][]{Dai2015}.    

In PPTA-DR2, PSR~J0437$-$4715 has the longest data span because of the availability of early CPSR2 data.
The timing data contains two main important features that we present and try to account for.

The first one is related to the evolution of pulse profiles.
After least-squares fitting of the data to the timing model of DR2, we identified a clear linear dependence of timing residuals on radio-frequency for observing systems, as well as sub-systems, described in Section~\ref{sec:fdlin}.
We conclude that regular \texttt{FD} parameters in the timing model for PSR~J0437$-$4715 do not allow this effect to be properly taken into account, and we introduced the linear residual-frequency model in Equation~\ref{eq:fdlin} to serve this purpose.
Before performing model selection for band and system noise, we perform model selection for system-specific frequency dependence.
We find evidence for linear dependence of residuals on radio frequency in these systems: \texttt{PDFB\_20CM}, \texttt{CPSR2\_50CM}, \texttt{CASPSR\_40CM}, \texttt{PDFB1\_early\_20CM}, \texttt{WBCORR\_10CM}, \texttt{PDFB1\_1433}, \texttt{PDFB1\_10CM}, \texttt{PDFB\_40CM}. 
Therefore, we include linear dependence of timing residuals on radio frequency for the above systems into the noise model of PSR~J0437$-$4715.


\begin{figure*}

    \centering
    \begin{subfigure}[b]{0.7\textwidth}
        \includegraphics[width=\textwidth]{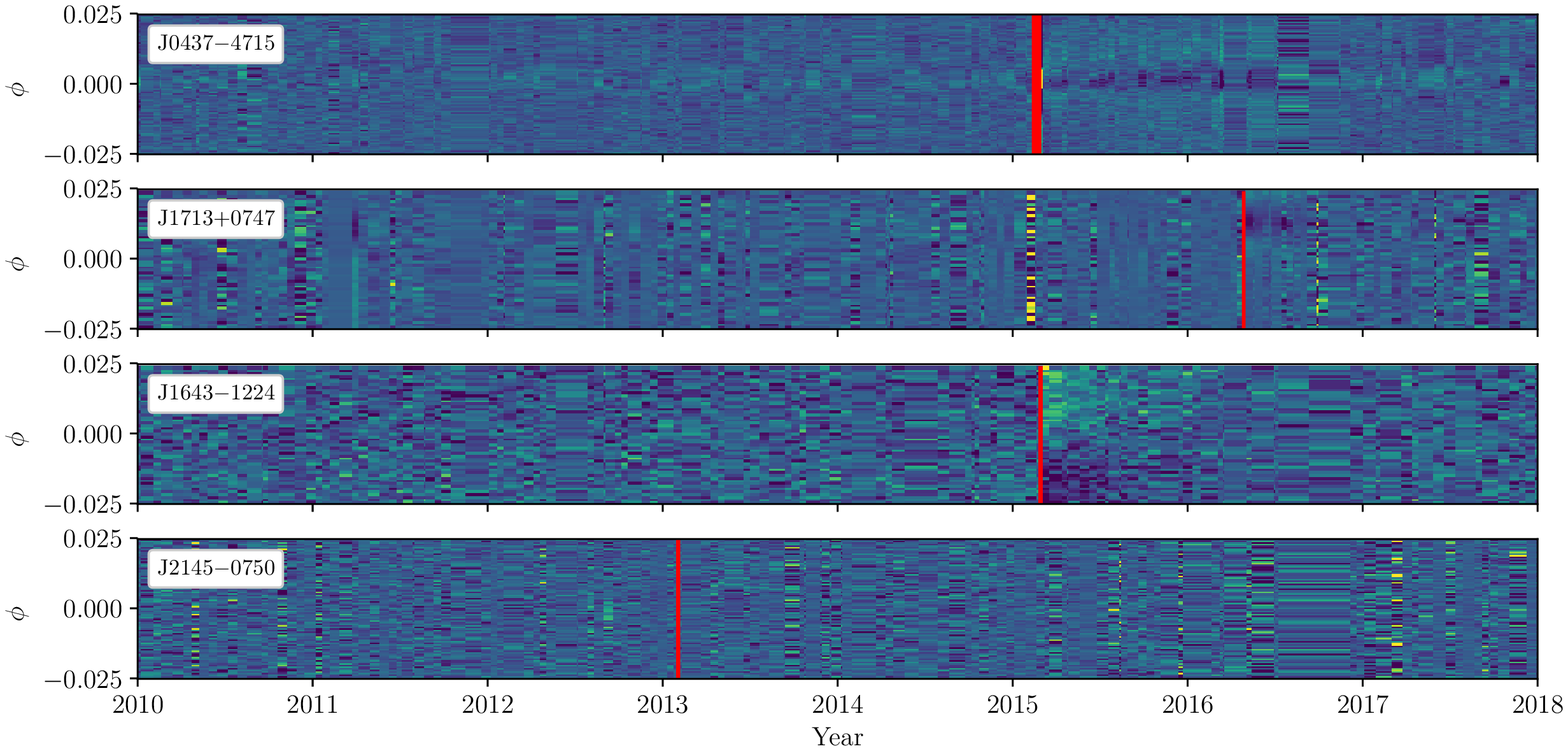}
       \caption{Pulse profile residuals\\~}
        \label{fig:pulse_profiles}
    \end{subfigure}

    \begin{subfigure}[b]{0.36\textwidth}
        \includegraphics[width=\textwidth]{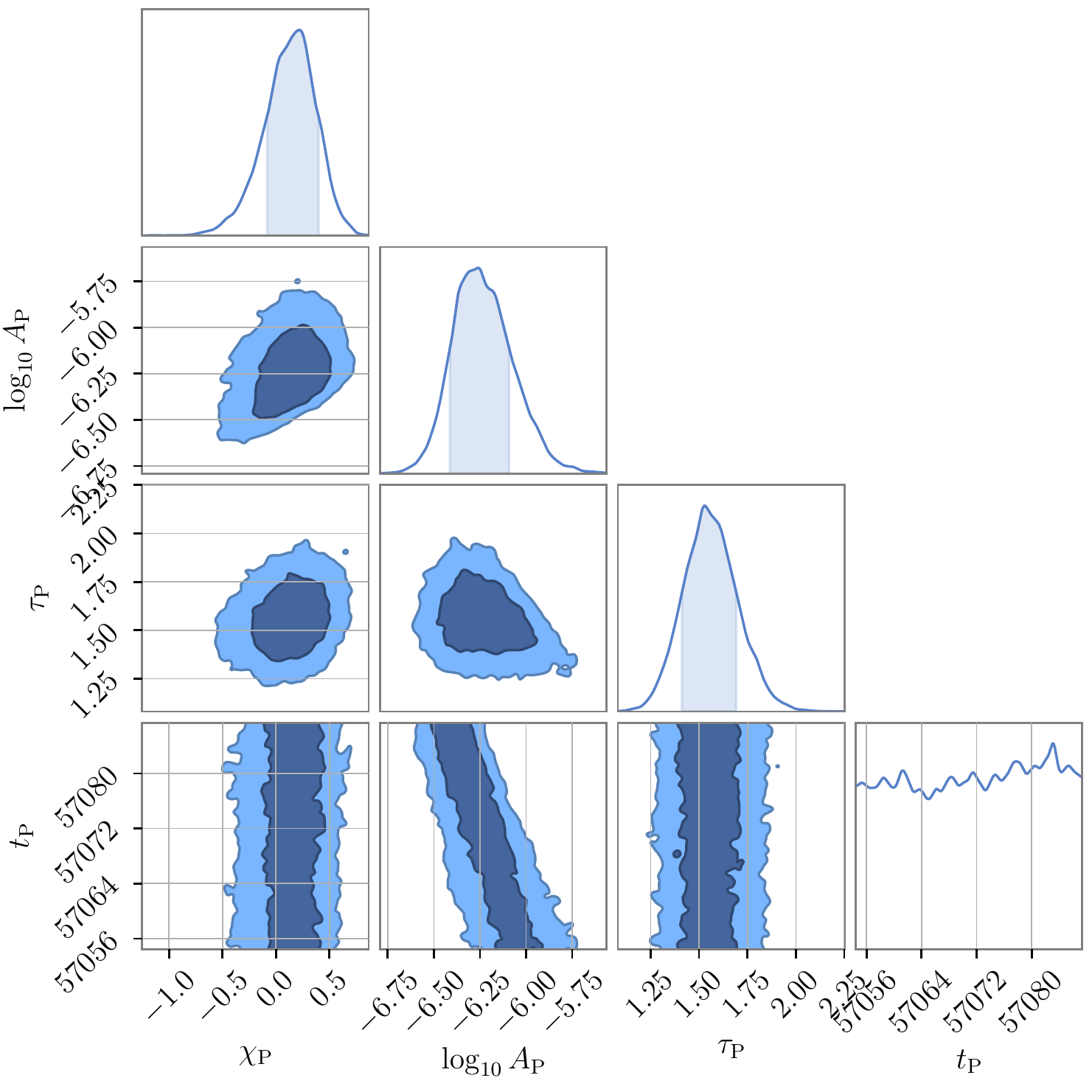}
       \caption{PSR~J0437$-$4715}
        \label{fig:event_j0437}
    \end{subfigure}
    \begin{subfigure}[b]{0.36\textwidth}
        \includegraphics[width=\textwidth]{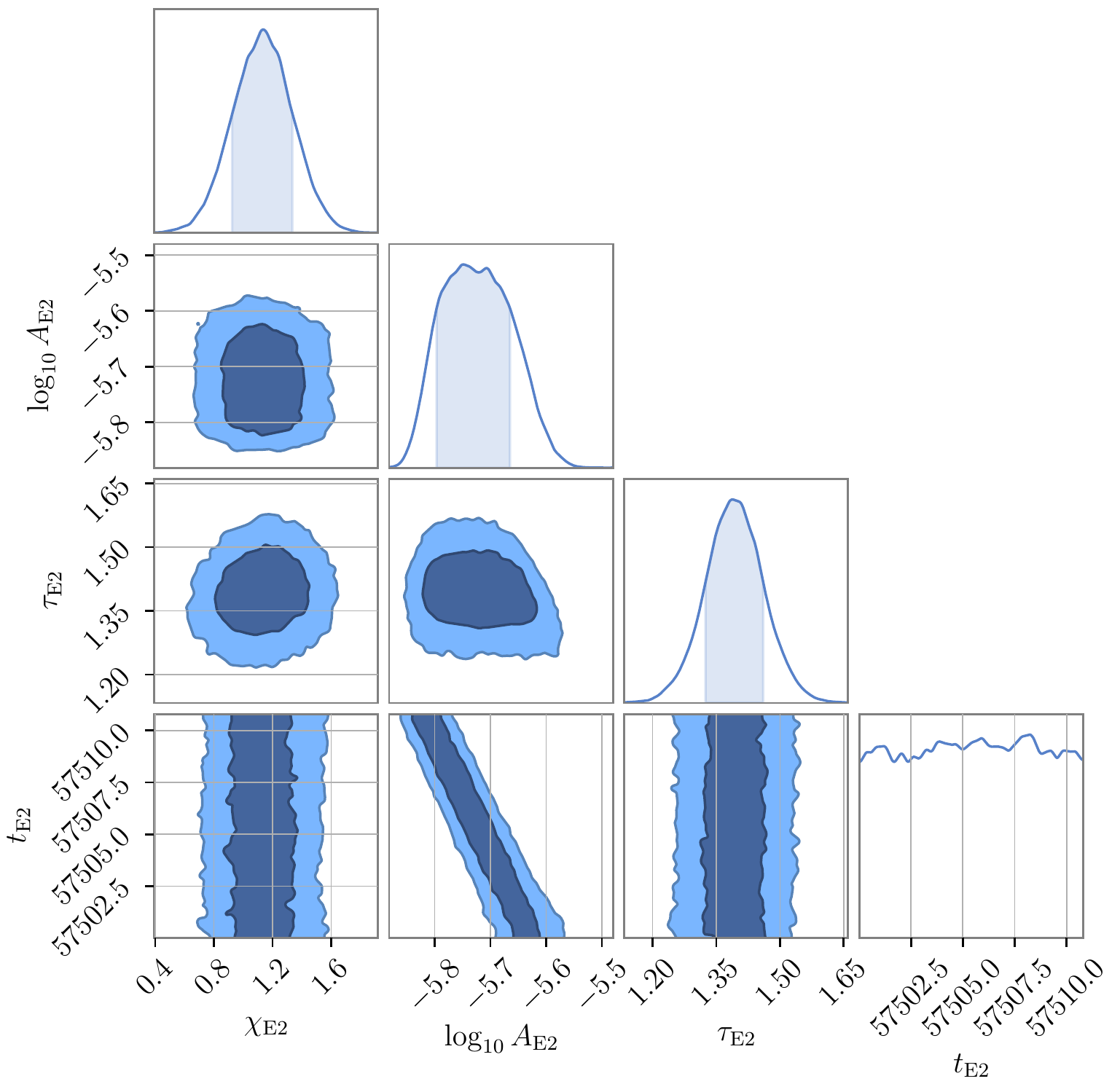}
        \caption{PSR~J1713$+$0747}
        \label{fig:event_j1713}
    \end{subfigure}
    
    \begin{subfigure}[b]{0.36\textwidth}
        \includegraphics[width=\textwidth]{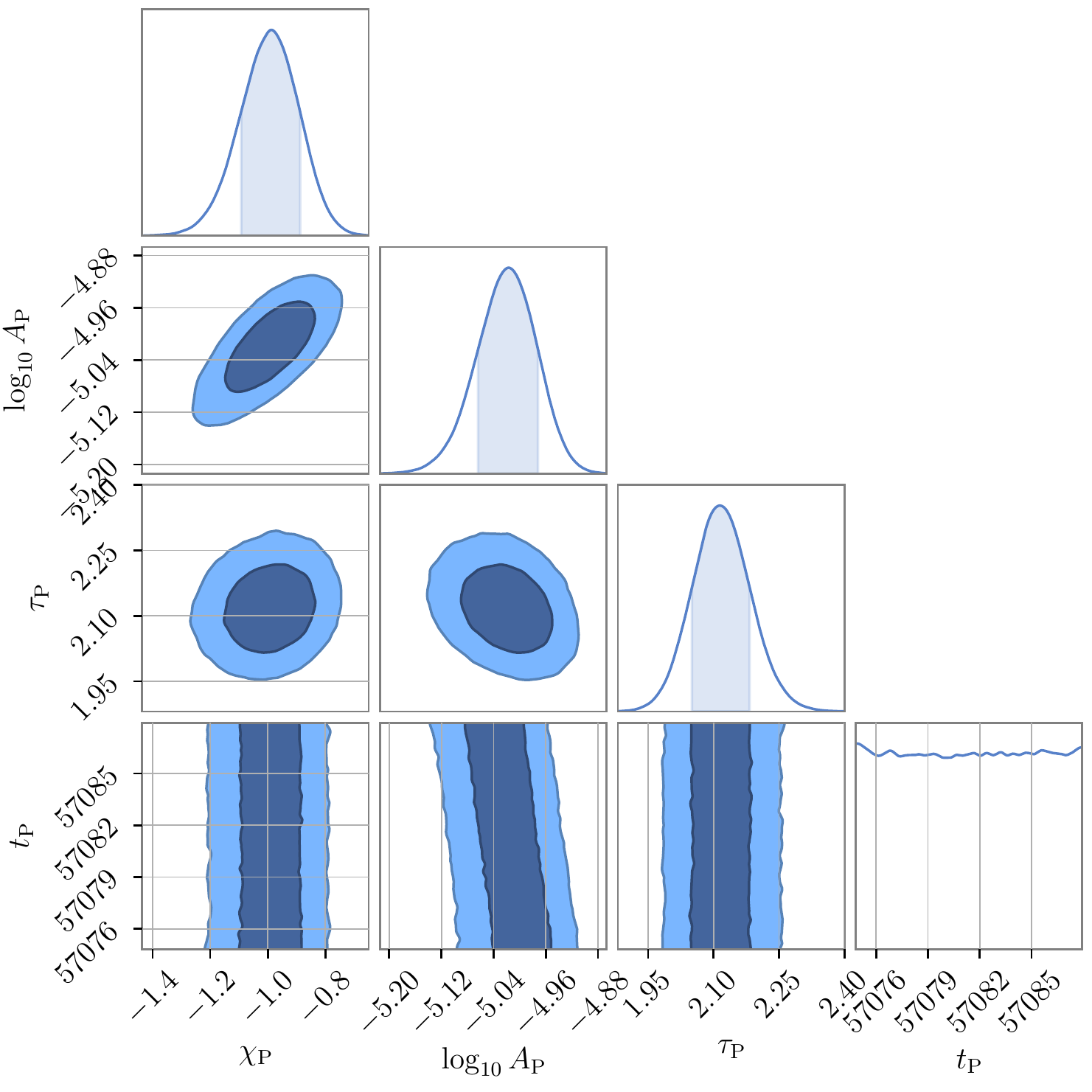}
       \caption{PSR~J1643$-$1224}
        \label{fig:event_j1643}
    \end{subfigure}
    \begin{subfigure}[b]{0.36\textwidth}
        \includegraphics[width=\textwidth]{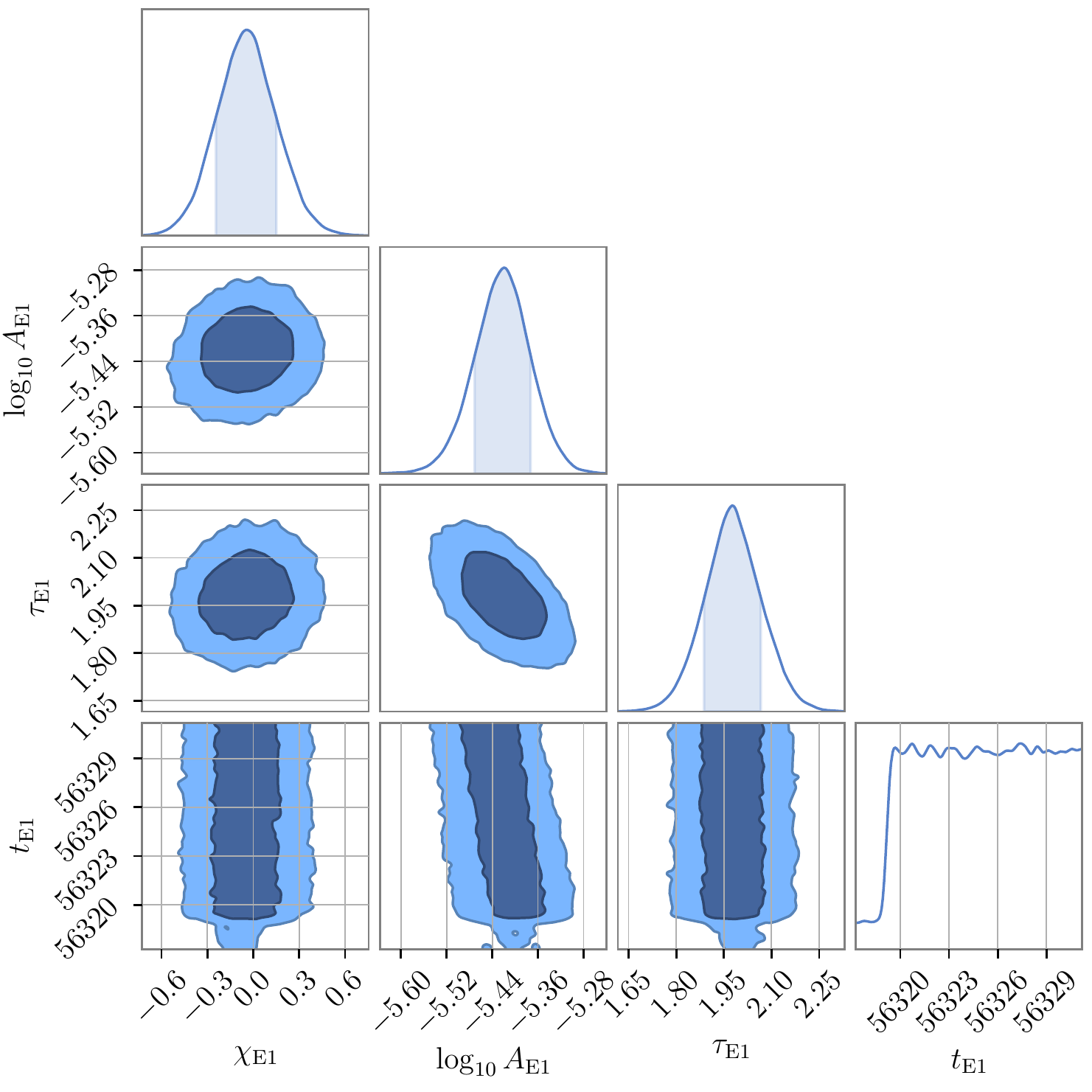}
        \caption{PSR~J2145$-$0750}
        \label{fig:event_j2145}
    \end{subfigure}
    
    \caption{Exponential dips and profile shape events. Top panel (\ref{fig:pulse_profiles}): profile residuals (colour) as a function of pulse phase ($\phi$) and year from \textsc{PDFB4} observations for PSRs J0437$-$4715 (10\,cm), J1713+0747 (20\,cm), and J1643$-$1224 (10\,cm). Red vertical lines correspond to the 1-$\sigma$ credible intervals of inferred start times of chromatic exponential dips in the timing residuals. We also include PSR~J2145$-$0750, where we identified an apparent exponential dip with $\chi$ consistent with zero. Yellow outliers can be caused by RFI missed by the data-processing pipeline, pulse signal-to-noise ratio (i.e., for J1713+0747, in early 2015). For J0437$-$4715, there is also a large profile distortion associated with the exponential event (discussed in Section~\ref{sec:j0437}). In panels below (\ref{fig:event_j0437},\ref{fig:event_j1713},\ref{fig:event_j1643},\ref{fig:event_j2145}), we provide posterior distributions for inferred exponential dip parameters: the time of the event $t$, the chromatic index $\chi$, the duration $\tau$, the amplitude $A$ [s]. The posterior distributions for event times look uniform, but the event times are, in fact, well constrained to within a few ToAs. The shaded regions correspond to 1-dimensional 1-$\sigma$ credible levels and 2-dimensional 1-$\sigma$ and 2-$\sigma$ credible levels.}
    \label{fig:profile_events}
\end{figure*}

The second feature is the sudden change in the timing residuals of PSR~J0437$-$4715 around MJD 57070, which is correlated with a change in the pulse profile. 
In Figure~\ref{fig:profile_events}, we plot the evolution of timing residuals and profile residuals, to demonstrate the clear connection between the two.
In our model selection, we include profile event, modelled by Equation~\ref{eq:model_expdip}.
Otherwise, the profile event will be absorbed in stochastic component of the noise.
We find the chromaticity of the event to be consistent with zero.
In Figure~\ref{fig:pulse_profiles} we added an additional observation for J0437$-$4715 at around the time of the event, which was initially flagged as an RFI and looks like a hint of a yellow dot behind the red line. Later we found that this observation causes an advance in pulse arrival time across all three radio bands, which is not consistent with a typical RFI.
The timing residuals for this observation also line up with the exponential event.
Thus, there is a reason to consider this observation the earliest observation of the exponential dip, associated with the profile event.
The additional observation, which took place at MJD 57073, also separates exponential events for J0437$-$4715 and J1643$-$1224 in time.
Initially, errors for event time estimates were overlapping, which may seem like a spurious coincidence.
The additional observation for J0437$-$4715 narrows down the uncertainty, whereas MJD 57073 is ruled out by the data from PSR~J1643$-$1224.

Compared to the other PPTA pulsars, PSR~J0437$-$4715 shows the largest number of red noise processes.
In the IPTA DR1 data set, \cite{lentati2016iptanoise} found evidence of band noise in all bands and system noise in \texttt{CPSR2\_20CM}.
In PPTA-DR2, consistent with \cite{lentati2016iptanoise}, we find evidence of red noise in \texttt{CPSR2\_20CM} system, and we also find red noise in three additional systems: \texttt{WBCORR\_10CM}, \texttt{CPSR2\_10CM}, \texttt{CPSR2\_50CM}.
We also find evidence of band noise in both 20-cm and joint 40-50-cm data.
On top of that, we find excess chromatic noise with $\chi$ consistent with 4.
Maximum-likelihood reconstruction of the red noise in PSR~J0437$-$4715, as well as the posterior distribution for power-law parameters of the chromatic noise, are provided in Figure~\ref{fig:j0437}.
PSR~J0437$-$4715 has the lowest DM in the PPTA data set.
It is surprising to identify chromatic noise with a chromactic index close to what would be expected from scattering variations. 
However this frequency scaling is consistent with what would be predicted from angle of arrival variations \citep{romaniaoa,cs2010}.
An alternative possible origin of the chromatic noise could be in the evolution of pulse profile.
Our red noise analysis was conducted after fitting for a second derivative of pulsar spin frequency, \texttt{F2}. 

Given its high timing precision, PSR~J0437$-$4715 provides great sensitivity to gravitational waves.
Future improvements to the noise model for PSR~J0437$-$4715 and profile-domain timing analyses would provide further answers about the origin of chromatic noise in this pulsar.




\begin{table*}
\caption{\label{tab:results_band_system_noise}Results for band noise and system noise parameter estimation and model selection. The second column contains flags and the corresponding flag values to select pulse times of arrival, which are affected by the red noise. The third and the fourth columns contain red noise power-law parameters. The fifth column is the $\ln \mathcal{B}$ in favour of the band/system noise in addition to white noise, spin noise and DM noise, against the same model without the band/system noise term. The last column represents the number of Fourier components, explained in Section~\ref{sec:rednoise}. Parameters with asterisks are estimated from the median marginalised posterior values, while other parameter estimates are calculated from the maximum-posterior values.}
\renewcommand{\arraystretch}{1.4}
\begin{tabular}{| l c c c c c | }
\hline

PSR & Flag and value & $\log_{10}A$ & $\gamma$ & $\text{ln} \mathcal{B}$ & $n_\text{c}$ \\ [1ex] \hline \hline 
\multirow{2}{4em}{J0437$-$4715} & \texttt{-group CPSR2\_10CM} & $-13.40^{+0.09}_{-0.10}$ & $1.13^{+0.38}_{-0.44}$ & 156.9 &  36 \\
 & \texttt{-group CPSR2\_20CM} & $-13.28^{+0.11}_{-0.07}$ & $1.18^{+0.54}_{-0.34}$ & 284.9 &  39 \\
 & \texttt{-group WBCORR\_10CM} & $-13.47^{+0.19}_{-0.10}$ & $0.41^{*+0.57}_{-0.30}$ & 85.9 &  10 \\
 & \texttt{-group CPSR2\_50CM} & $-12.85^{+0.08}_{-0.08}$ & $2.43^{+0.47}_{-0.24}$ & 323.6 &  36 \\
 & \texttt{-B 40CM -B 50CM} & $-13.51^{+0.10}_{-0.16}$ & $1.88^{+0.87}_{-0.20}$ & 356.8 & 90  \\
 & \texttt{-B 20CM} & $-13.80^{+0.08}_{-0.07}$ & $2.90^{+0.43}_{-0.23}$ & 353.8 & 90  \\\hline 
J1600$-$3053 & \texttt{-B 40CM -B 50CM} & $-12.61^{+0.08}_{-0.10}$ & $1.85^{+0.36}_{-0.28}$ & 27.9 &  86 \\\hline 
J1643$-$1224 & \texttt{-B 40CM -B 50CM} & $-12.06^{+0.04}_{-0.04}$ & $2.33^{+0.30}_{-0.19}$ & 105.4 & 85  \\\hline 
\multirow{2}{4em}{J1713+0747} & \texttt{-B 10CM -B 20CM} & $-14.49^{+0.23}_{-0.65}$ & $4.18^{+1.22}_{-0.91}$ & 4.4 &  86 \\
 & \texttt{-group CPSR2\_20CM} & $-13.57^{+0.21}_{-0.19}$ & $1.60^{+1.35}_{-0.22}$ & 21.9 & 36  \\\hline 
J1744$-$1134 & \texttt{-B 10CM -B 20CM} & $-13.56^{+0.17}_{-0.02}$ & $1.00^{+0.54}_{-0.20}$ & 12.7 & 86  \\\hline 
J1824$-$2452A & \texttt{-B 40CM -B 50CM} & $-12.06^{+0.07}_{-0.06}$ & $1.01^{+0.32}_{-0.19}$ & 78.2 & 83  \\\hline 
J1909$-$3744 & \texttt{-B 40CM -B 50CM} & $-13.42^{+0.10}_{-0.08}$ & $0.64^{+0.47}_{-0.21}$ & 62.3 & 85  \\\hline 
J1939+2134 & \texttt{-B 40CM -B 50CM} & $-13.16^{+0.10}_{-3.29}$ & $3.13^{*,+4.33}_{-1.57}$ & 135.1 & 84  \\\hline 
J2124$-$3358 & \texttt{-B 20CM} & $-16.07^{+1.81}_{-0.87}$ & $8.40^{+0.56}_{-4.26}$ & 1.9 &  84 \\\hline 
J2145$-$0750 & \texttt{-group CPSR2\_50CM} & $-14.74^{+1.14}_{-0.43}$ & $7.18^{*,+1.97}_{-2.60}$ & 16.9 & 65  \\\hline 

\end{tabular}

\end{table*}

\subsection{Evaluating the performance of the noise models}\label{sec:testmodels}

We can further test our models by analyzing the distribution of timing residuals after the subtraction of maximum-likelihood realisations of red noise. 
Because ECORR noise is difficult to subtract, for pulsars with ECORR we perform the tests after weighted-averaging sub-banded times of arrivals.
Because the ECORR terms are analytically marginalized over arrival times, the procedure may result in excess white noise after the subtraction.
We normalise timing residuals by dividing them by the corresponding ToA errors.
If the models well describe the data we would expect these residuals to be consistent with Gaussian distribution with zero mean and unit variance.
We perform three tests to determine how Gaussian, uncorrelated and variance-stationary are whitened residuals of PPTA-DR2 pulsars.
The results are summarized in Table~\ref{tab:results_whiteness}.
We carry out the \cite{anderson1952asymptotic} test to determine if the distribution of timing residuals is consistent with Gaussian distribution with zero mean and unit variance.
We find significant non-Gaussianity for PSRs J0437$-$4715, J1022$+$1001, J1909$-$3744, J2241$-$5236, with a probability of whitened timing residuals being drawn from such distribution of less than 1\%.
All of the above pulsars contain ECORR.
The distribution of whitened residuals of PSR~J0437$-$4715 has a mean of $0.05$ and standard deviation of $1.12$.
The increased standard deviation could mean that EFAC, EQUAD and ECORR parameters are insufficient to describe the white noise, or that the weighted averaging of sub-banded observations does not represent ECORR sufficiently well.
We also test how white are the actual whitened timing residuals with the help of the statistic derived by~\cite{ljung1978whiteness}.
Four PSRs, J0437$-$4715, J1713$+$0747, J1939$+$2134, J2124$-$3358, have a probability of being uncorrelated in time of less than 1\%.
Moreover, with the~\cite{breusch1979heteroscedasticity} test, we find that whitened residuals for PSRs J0437$-$4715 and J2241$-$5236 have a probability of having temporally stationary variance\footnote{also termed homoskedasticity} of less than 1\%.
Overall, six pulsars did not pass the three tests we discussed in this paragraph.
Five of these pulsars have band or system noise, described in Table~\ref{tab:results_band_system_noise} and Section~\ref{sec:bandsystem}, with $\gamma < 0.5$.
For PSR~J2124$-$3358, we found evidence for band noise in 20-cm data with $\gamma \approx 8$, although we do not rule out that this low-frequency red noise is pulsar spin noise (see discussion in Section~\ref{sec:results_spin}).
We defer the further improvement of noise models for the above nine pulsars to future work.
The upcoming publication by Reardon et al. (in prep.) will update the data set for J0437--4715, which results in improved noise modelling.


\begin{table}
\caption{\label{tab:results_whiteness}Tests of the noise models.
Pulsars with ECORR parameters are marked by asterisks.
The second column shows Anderson-Darling statistic (ADS), with the null hypothesis that the whitened timing residuals are described by a normal distribution. Values smaller than $2.5$ are within 95-\% confidence intervals and values smaller than $3.9$ are within 99-\% confidence intervals~\citep{stephens1974edf}. PSRs J0437$-$4715, J1022$+$1001, J1909$-$3744, J2241$-$5236 have the statistic value greater than $3.9$. The third column shows Ljung–Box statistic (LBS), with the null hypothesis that the whitened timing residuals are uncorrelated in time. The fourth column reports the  p-values  (LBp) that correspond to the Ljung–Box statistic values. We choose a number of Ljung–Box lags to be equal to $n_\text{c}$, listed in Table~\ref{tab:results_wrd}. For four PPTA PSRs we obtain a probability of whitened residuals of being uncorrelated in time of less than 1\%: J0437$-$4715, J1713$+$0747, J1939$+$2134, J2124$-$3358. The last two columns contain Breusch-Pagan statistic and corresponding p-values, with the null hypothesis that variance of the whitened timing residuals is constant in time. Only two PSRs, J0437$-$4715 and J2241$-$5236, have a statistically-significant probability of having non-stationary excess variance.}
\renewcommand{\arraystretch}{1.4}
\begin{tabular}{ | l c c c c c | }
\hline

PSR & ADS & LBS & LBp & BPS & BPp\\ [1ex] \hline \hline 
J0437$-$4715$^*$ & $27.4$ & $997$ & $2 \times 10^{-152}$ & $339.08$ & $8 \times 10^{-73}$ \\
J0613$-$0200$^*$ & $2.4$ & $124$ & $5 \times 10^{-3}$ & $2.16$ & $1 \times 10^{-1}$ \\ 
J0711$-$6830 & $0.4$ & $81$ & $6 \times 10^{-1}$ & $0.00$ & $1 \times 10^{0}$ \\ 
J1017$-$7156$^*$ & $1.5$ & $69$ & $2 \times 10^{-2}$ & $0.88$ & $3 \times 10^{-1}$ \\ 
J1022+1001$^*$ & $11.3$ & $82$ & $6 \times 10^{-1}$ & $1.56$ & $2 \times 10^{-1}$ \\ 
J1024$-$0719 & $0.3$ & $109$ & $4 \times 10^{-2}$ & $0.02$ & $9 \times 10^{-1}$ \\ 
J1045$-$4509 & $1.2$ & $84$ & $6 \times 10^{-1}$ & $0.03$ & $9 \times 10^{-1}$ \\ 
J1125$-$6014 & $0.2$ & $78$ & $4 \times 10^{-1}$ & $4.56$ & $3 \times 10^{-2}$ \\ 
J1446$-$4701$^*$ & $0.4$ & $56$ & $1 \times 10^{-1}$ & $1.86$ & $2 \times 10^{-1}$ \\ 
J1545$-$4550 & $0.4$ & $41$ & $5 \times 10^{-1}$ & $0.64$ & $4 \times 10^{-1}$ \\ 
J1600$-$3053$^*$ & $3.9$ & $122$ & $6 \times 10^{-3}$ & $0.14$ & $7 \times 10^{-1}$ \\ 
J1603$-$7202$^*$ & $3.8$ & $80$ & $7 \times 10^{-1}$ & $9.46$ & $2 \times 10^{-3}$ \\ 
J1643$-$1224 & $0.4$ & $100$ & $1 \times 10^{-1}$ & $0.15$ & $7 \times 10^{-1}$ \\ 
J1713+0747$^*$ & $1.8$ & $164$ & $9 \times 10^{-7}$ & $2.98$ & $8 \times 10^{-2}$ \\ 
J1730$-$2304 & $0.5$ & $86$ & $5 \times 10^{-1}$ & $1.37$ & $2 \times 10^{-1}$ \\ 
J1732$-$5049 & $0.9$ & $46$ & $4 \times 10^{-1}$ & $0.69$ & $4 \times 10^{-1}$ \\ 
J1744$-$1134$^*$ & $1.5$ & $94$ & $3 \times 10^{-1}$ & $0.26$ & $6 \times 10^{-1}$ \\ 
J1824$-$2452A$^*$ & $3.6$ & $93$ & $2 \times 10^{-1}$ & $9.74$ & $2 \times 10^{-3}$ \\ 
J1832$-$0836 & $0.4$ & $50$ & $2 \times 10^{-2}$ & $0.03$ & $9 \times 10^{-1}$ \\ 
J1857+0943 & $1.1$ & $79$ & $7 \times 10^{-1}$ & $0.26$ & $6 \times 10^{-1}$ \\ 
J1909$-$3744$^*$ & $5.6$ & $93$ & $3 \times 10^{-1}$ & $1.00$ & $3 \times 10^{-1}$ \\ 
J1939+2134$^*$ & $1.7$ & $153$ & $9 \times 10^{-6}$ & $2.27$ & $1 \times 10^{-1}$ \\ 
J2124$-$3358 & $3.7$ & $199$ & $7 \times 10^{-11}$ & $0.47$ & $5 \times 10^{-1}$ \\ 
J2129$-$5721 & $0.3$ & $55$ & $1 \times 10^{0}$ & $1.88$ & $2 \times 10^{-1}$ \\ 
J2145$-$0750$^*$ & $3.0$ & $89$ & $4 \times 10^{-1}$ & $1.27$ & $3 \times 10^{-1}$ \\ 
J2241$-$5236$^*$ & $4.0$ & $68$ & $4 \times 10^{-2}$ & $51.49$ & $2 \times 10^{-12}$ \\

\hline
\end{tabular}
\end{table}

\section{Conclusion}
\label{sec:conclusion}

We have robustly determined noise models for the PPTA DR2 pulsars, which include models for red noise processes (Tables \ref{tab:results_chromatic}, \ref{tab:results_wrd}, and \ref{tab:results_band_system_noise}) and specific models for deterministic signals, described throughout Section~\ref{sec:results}.
These models can be used in timing analyses of the pulsars.

In our analysis we considered each pulsar independently.
Therefore, our red noise models may have absorbed spatially correlated signals, such as the  gravitational-wave background or errors in Solar System ephemeris.
For example, the low fluctuation frequency red noise observed in PSRs J1713$+$0747 and J1909$-$3744 can possibly be attributed to this. 
Although the processes could still be detected with our noise models, either because of the fact that they are correlated between pulsars or because of the non-power-law model, some of our red noise terms may decrease the sensitivity to these signals one is interested in studying.
This effect can be mitigated by the additional model comparison between the desired signal and such red noise terms in our noise models that are likely to be co-variant with the signal.

Due to the exceptional brightness and hence the high individual pulse signal-to-noise ratio of PSR~J0437$-$0715, pulse profile evolution becomes a significant noise component.
We expect this source of noise will become even more important for data from high-sensitivity radio telescopes.
An effective way to account for this noise is to perform profile-domain timing.
The methodology has been outlined in~\cite{lentati2017profiledomaintim}.

We measured some band and system noise power-law indices to be nearly zero, which indicates that the power-law model might not be the best one to describe these noise processes.

Maximum-likelihood noise reconstructions and the tests, described in Table~\ref{tab:results_whiteness}, complement Bayesian inference and provide an opportunity to validate noise models for current \cite[][]{iptadr2} and future IPTA data releases.

Chromatic effects will become more apparent with the deployment of future wide-band receiver systems.
At Parkes, pulsar timing array observations are being undertaken with the ultrawide-band (low) receiver at Parkes \cite[][]{hobbsuwl}, which records data over a contiguous band from 700\,MHz to 4.2\,GHz.
The MeerTime project (Bailes et al., PASA, submitted) is currently conducting sensitive observations of millisecond pulsars over an octave bandwidth.
Wide-band systems are planned or proposed for many additional telescopes as well.

\section*{Data availability}
\label{sec:dataavail}
The second data release of the PPTA underlying this article is available
at \href{https://doi.org/10.25919/5db90a8bdeb59}{doi.org/10.25919/5db90a8bdeb59}. The list of pulsar noise models and the software to reproduce the results of our parameter estimation is available at \href{https://github.com/bvgoncharov/ppta_dr2_noise_analysis}{github.com/bvgoncharov/ppta\_dr2\_noise\_analysis}.
Other data will be shared on a reasonable request to the corresponding author.

\section*{Acknowledgements}
\label{sec:acknowledgements}
We thank William A. Coles for useful comments on the paper.
The Parkes radio telescope is part of the Australia Telescope, which is funded by the Commonwealth Government for operation as a National Facility managed by CSIRO. This paper includes archived data obtained through the Australia Telescope Online Archive and the CSIRO Data Access Portal (\href{http://data.csiro.au}{data.csiro.au}).
E.T. acknowledges Australian Research Council grant FT150100281.
M.B., R.M.S, and R.S. acknowledge Australian Research Council grant FL150100148.
R.M.S. also acknowledges funding support through Australian Research Council Future Fellowship FT190100155.
Parts of this research were conducted by the Australian Research Council Centre of Excellence for Gravitational Wave Discovery (OzGrav), through project number CE170100004.
J.W. is supported by the Youth Innovation Promotion Association of Chinese Academy of Sciences. Work at NRL is supported by NASA.  This research has made use of NASA's Astrophysics Data System.






\bibliographystyle{mnras}
\bibliography{mybib} 





\appendix

\section{Prior probability distributions for the PPTA DR2 noise model selection}
\label{sec:priors}
\begin{table}
\caption{\label{tab:priors}Priors used for our model selection study. In the top five rows of the table we list priors for stochastic signals, described in Section~\ref{sec:rednoise}. In the remaining rows, we list priors for deterministic signals, described in Section~\ref{sec:determ}.}
\renewcommand{\arraystretch}{1.4}
\begin{tabular}{| l c c | }
\hline
Parameter $\bm{\theta}$ [unit] & PSR & Prior $\pi(\bm{\theta})$ \\ [1ex] \hline
\hline
EFAC & all & $\mathcal{U}(0.01,10)$ \\ [2pt] 
EQUAD [s] & all & $\text{log}_{10}~\mathcal{U}(10^{-8.5},10^{-5})$ \\ [2pt]
$A$ & all & $\text{log}_{10}~\mathcal{U}(10^{-20},10^{-8})$ \\ [2pt]
$\gamma$ & all & $\mathcal{U}(0,10)$ \\ [2pt] 
$\chi$ & all & $\mathcal{U}(0,6)$ \\ [2pt] \hline
$A_{\text{E}}$ [s] & all & $\text{log}_{10}~\mathcal{U}(10^{-10},10^{-2})$ \\ [2pt]
$t_{\text{E}}$ [MJD] & J1713+0747 (1) & $\mathcal{U}(54500,54900)$ \\ [2pt]
~ & J1713+0747 (2) & $\mathcal{U}(57500,57520)$ \\ [2pt]
~ & J1643$-$1224 & $\mathcal{U}(57050,57150)$ \\ [2pt]
~ & J0437$-$4715 & $\mathcal{U}(57050,57150)$ \\ [2pt]
~ & J2145$-$0750 & $\mathcal{U}(56100,56500)$ \\ [2pt]
$\chi_\text{E}$ & all & $\mathcal{U}(-7,7)$ \\ [2pt]
$\log_{10} \tau_{\text{E}}$ [MJD] & J1713+0747 (1) & $\mathcal{U}(\log_{10}5,3)$ \\ [2pt]
~ & all & $\mathcal{U}(\log_{10}5,2)$ \\ [2pt]
$A_{\text{G}}$ [s] & J1603$-$7202 & $\text{log}_{10}~\mathcal{U}(10^{-6},10^{-1})$ \\ [2pt]
$t_{\text{G}}$ [MJD] & J1603$-$7202 & $\mathcal{U}(53710,54070)$ \\ [2pt]
$\sigma_G$ [MJD] & J1603$-$7202 & $\mathcal{U}(20,140)$ \\ [2pt]
$A_\text{Y}$ [s] & all & $\text{log}_{10}~\mathcal{U}(10^{-10},10^{-2})$ \\ [2pt]
$\phi_\text{Y}$ & all & $\mathcal{U}(0,2 \pi)$ \\ [2pt] 
\hline
\end{tabular}
\end{table}

\section{Consistency between inferences}
\label{sec:consistency}
\begin{figure}
\centering
\includegraphics[width=0.87\linewidth]{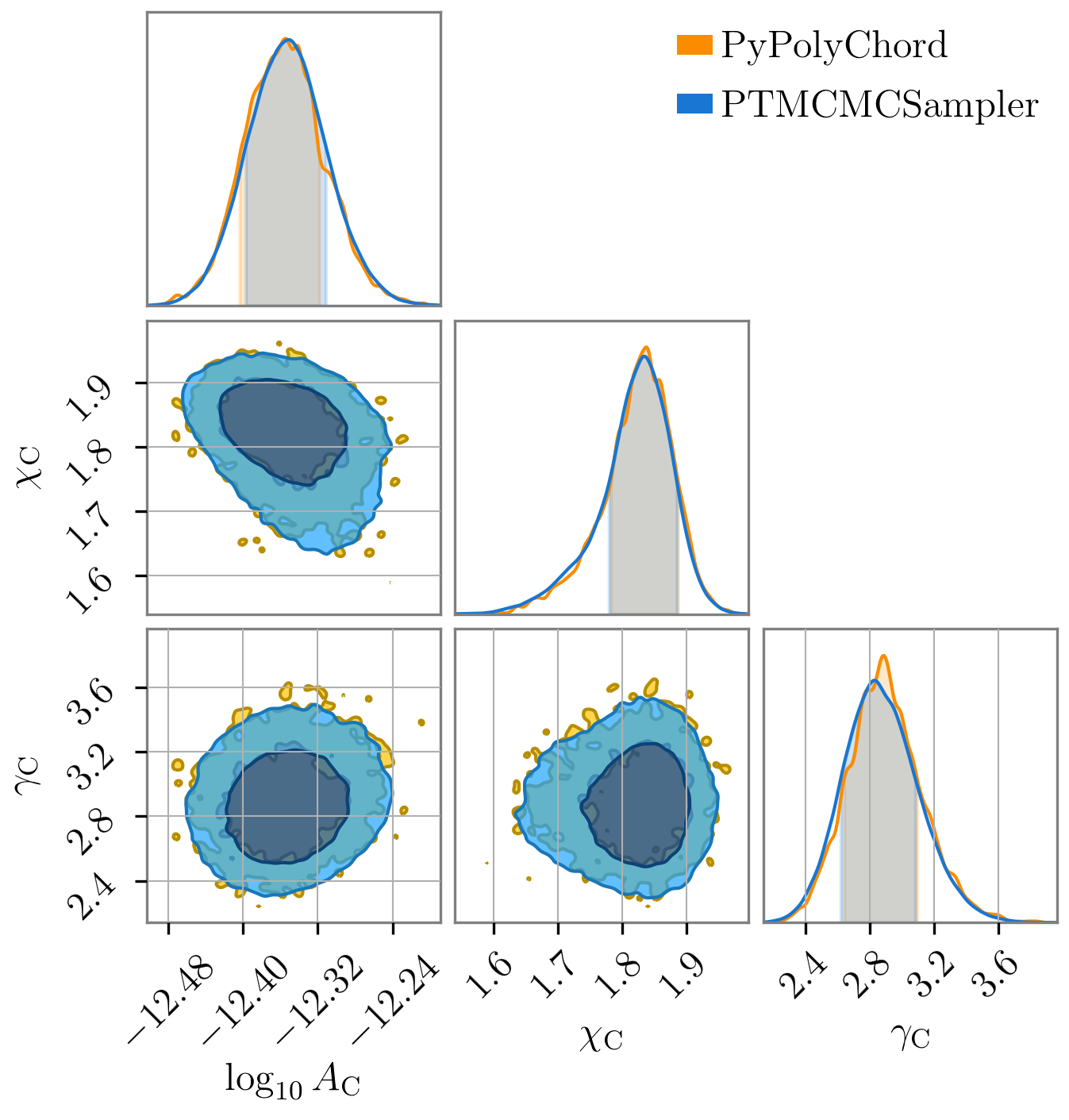}
\caption{
Comparison of different methods of estimating noise parameters for PSR~J1045$-$4509.
One set of posterior samples is obtained with \textsc{ptmcmcsampler} and white noise parameters were set as free parameters.
Another set is obtained with \textsc{polychordlite} and white noise parameters were fixed at their maximum-posterior values, obtained as described in Section~\ref{sec:results}.
The two methods yield the same posterior distribution.
}
\label{fig:consistency}
\end{figure}

We discussed different posterior sampling methods and Bayes factor calculations methods throughout the paper.
In this Section, we comment on the consistency of these methods.
We checked that parameter estimation with \textsc{polychordlite} and \textsc{ptmcmcsampler} yield consistent results.
This is confirmed in Figure~\ref{fig:consistency}, where we show the posterior distributions derived for the chromatic noise in PSR~J1045$-$4509.
We also checked that Bayes factors, calculated using the product-space method using \textsc{ptmcmcsampler} are consistent with those, calculated from evidences, obtained with nested sampling and \textsc{polychordlite}.
For example, the reported $\text{ln} \mathcal{B}$ for 20-cm band noise in PSR~J2241$-$5236 is 6.3 was obtained from evidences with \textsc{polychordlite}.
Using the product-space sampling method and \textsc{ptmcmcsampler}, we obtain $\text{ln} \mathcal{B}$ of 6.5, with 717576 samples in favour of 20-cm band noise and 1074 samples of the null hypothesis.
The equivalence between nested sampling and the product-space method has also been discussed in~\cite{goncharov2019turnover}.

\newpage

\section{Red noise reconstruction for the remaining PPTA DR2 pulsars}
\label{sec:noiserealizations}

\begin{figure*}
    \centering
    \begin{subfigure}[b]{0.32\textwidth}
        \includegraphics[width=\textwidth]{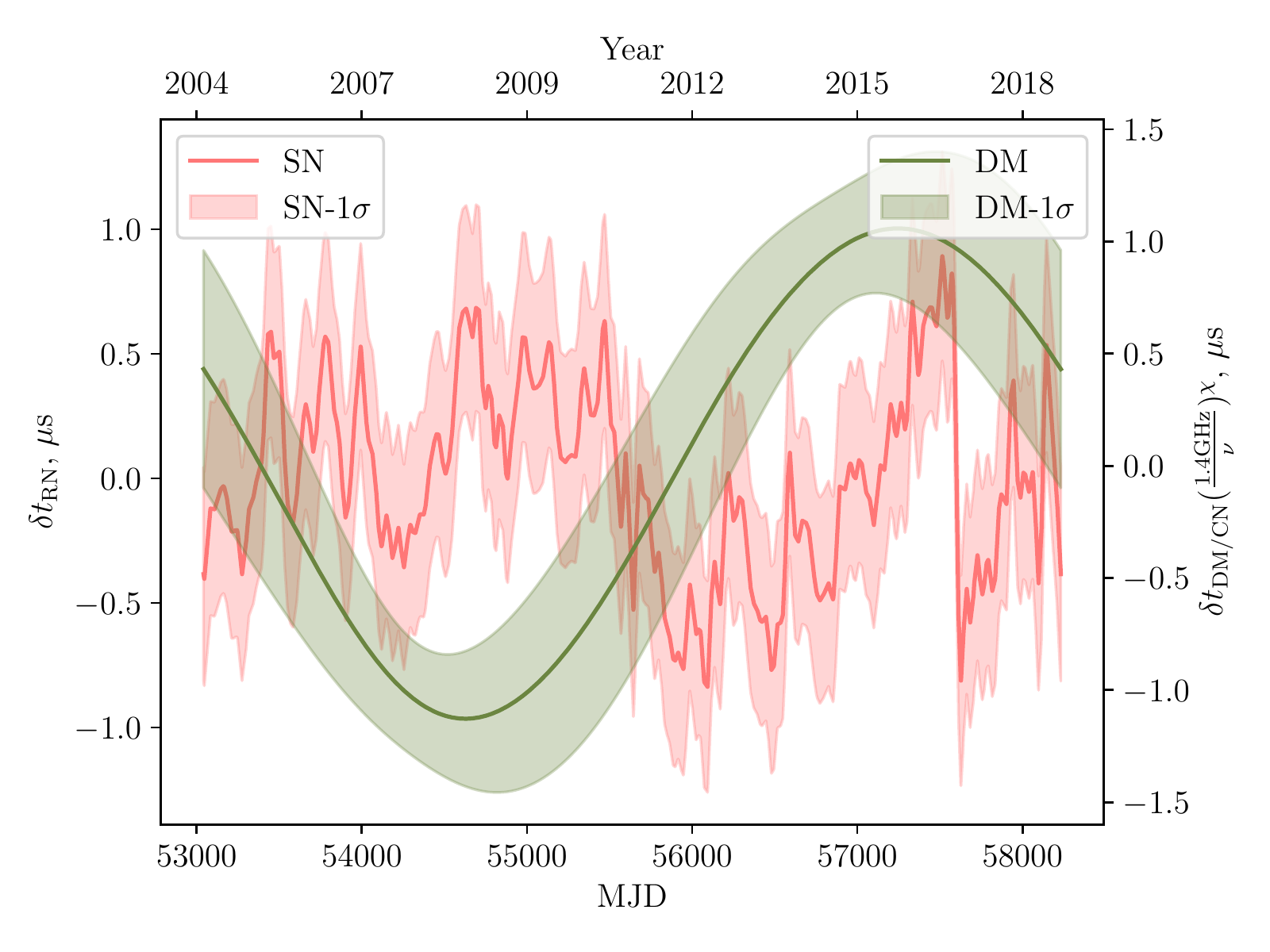}
       \caption{PSR~J0711-6830}
        \label{fig:J0711}
    \end{subfigure}
    \begin{subfigure}[b]{0.32\textwidth}
        \includegraphics[width=\textwidth]{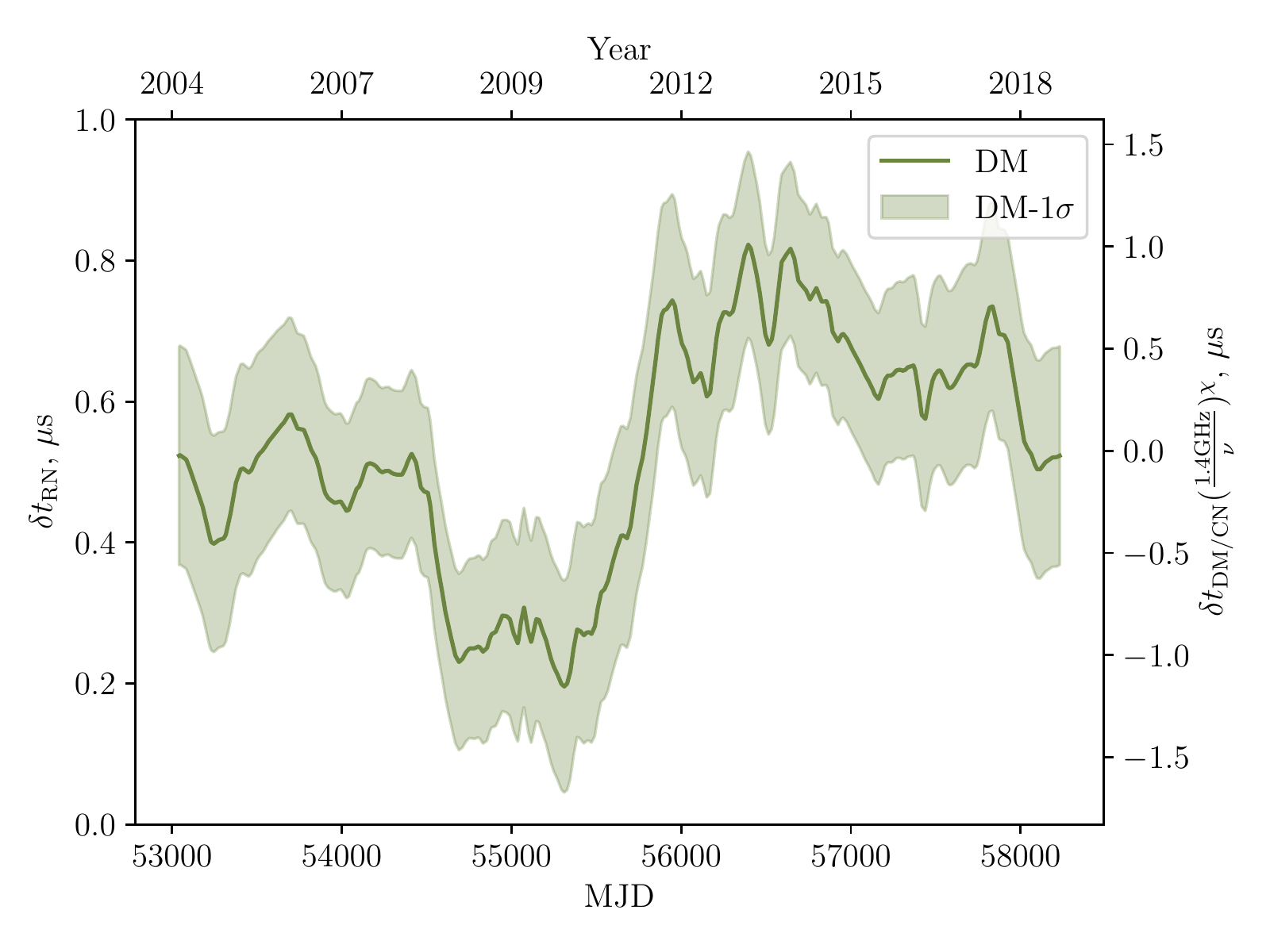}
        \caption{PSR~J1022+1001}
        \label{fig:J1022}
    \end{subfigure}
    \begin{subfigure}[b]{0.32\textwidth}
        \includegraphics[width=\textwidth]{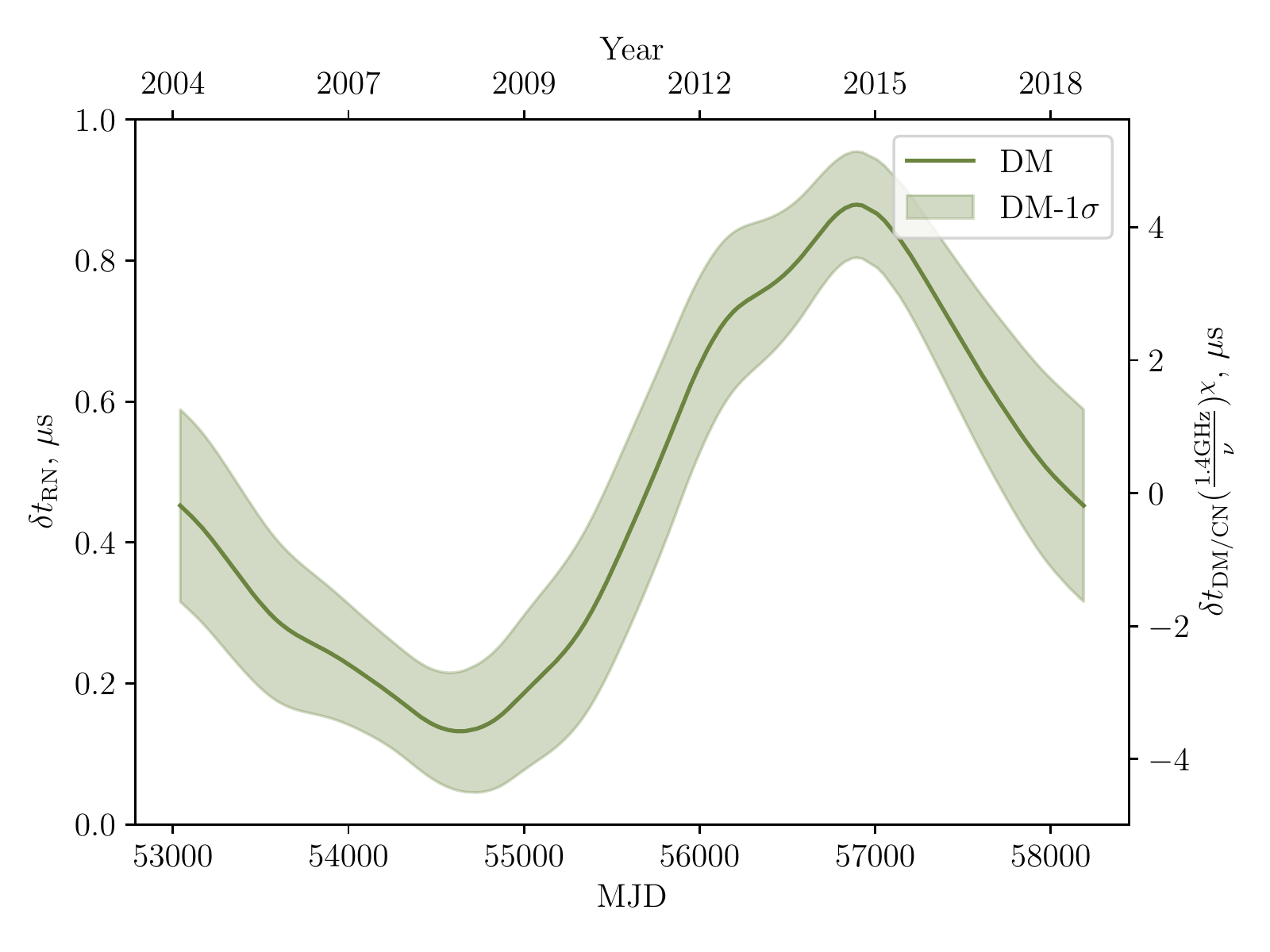}
        \caption{PSR~J1024-0719}
        \label{fig:J1024}
    \end{subfigure}
    \begin{subfigure}[b]{0.32\textwidth}
        \includegraphics[width=\textwidth]{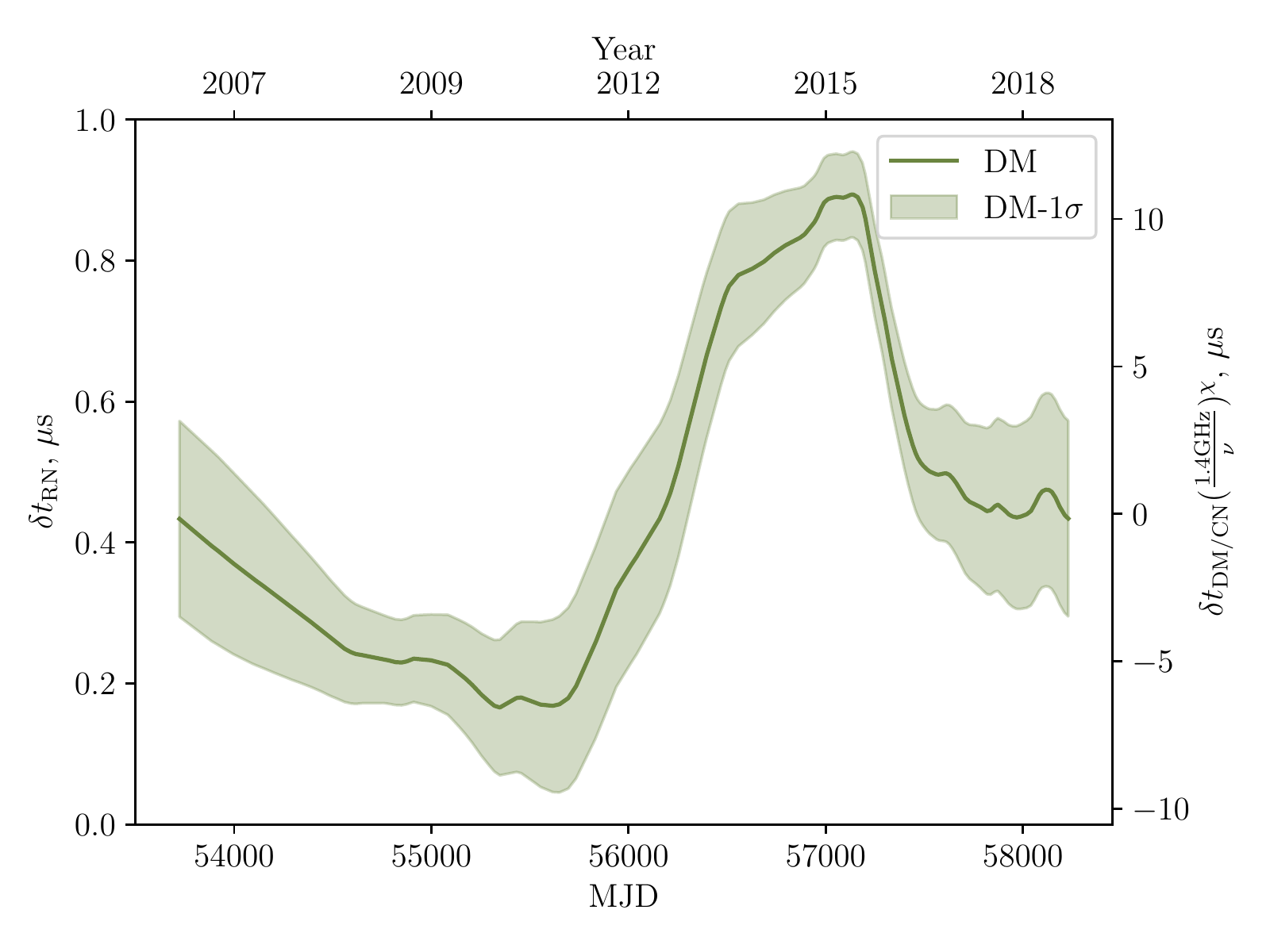}
        \caption{PSR~J1125$-$6014}
        \label{fig:J1125}
    \end{subfigure}
    \begin{subfigure}[b]{0.32\textwidth}
        \includegraphics[width=\textwidth]{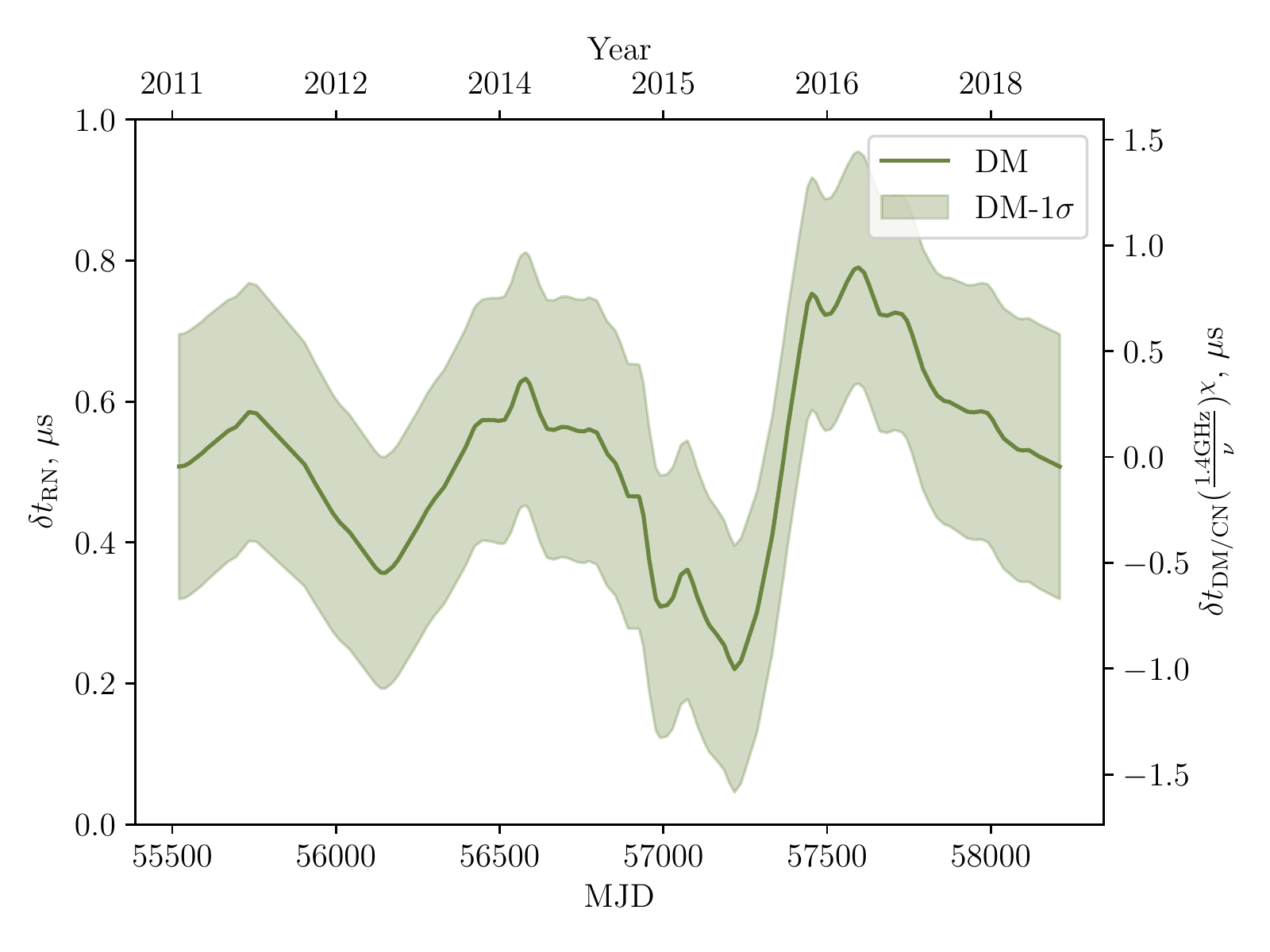}
        \caption{PSR~J1446$-$4701}
        \label{fig:J1446}
    \end{subfigure}
    \begin{subfigure}[b]{0.32\textwidth}
        \includegraphics[width=\textwidth]{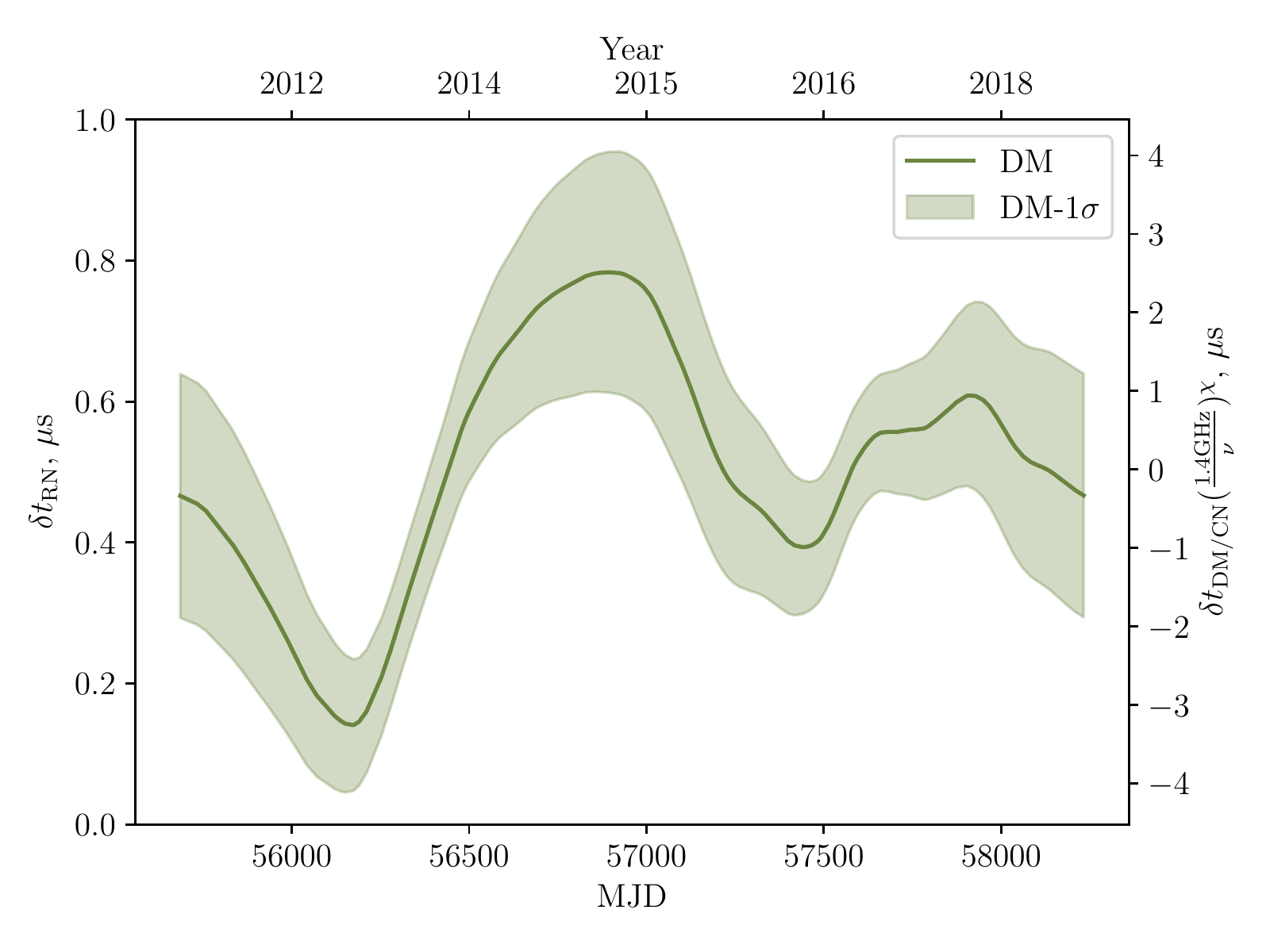}
        \caption{PSR~J1545$-$4550}
        \label{fig:J1545}
    \end{subfigure}
    \begin{subfigure}[b]{0.32\textwidth}
        \includegraphics[width=\textwidth]{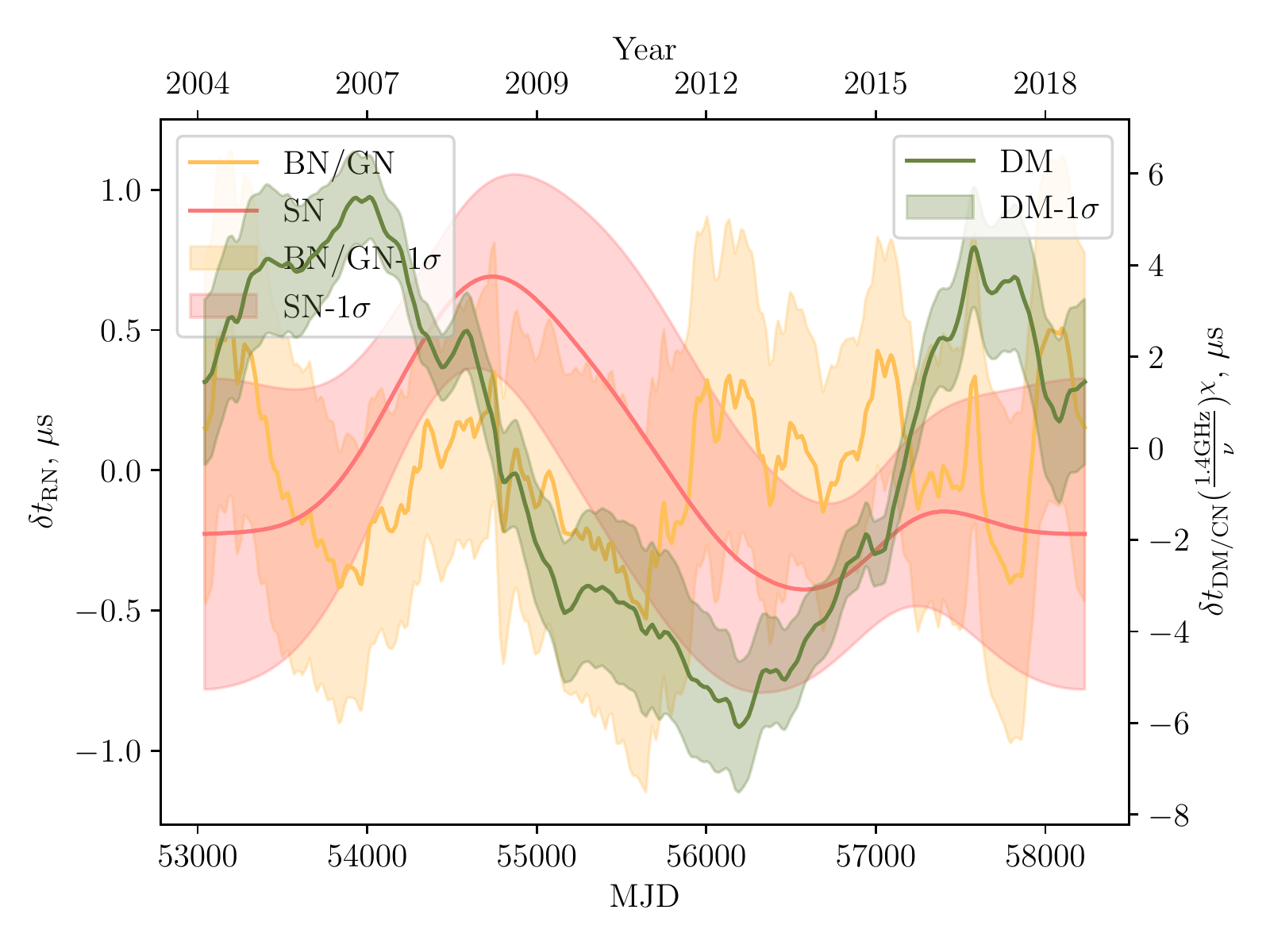}
        \caption{PSR~J1600$-$3053}
        \label{fig:J1600}
    \end{subfigure}
    \begin{subfigure}[b]{0.32\textwidth}
        \includegraphics[width=\textwidth]{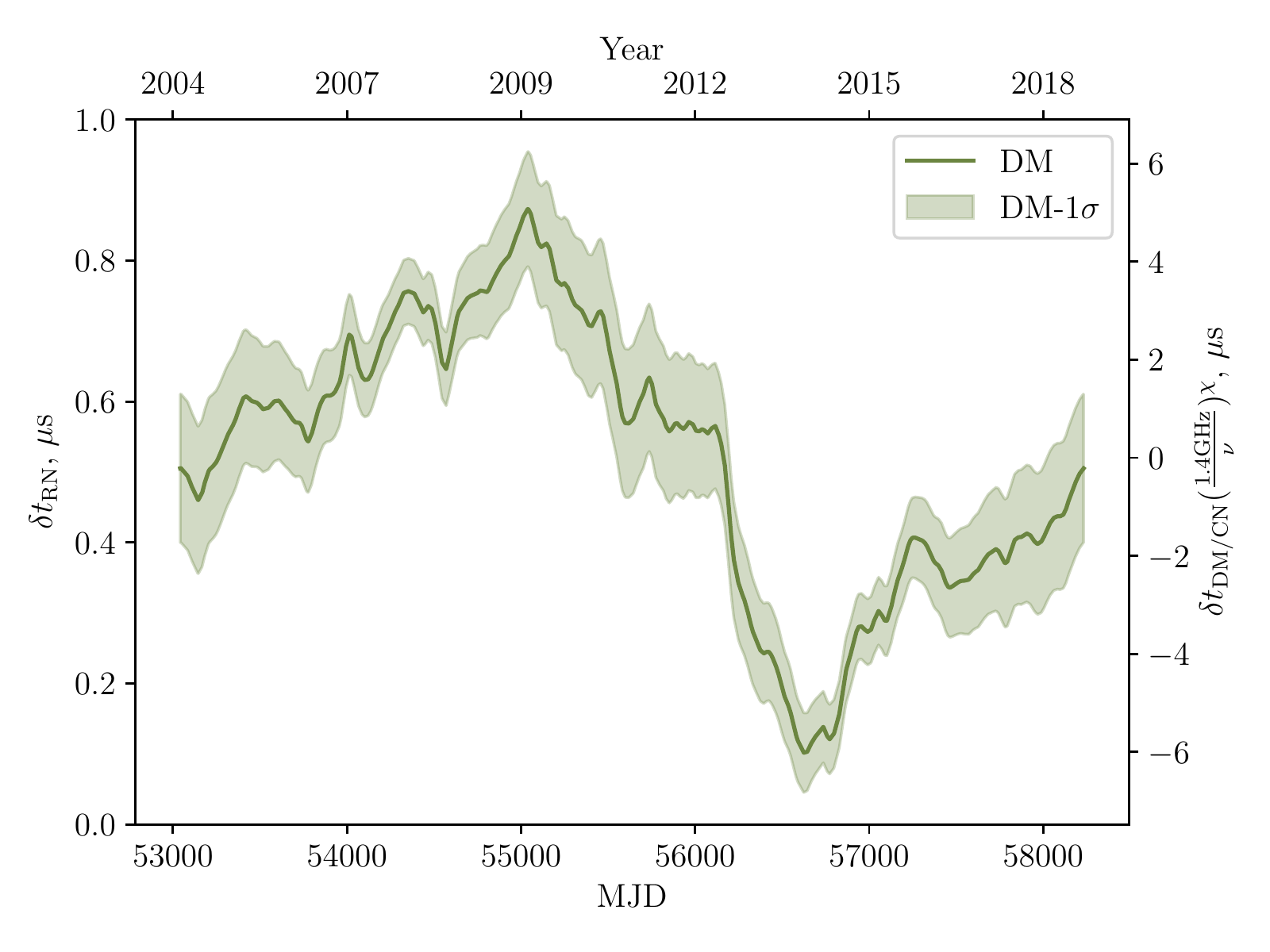}
        \caption{PSR~J1603$-$7202}
        \label{fig:J1603}
    \end{subfigure}
    \begin{subfigure}[b]{0.32\textwidth}
        \includegraphics[width=\textwidth]{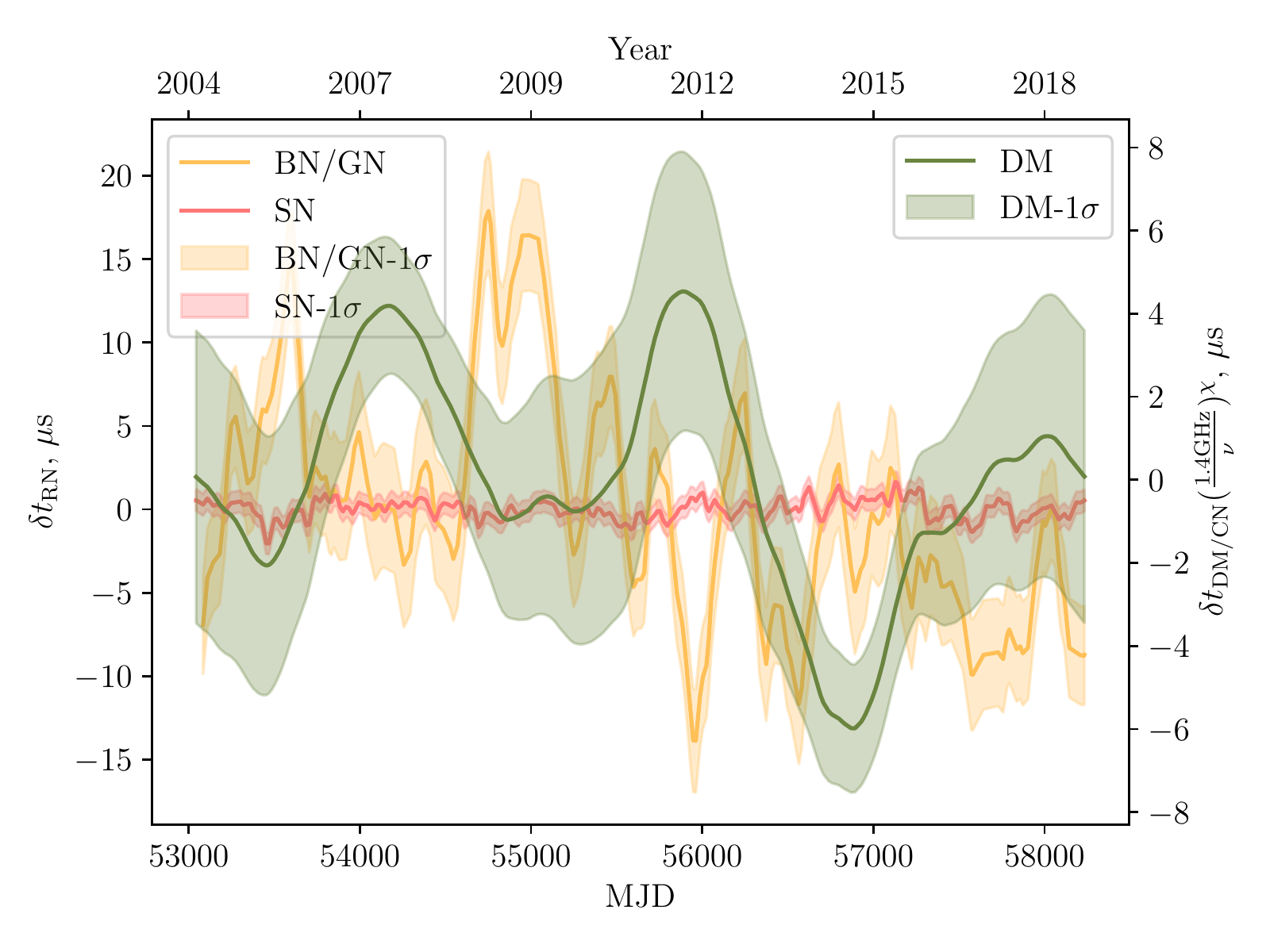}
        \caption{PSR~J1643$-$1224}
        \label{fig:J1644}
    \end{subfigure}
    \begin{subfigure}[b]{0.32\textwidth}
        \includegraphics[width=\textwidth]{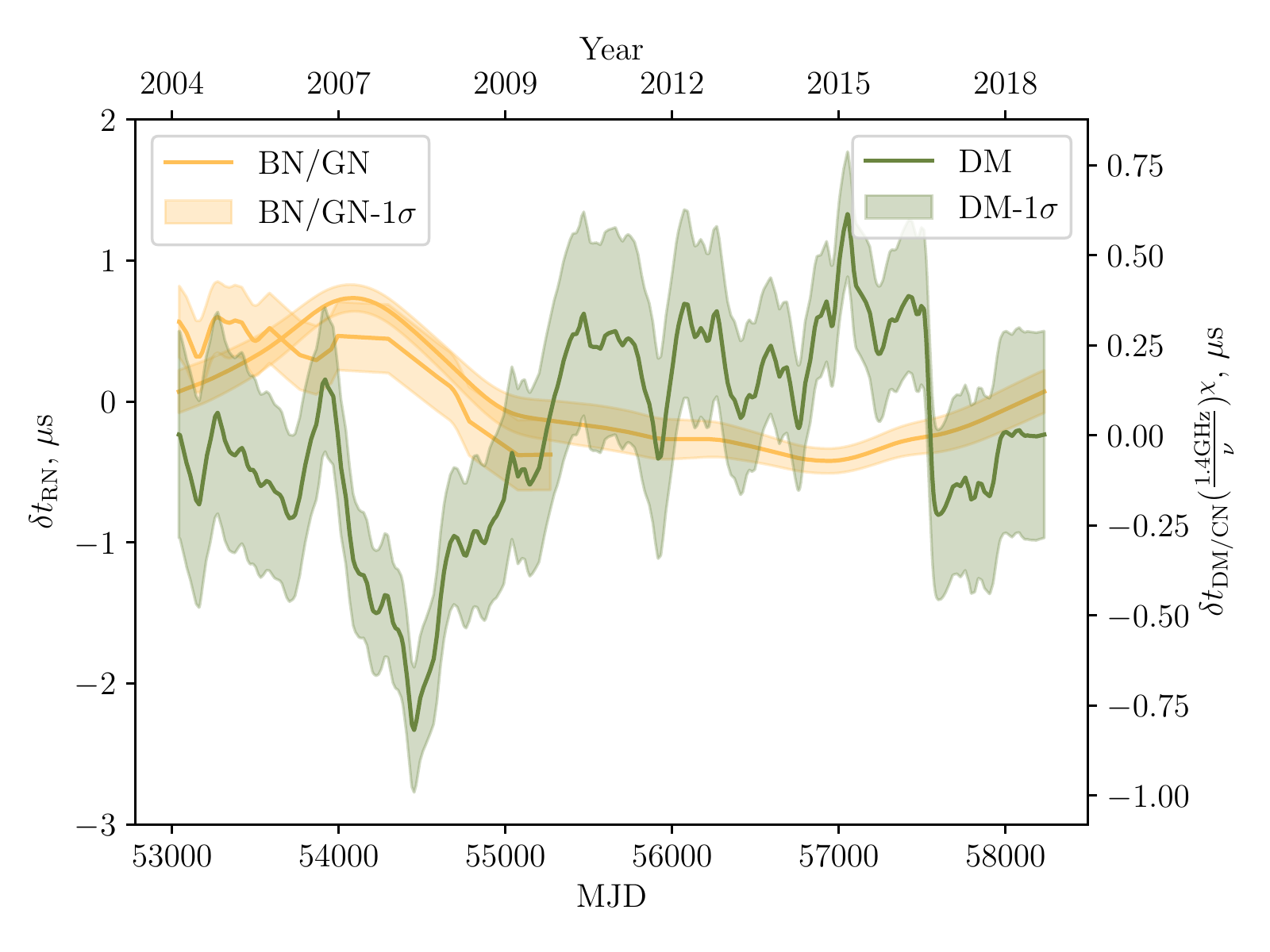}
        \caption{PSR~J1713+0747}
        \label{fig:J1713}
    \end{subfigure}
    \begin{subfigure}[b]{0.32\textwidth}
        \includegraphics[width=\textwidth]{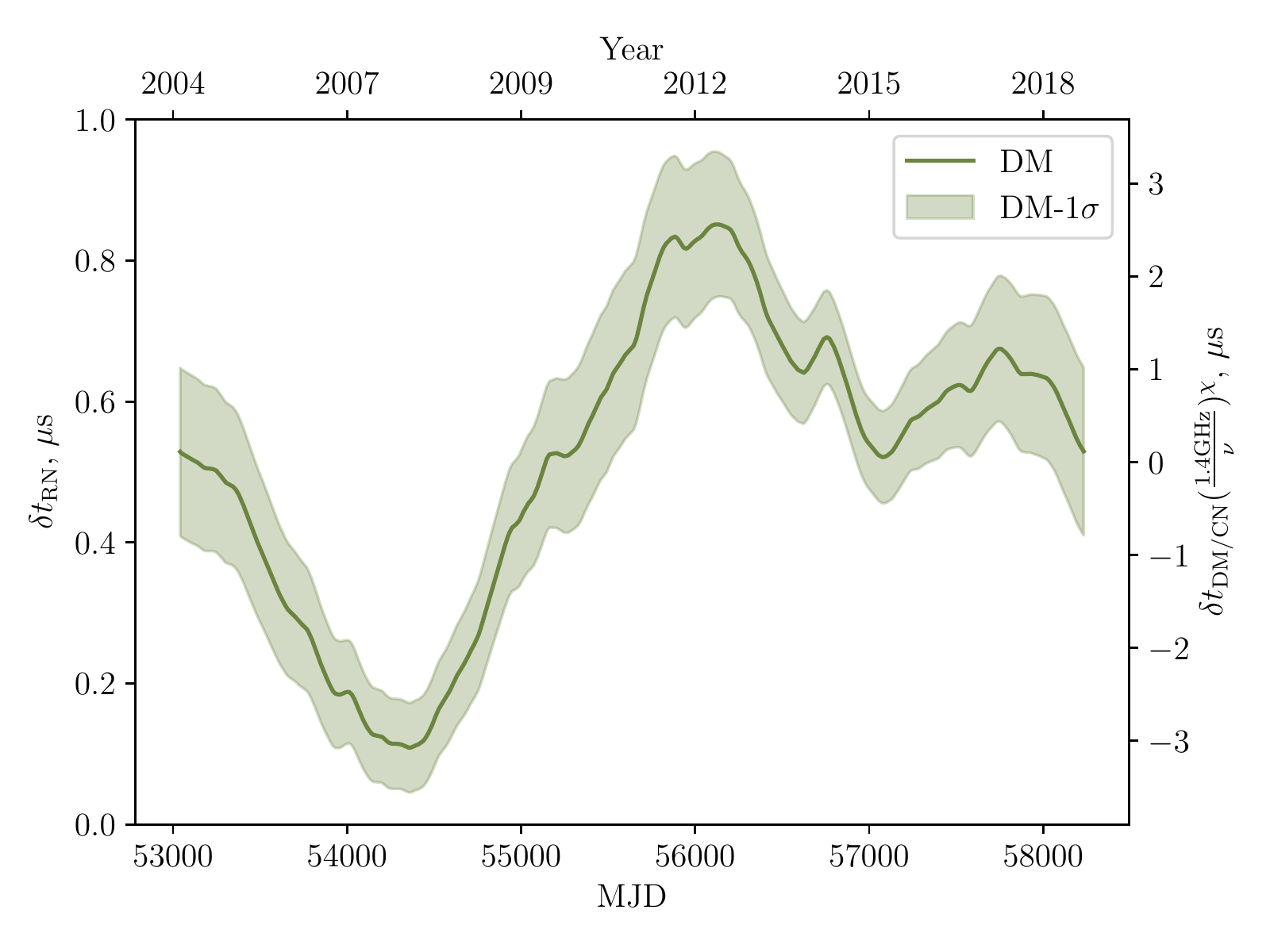}
        \caption{PSR~J1730$-$2304}
        \label{fig:J1730}
    \end{subfigure}
    \begin{subfigure}[b]{0.32\textwidth}
        \includegraphics[width=\textwidth]{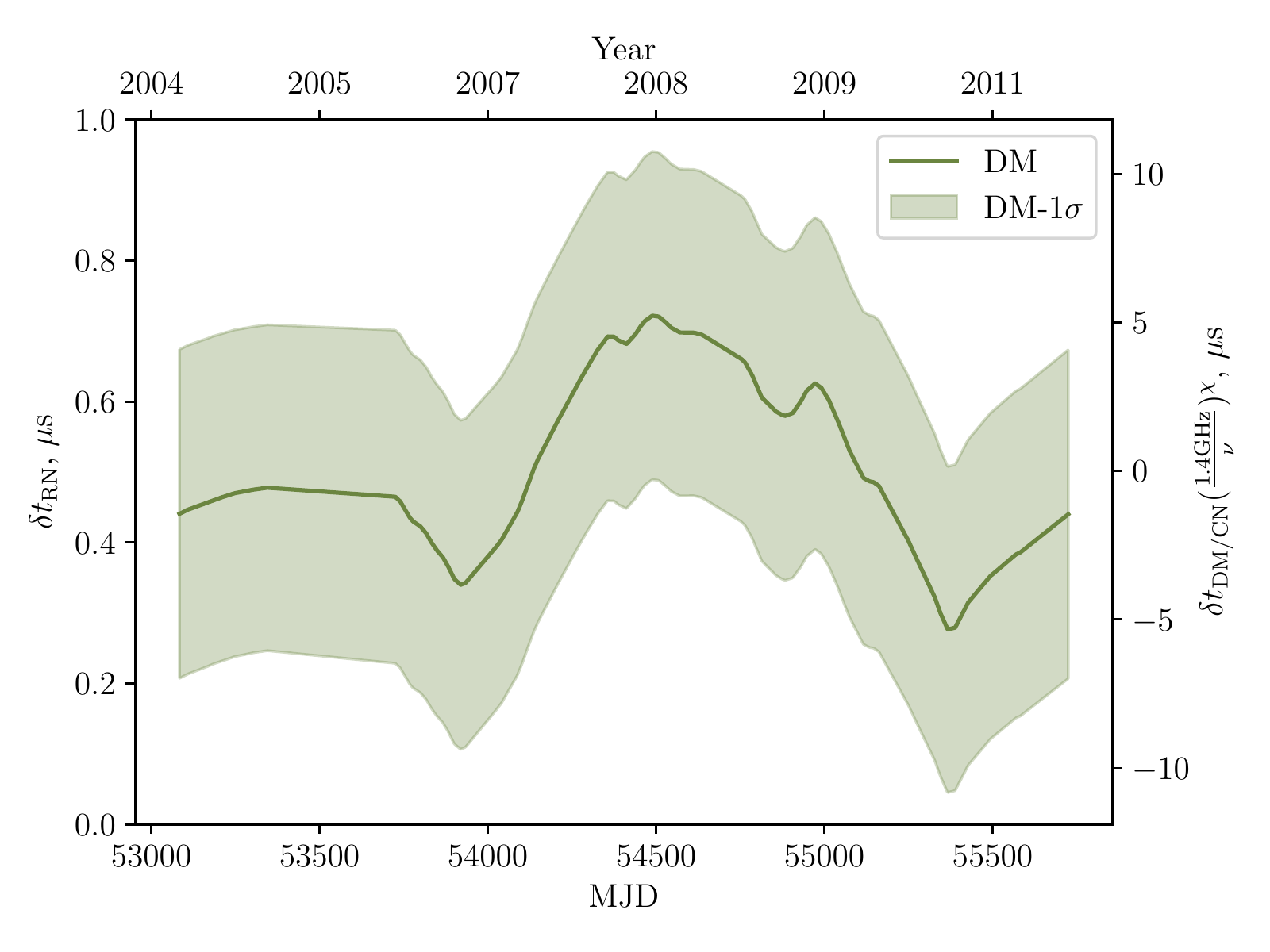}
        \caption{PSR~J1732$-$5049}
        \label{fig:J1732}
    \end{subfigure}
\end{figure*}

\begin{figure*}\ContinuedFloat
    \centering
    \begin{subfigure}[b]{0.32\textwidth}
        \includegraphics[width=\textwidth]{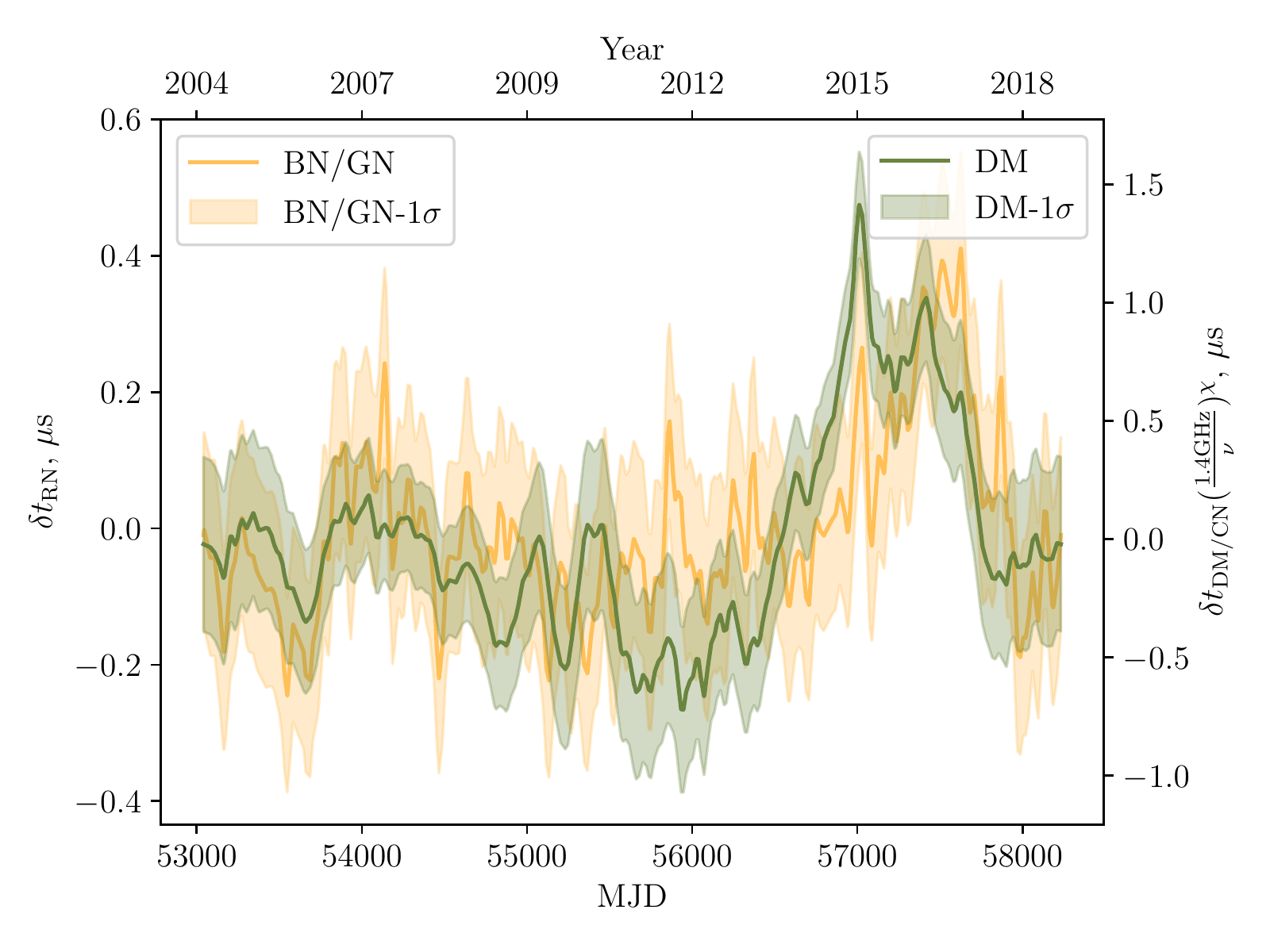}
        \caption{PSR~J1744$-$1134}
        \label{fig:J1744}
    \end{subfigure}
    \begin{subfigure}[b]{0.32\textwidth}
        \includegraphics[width=\textwidth]{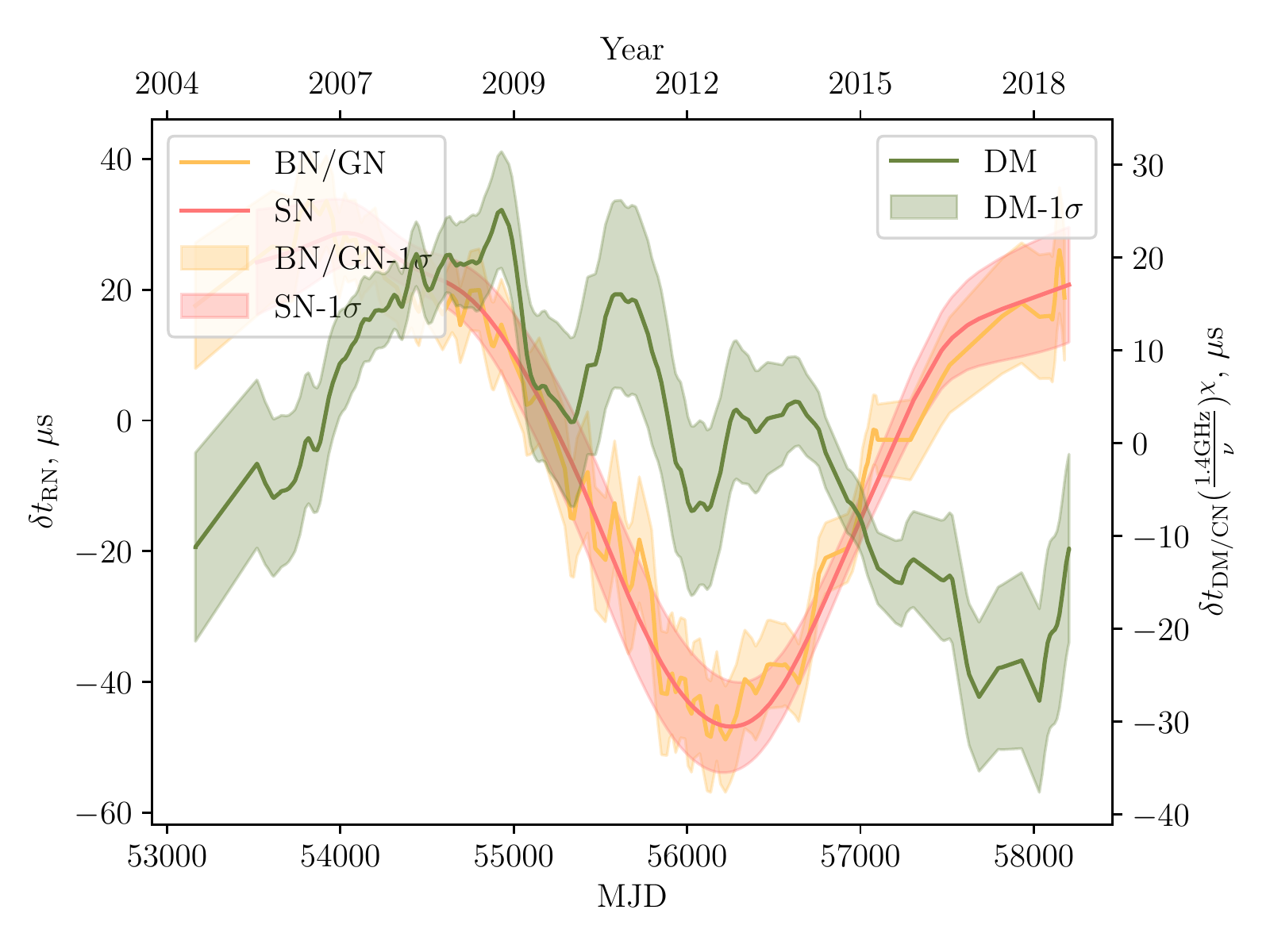}
        \caption{PSR~J1824$-$2452A}
        \label{fig:J1824}
    \end{subfigure}
    \begin{subfigure}[b]{0.32\textwidth}
        \includegraphics[width=\textwidth]{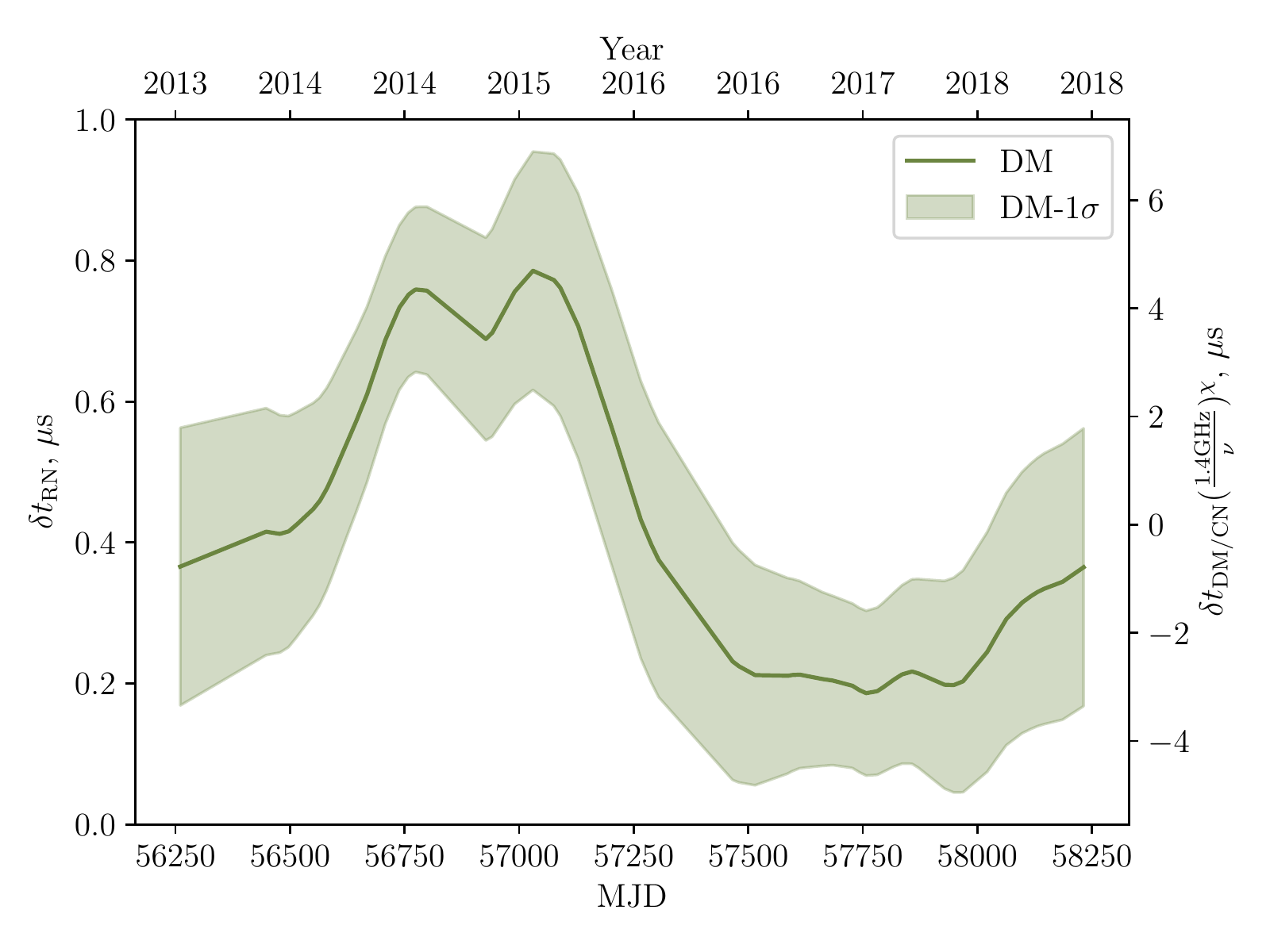}
        \caption{PSR~J1832$-$0836}
        \label{fig:J1832}
    \end{subfigure}
    \begin{subfigure}[b]{0.32\textwidth}
        \includegraphics[width=\textwidth]{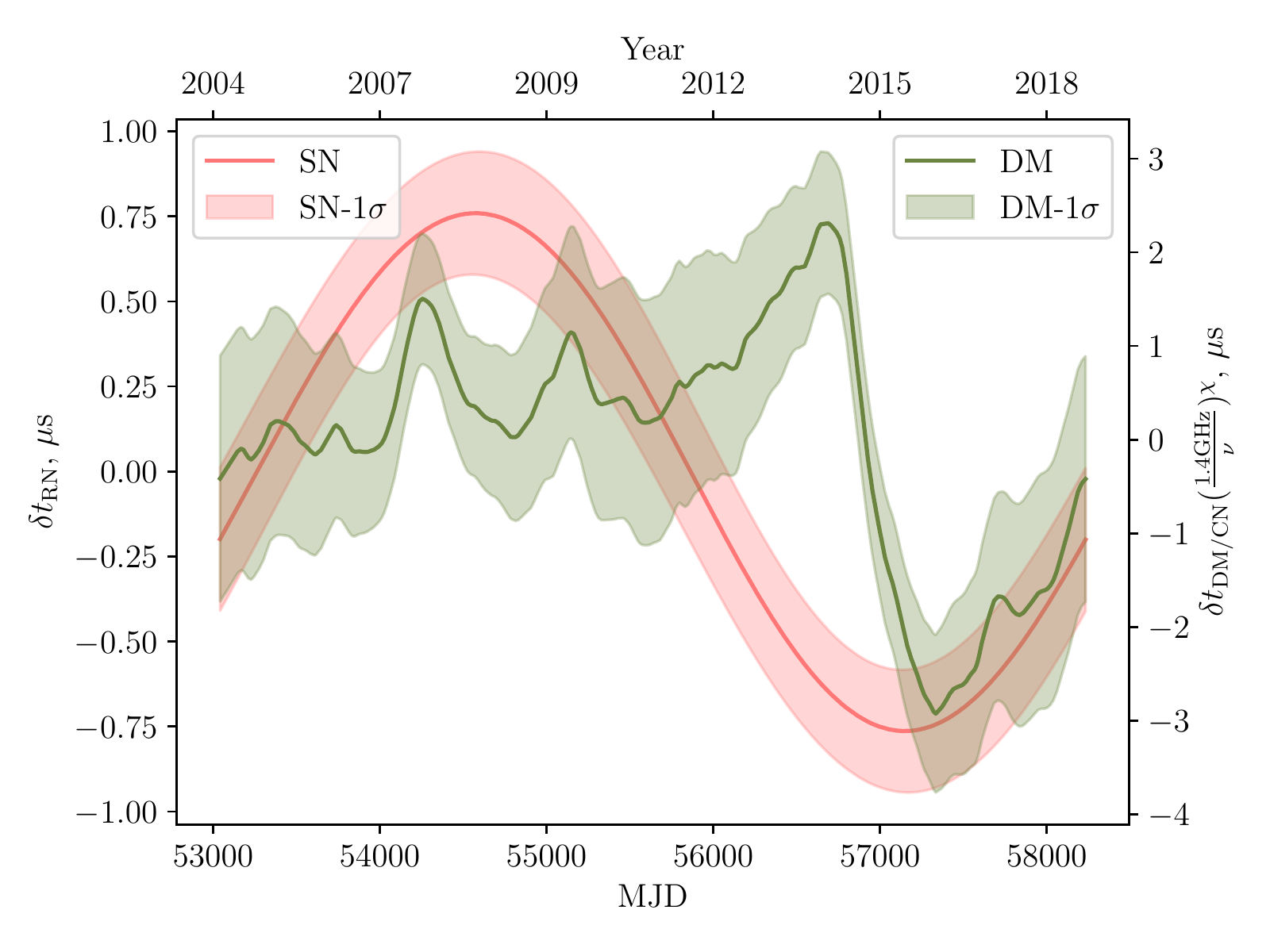}
        \caption{PSR~J1857+0943}
        \label{fig:J1857}
    \end{subfigure}
    \begin{subfigure}[b]{0.32\textwidth}
        \includegraphics[width=\textwidth]{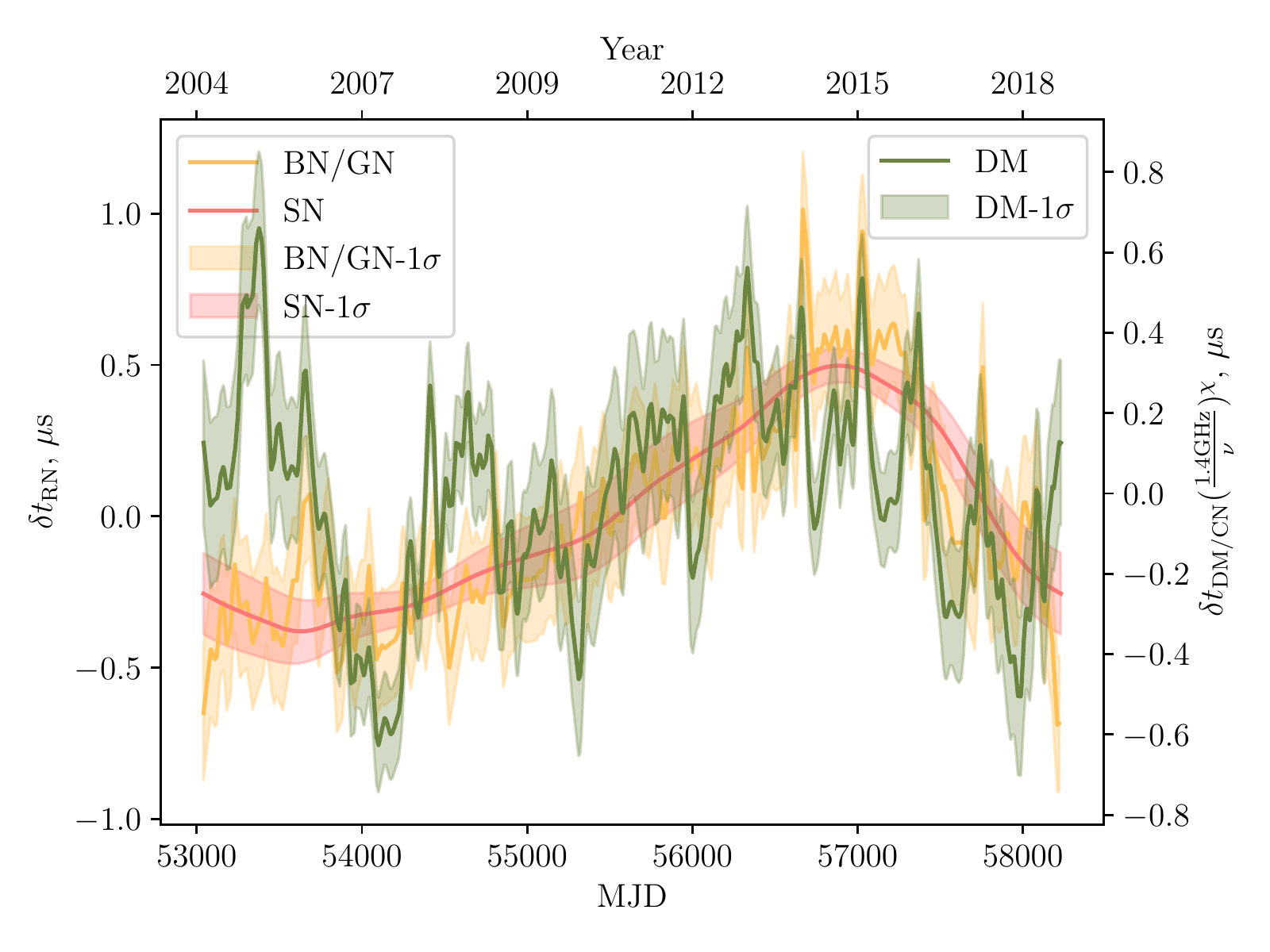}
        \caption{PSR~J1909$-$3744}
        \label{fig:J1909}
    \end{subfigure}
    \begin{subfigure}[b]{0.32\textwidth}
        \includegraphics[width=\textwidth]{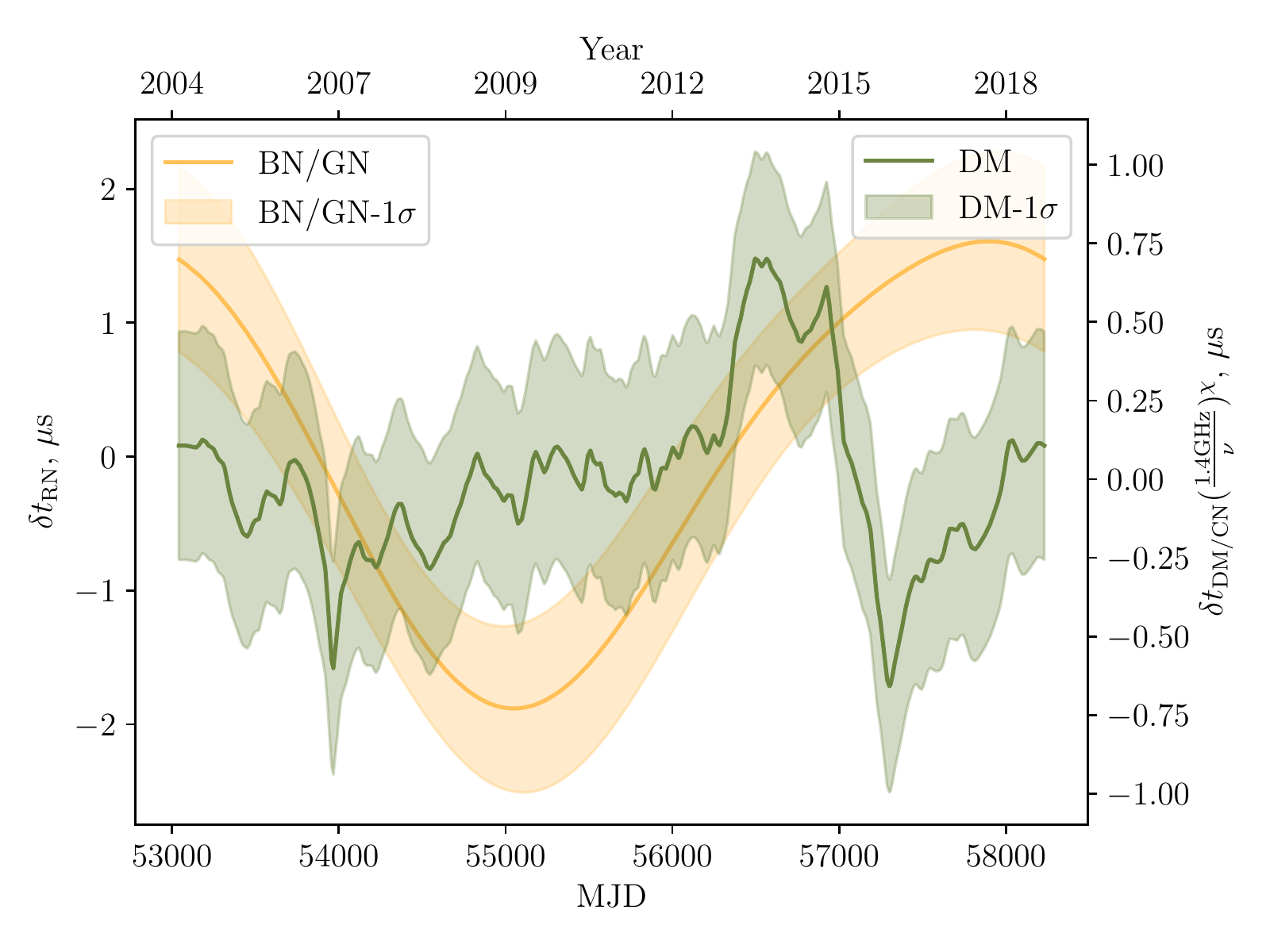}
        \caption{PSR~J2124$-$3358}
        \label{fig:J2124}
    \end{subfigure}
    \begin{subfigure}[b]{0.32\textwidth}
        \includegraphics[width=\textwidth]{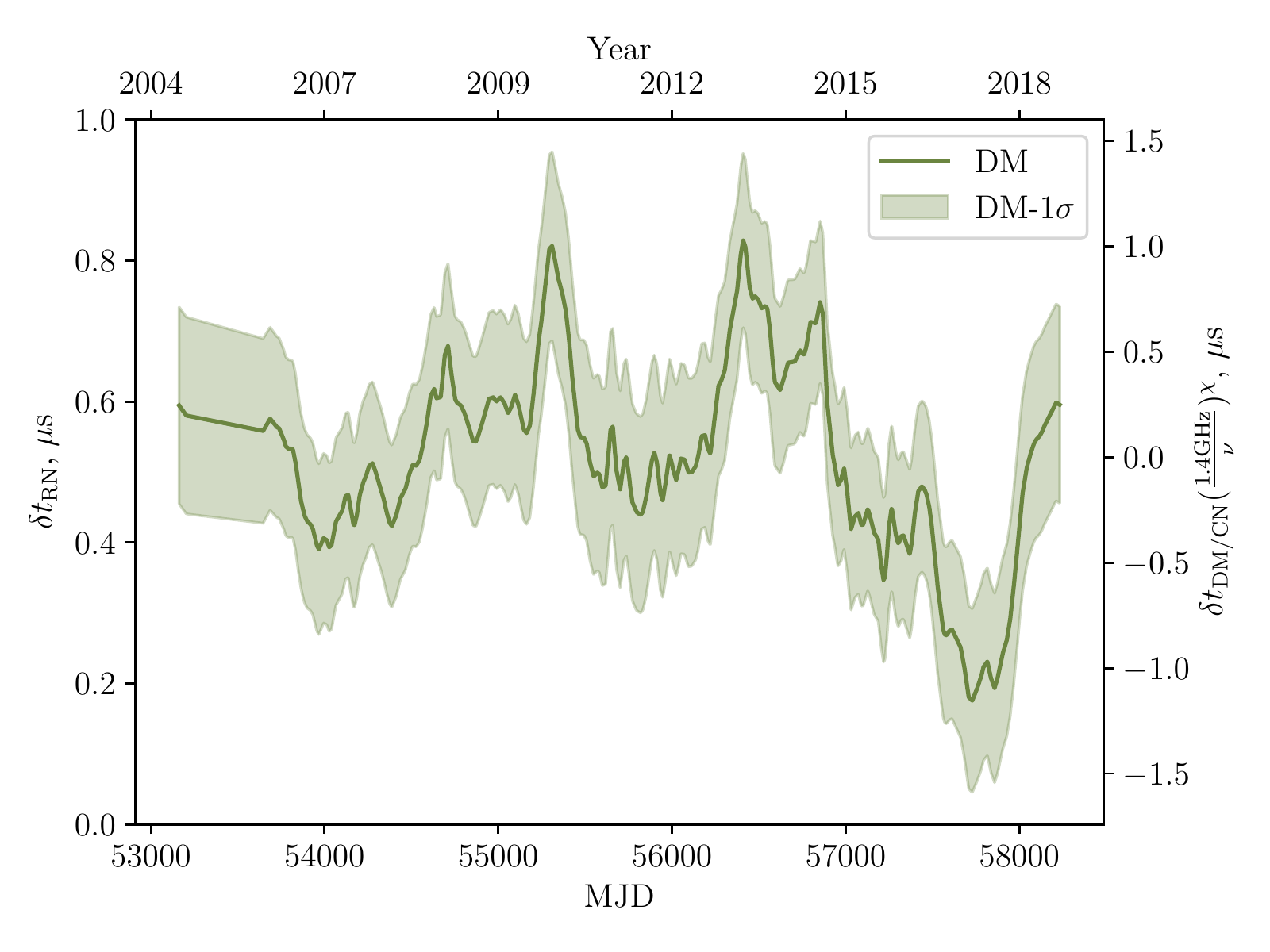}
        \caption{PSR~J2129$-$5721}
        \label{fig:J2129}
    \end{subfigure}
    \begin{subfigure}[b]{0.32\textwidth}
        \includegraphics[width=\textwidth]{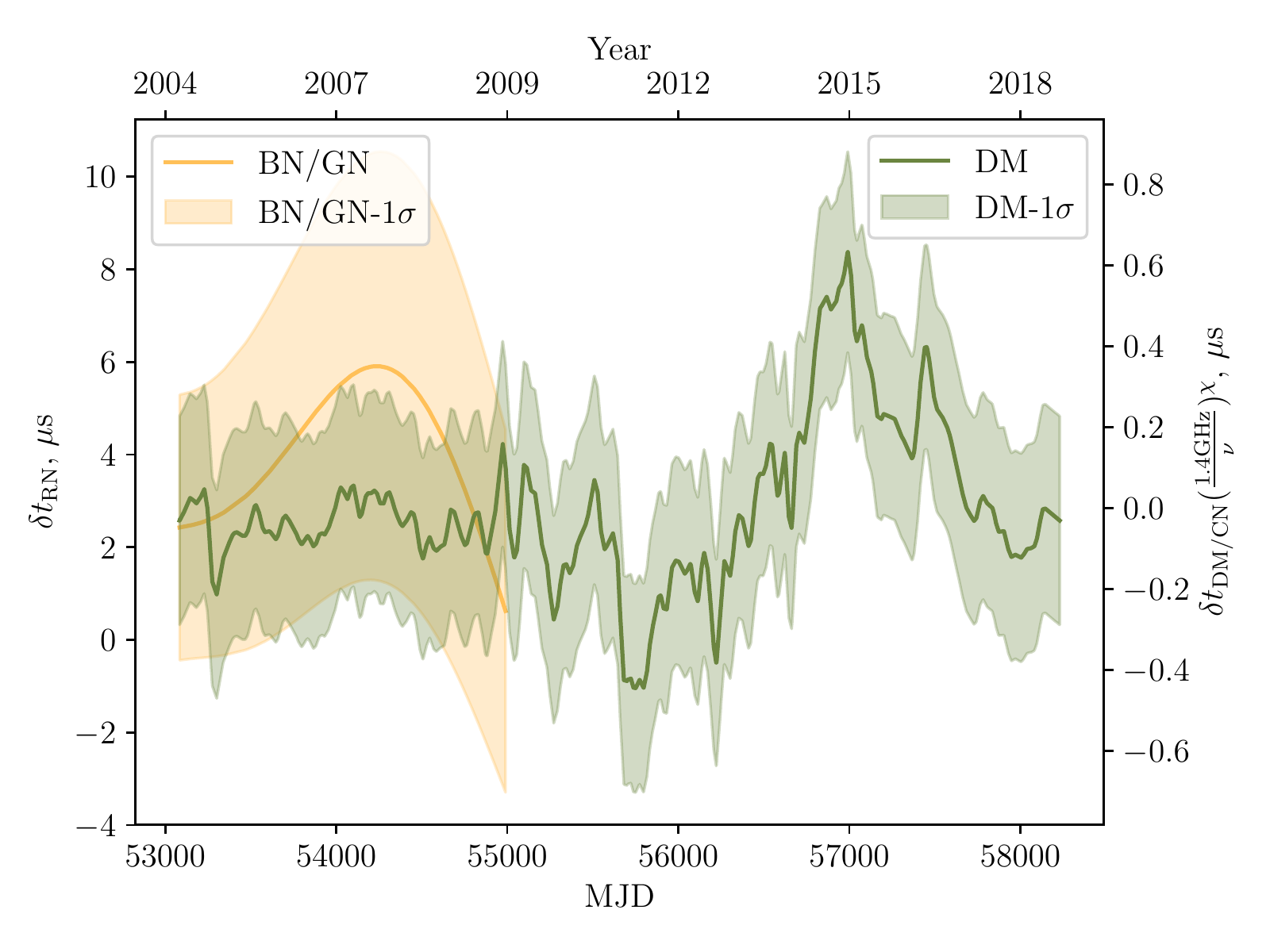}
        \caption{PSR~J2145$-$0750}
        \label{fig:J2145}
    \end{subfigure}
    \begin{subfigure}[b]{0.32\textwidth}
        \includegraphics[width=\textwidth]{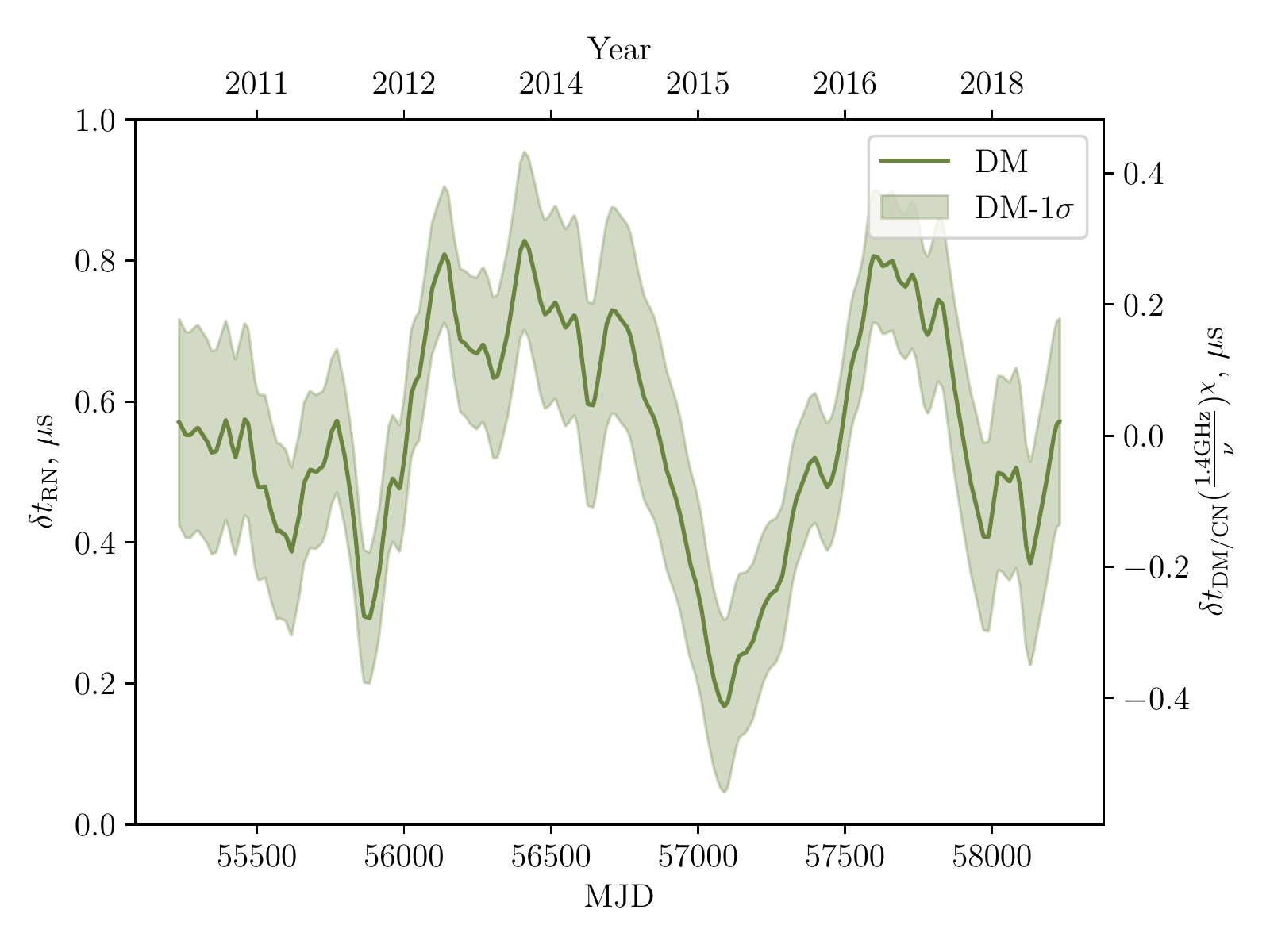}
        \caption{PSR~J2241$-$5236}
        \label{fig:J2241}
    \end{subfigure}
    \caption{Maximum-likelihood realizations of time-correlated stochastic noise in pulsars. SN is the spin noise, BN is the band noise, GN  is the system noise (group noise), DM is the stochastic dispersion measure variations, and CN is the chromatic noise. Horizontal axes determine pulse arrival time in years (top) and MJD (bottom), vertical axes determine timing residuals in $\mu$s (left) with reference to 1400 MHz (right).}
    \label{fig:noiserealizations}
\end{figure*}

\bsp	
\label{lastpage}
\end{document}